\documentclass[iop,numberedappendix]{emulateapj}
\pdfoutput=1

\usepackage[linktoc=all]{hyperref}
\usepackage{graphicx}
\usepackage{color}
\usepackage{transparent}
\usepackage{pdfpages}
\usepackage{rotating}
\usepackage[squaren, Gray, mediumspace]{SIunits}
\usepackage{MnSymbol}

\emergencystretch=\maxdimen
\slugcomment{Draft 2016/09/04}

\shorttitle{Confirmation of two $z = 2$ Structures}
\shortauthors{Noirot et al.}

\begin{document}

\title{{\it HST} Grism Confirmation of Two $\MakeLowercase{z} \sim 2$ Structures\\ from the Clusters Around Radio-Loud AGN (CARLA) Survey}

\author{Ga\"el Noirot\altaffilmark{1,2,3}, Jo\"el Vernet\altaffilmark{2}, Carlos De Breuck\altaffilmark{2}, Dominika Wylezalek\altaffilmark{4}, Audrey Galametz\altaffilmark{5}, Daniel Stern\altaffilmark{3}, Simona Mei\altaffilmark{1,6,7}, Mark Brodwin\altaffilmark{8}, Elizabeth A. Cooke\altaffilmark{9}, Anthony H. Gonzalez\altaffilmark{10}, Nina A. Hatch\altaffilmark{9}, Alessandro Rettura\altaffilmark{11}, Spencer Adam Stanford\altaffilmark{12}
}

\altaffiltext{1}{Universit\'e Paris-Diderot Paris VII, Universit\'e de Paris Sorbonne Cit\'e (PSC), 75205 Paris Cedex 13, France}
\altaffiltext{2}{European Southern Observatory, Karl-Schwarzschildstrasse 2, 85748 Garching, Germany}
\altaffiltext{3}{Jet Propulsion Laboratory, California Institute of Technology, 4800 Oak Grove Drive, Pasadena, CA 91109, USA}
\altaffiltext{4}{Johns Hopkins University, Zanvyl Krieger School of Arts \& Sciences, 3400 N. Charles St, Baltimore, MD 21218, USA}
\altaffiltext{5}{Max-Planck-Institut f\"ur extraterrestrische Physik, Giessenbachstrasse 1, 85748 Garching, Germany}
\altaffiltext{6}{GEPI, Observatoire de Paris, PSL Research University,  CNRS, University of Paris Diderot, 61, Avenue de l'Observatoire, 75014 Paris, France}
\altaffiltext{7}{California Institute of Technology, Pasadena, CA 91125, USA}
\altaffiltext{8}{Department of Physics, University of Missouri, 5110 Rockhill Road, Kansas City, MO 64110, USA}
\altaffiltext{9}{School of Physics and Astronomy, University of Nottingham, University Park, Nottingham NG7 2RD, UK}
\altaffiltext{10}{Department of Astronomy, University of Florida, Gainesville, FL 32611-2055, USA}
\altaffiltext{11}{Infrared Processing and Analysis Center, California Institute of Technology, KS 314-6, Pasadena, CA 91125, USA}
\altaffiltext{12}{Department of Physics, University of California, One Shields Avenue, Davis, CA 95616, USA}

\begin{abstract}
Using {\it HST} slitless grism data, we report the spectroscopic confirmation of two distant structures at $z \sim 2$ associated with powerful high-redshift radio-loud AGN. These rich structures, likely (forming) clusters, are among the most distant currently known and were identified on the basis of {\it Spitzer}/IRAC $[3.6] - [4.5]$ color. We spectroscopically confirm 9 members in the field of MRC~2036$-$254, comprising eight star-forming galaxies and the targeted radio galaxy. The median redshift is $z=2.000$. We spectroscopically confirm 10 members in the field of B3~0756+406, comprising eight star-forming galaxies and two AGN, including the targeted radio-loud quasar. The median redshift is $z = 1.986$. All confirmed members are within $500$ kpc ($1$ arcmin) of the targeted AGN. We derive median (mean) star-formation rates of $\sim 35~M_{\odot}\rm ~ yr^{-1}$ ($\sim 50~M_{\odot}\rm ~ yr^{-1}$) for the confirmed star-forming members of both structures based on their [\ion{O}{3}]$\lambda5007$ luminosities, and estimate average galaxy stellar masses $\la 1 \times 10^{11} ~M_{\odot}$ based on mid-infrared fluxes and SED modeling. Most of our confirmed members are located above the star-forming main-sequence towards starburst galaxies, consistent with clusters at these early epochs being the sites of significant levels of star formation.
The structure around MRC~2036$-$254 shows an overdensity of IRAC-selected candidate galaxy cluster members consistent with being quiescent galaxies, while the structure around B3~0756+406 shows field values, albeit with many lower limits to colors that could allow an overdensity of faint red quiescent galaxies. The structure around  MRC~2036$-$254 shows a red sequence of passive galaxy candidates.\\
\end{abstract}


\section{Introduction}\label{sec:intro}

At low to intermediate redshifts ($z \la 1.4$), massive early-type galaxies dominate galaxy cluster cores and form a tight red sequence (e.g., \citealp{Lidman08}, \citealp{Mei09}). The few studies at higher redshifts suggest that clusters at $z>1.5$ are still in the process of forming (\citealp{Snyder12}, \citealp{Mei15}), and that although clusters at these redshifts show a mixed population of both star-forming (SF) and quiescent galaxies, even the reddest (early-type) galaxies show on-­going star-formation (\citealp{Mei15}). Star-formation activity has also been observed in the cores of massive galaxy clusters at $z > 1.4$ (e.g., \citealp{Tran10},  \citealp{Hayashi10}, \citealp{Zeimann13}, \citealp{Alberts14}, \citealp{Bayliss14}). For example, based on a sample of 16 spectroscopically confirmed clusters at $1 < z < 1.5$, \cite{Brodwin13} showed that at $z > 1.3$ the fraction of SF cluster members increases towards the cluster centers. These results suggest that the majority of star formation actually occurs in high-density environments at early epochs, implying that environmental-dependent quenching has not yet been established at $z > 1.3$. \cite{Brodwin13} predicted that this transition redshift should be a function of halo mass, with more massive halos transitioning earlier. This is consistent with the findings of \cite{Wylezalek14} of a $z\sim3$ transition period for clusters around radio-loud active galactic nuclei (RLAGN), which are extreme objects that tend to reside in the most massive dark matter halos (e.g., \citealp{Mandelbaum09}, \citealp{Hatch14}, \citealp{Orsi16}).

To better understand the dependence of the formation mechanisms of massive galaxies on environment, we must focus on clusters at the relatively unexplored redshift range $z > 1.5$ where major assembly is in progress (e.g., \citealp{Mancone10}). Various selection methods are used to find (proto)cluster candidates, e.g., the red sequence (\citealp{GladdersYee00}, \citealp{Rykoff14}, \citealp{Licitra16}), a mid-infrared adaptation of the red sequence (\citealp{Muzzin13}, \citealp{Webb15}), photometric redshifts of infrared-selected samples (\citealp{Eisenhardt08}, \citealp{Stanford12}, \citealp{Zeimann12}), {\it Spitzer}/IRAC color selection (\citealp{Papovich08}, \citealp{Rettura14}), overdensities of sub-millimeter sources (\citealp{Smail14}, \citealp{Planck15}), X-ray emission (\citealp{Rosati98}, \citealp{Tozzi15}), and the Sunyaev Zel'dovich (SZ) effect (\citealp{Vanderlinde10}, \citealp{Bleem15}). These methods mostly rely on wide-field surveys. Discovering larger samples of galaxy clusters at high redshifts using these techniques therefore requires prohibitive amounts of telescope time over yet wider areas. Moreover, X-ray detections are limited by the surface brightness of the sources\footnote{With the caveat that based on the low-redshift \cite{Vikhlinin09} scaling relations, \cite{Churazov15} showed that at higher redshifts ($z \simeq 1-2$), clusters as massive as $z\sim0$ clusters should be as easily detectable.} dimming as $(1 + z)^4$. Both X-ray and SZ selections are also only able to detect very massive structures via their hot intra-cluster medium, which requires mature, collapsed clusters. Additionally, AGN activity increases for higher redshift clusters (e.g., \citealp{Galametz10}, \citealp{Martini13}), adding a complication for X-ray and SZ selections. These issues all conspire against finding galaxy clusters at high redshifts, although \cite{Mantz14} recently reported a massive cluster candidate ($M_{500}\sim(1 - 2)\times 10^{14}~M_{\odot}$) at $z\simeq1.9$ via a weak X-ray detection, the SZ decrement and photometric redshifts. Currently, only about ten galaxy clusters have been spectroscopically confirmed at $z > 1.5$ (e.g., \citealp{Papovich10} -- independently reported in \citealp{Tanaka10}, \citealp{Stanford12}, \citealp{Zeimann12}, \citealp{Gobat13}, \citealp{Muzzin13}, \citealp{Newman14}, \citealp{Mei15}).
All these confirmed clusters are at $z \leq 2.0$. A few confirmed clusters and cluster candidates at $1.5 < z < 2$ have significant X-ray detections, implying they are likely virialized (e.g., \citealp{Santos11}, \citealp{Mantz14}, \citealp{Newman14}). This relatively small number of high-redshift confirmed clusters makes it challenging to draw a clear picture of cluster formation and evolution.

Powerful high-redshift RLAGN are known to preferentially lie in overdense fields (with literature stretching back more than 50 years; e.g., \citealp{Matthews64}) and are efficient beacons for identifying large-scale structures and (proto)clusters. Indeed, targeted searches around RLAGN are a proven technique for identifying galaxy clusters at high redshifts (e.g., \citealp{Stern03}, \citealp{Venemans07}, \citealp{Galametz10}, \citealp{Hatch11}). Our team has made a major contribution to this effort with a targeted 400-hour {\it Warm} {\it Spitzer Space Telescope} program surveying 420 radio-loud AGN at $1.3 < z < 3.2$ across the full sky: Clusters Around Radio-Loud AGN (CARLA, \citealp{Wylezalek13, Wylezalek14}). Using a simple mid-infrared color selection technique, we successfully identified nearly 200 promising cluster candidates at $z > 1.3$.

Ground-based observations are challenging for spectroscopically confirming high-redshift galaxy clusters because of atmospheric absorption and emission. In contrast, {\it Hubble Space Telescope} ({\it HST}) infrared spectra obtained with the Wide-Field Camera 3 (WFC3) slitless grism are free from atmospheric constraints and are thus ideal for obtaining spectra of high-redshift galaxies, albeit with a low dispersion ($46.5$ \AA/pix$^{-1}$) and a low resolving power ($ R = \lambda / \Delta \lambda = 130$; numbers for G141 grism, \citealp{Dressel14}).

Following up on the {\it Spitzer}/CARLA survey, our team is using the WFC3 slitless G141 grism to study our $20$ densest cluster candidates at $1.4 \leq z \leq 2.8$. This paper presents early results confirming structures around MRC 2036$-$254 and B3~0756+406, two of the first fields to have their {\it HST} observations completed.
Previous papers from the CARLA project include ground-based spectroscopic confirmation of two (proto)clusters, reported in \cite{Galametz13} and Rettura et al. (in prep.), ground-based imaging to study the
formation histories of CARLA clusters, reported in \cite{Cooke15a, Cooke16}, and a comparison of mass-matched samples of radio-loud and
radio-quiet galaxies at $z > 1.3$, showing that RLAGN indeed reside in
significantly denser environments (\citealp{Hatch14}).

We adopt here the \cite{Eisenhardt08} criteria of $z>1$ spectroscopic cluster confirmation: at least five galaxies within a physical radius of 2 Mpc whose spectroscopic redshifts are confined to within $\pm 2000(1+\left<z_{\rm spec}\right>) \rm~km~\sec^{-1}$. A significant concern is that this definition alone may also identify groups, protoclusters, sheets and filaments when applied to grism data. In the lower redshift universe, a more exacting definition for a confirmed galaxy cluster typically also requires: {\it (i)} detection of an extended X-ray emitting diffuse intracluster medium, {\it (ii)} a significant population of early-type (i.e., passive) galaxies, and {\it (iii)} a centrally concentrated distribution of galaxies. Current literature will often forego the first requirement, particularly for distant clusters, due to the challenges of acquiring such data --- and this is particularly problematic for our RLAGN targets due to Inverse Compton scattering of the cosmic microwave background by the hot plasma associated with AGN radio lobes into the X-ray regime. Nonetheless, we show that  
some CARLA systems have clear overdensities of passive galaxy candidates (see also \citealp{Cooke15a, Cooke16}, in prep.), and \citet{Wylezalek13} show that the CARLA cluster member candidates are, on average, centrally concentrated around the target RLAGN.

This paper is organized as follows. Section \ref{sec:obs} briefly presents the CARLA sample and the overall strategy of our {\it HST}/CARLA program. Section \ref{sec:red} presents the analysis strategy carried out on the {\it HST} data, including detection limits. Section \ref{sec:res} presents the results, including cluster membership, star-formation rates (SFRs), and stellar masses. In Section \ref{s:discussion} we compare the {\it HST} results with the CARLA selection method and discuss the results. We summarize our work in Section \ref{sec:con}. We also present details on the data analysis in Appendix \ref{ap:notesdataprep}. Appendices \ref{ap:membprop} and \ref{ap:nonmembprop} list the line properties of structure members and non-members, respectively, and we present notes on individual sources in Appendix \ref{ap:clmemb}. Throughout, all magnitudes are expressed in the AB photometric system, and we use a flat $\Lambda$CDM cosmology with $H_{0} = 70 ~\rm km ~Mpc^{-1}~ s^{-1}$, $\Omega_{\Lambda} = 0.7$, and $\Omega_{m} = 0.3$.\\

\section{Observations}\label{sec:obs}

\subsection{The CARLA Sample}

The CARLA sample consists of {\it Spitzer}/IRAC channel $1$ and $2$ (3.6 and 4.5 \micron~bands, respectively) observations of 420 fields around powerful RLAGN obtained during a 400-hour {\it Warm} {\it Spitzer} Cycle 7 and 8 snapshot program. Each image covers an area of $5.2 \times 5.2$ arc$\min^2$ with an original sampling of $1.22$ arc$\sec \, \rm pix^{-1}$ reprocessed to $0.61$ arc$\sec \, \rm pix^{-1}$. The data reach 95\% completeness at $22.8$ mag and $22.9$ mag for the $3.6 \micron$ and $4.5 \micron$ bands, respectively. The program imaged the fields of 209 high-redshift radio galaxies (HzRGs) and 211 radio-loud quasars (RLQs), uniformly selected in radio-luminosity over the redshift range $1.3 < z < 3.2$ (with rest-frame $L_{500{\rm MHz}} \geq 10^{27.5}$ W~Hz$^{-1}$). The HzRGs (type-$2$ RLAGN) were selected from the updated compendium of \cite{MileyDeBreuck08}, and the RLQs (type-$1$ RLAGN) were selected from the Sloan Digital Sky Survey (SDSS, \citealp{Schneider10}) and 2dF QSO Redshift Survey (2QZ, \citealp{Croom04}). Galaxy cluster candidates were then identified as IRAC color-selected galaxy overdensities in the fields of the targeted RLAGN. We applied the color cut $([3.6]-[4.5])_{\rm AB} > -0.1~ \rm mag$, which reliably identifies $z > 1.3$ sources (\citealp{Papovich08}). Galaxy cluster candidates are defined as fields showing a $> 2\sigma$ overdensity of IRAC color-selected sources compared to the blank field surface density of similarly selected sources, as measured in the {\it Spitzer} UKIDSS Ultra-Deep Survey (SpUDS; P.I. Dunlop). We measured the density of IRAC color-selected sources within $1\arcmin$ radius apertures centered on the RLAGN ($\sim 500$ kpc at $z \sim 2$). We identified 178 cluster candidates among the 420 fields. The galaxy density of these cluster candidates strongly peaks towards the position of the targeted RLAGN, suggesting the clusters are reliable and that the RLAGN reside at the cluster cores. Detailed descriptions of the sample and initial scientific results are presented in \cite{Wylezalek13} and \cite{Wylezalek14}.\\

\begin{table}
\caption{{\it HST} WFC3 observations}\smallskip
\label{table:obs}
\centering
\begin{tabular}{ lccc }
\hline \hline
\multicolumn{1}{p{1.25cm}}{Target}& \multicolumn{1}{p{1.55cm}}{\centering UT Date}& \multicolumn{1}{p{1.3cm}}{\centering Position angle\tablenotemark{a} ($\deg$)}&  \multicolumn{1}{p{1.7cm}}{\centering F140W/G141 exp. time ($\sec$)}\\ \hline 
MRC 2036$-$254&               	  2014 Oct 14& -62&        512/2012\\
\mbox{}& 		  	  	  2014 Oct 15& -39&   512/2012\\
\hline
B3~0756+406& 		   	  2014 Nov 03& +125&       537/2062\\
\mbox{}& 		   	 	  2014 Nov 07& +162&  537/2062\\
\hline
\end{tabular}
\tablenotetext{1}{East of north.}
\end{table}

\subsection{The CARLA {\it HST} Program}

Among the 178 CARLA cluster candidates, we selected the 20 richest fields as the most promising targets for additional study. These fields are $5.8\sigma$ to $9.0 \sigma$ overdense above the mean SpUDS density and the associated RLAGN at the center of each field cover the redshift range $1.4 < z < 2.8$. Ten of the twenty fields are associated with HzRGs and the other ten with RLQs. Our team was awarded 40 {\it HST} orbits in a Cycle 22 WFC3 program to image and obtain spectroscopy of these 20 CARLA fields with the primary goal of spectroscopically confirming the cluster candidates (Program ID: 13740). Each field is visited twice using different orientations to mitigate contamination from overlapping spectra. The first field was observed in October 2014 and the whole program was completed in April 2016. For each visit, our program obtains $0.5$ k$\sec$ F140W direct imaging and $2$ k$\sec$ G141 slitless grism spectroscopy.

Each image covers a field of view of $2 \times 2.3$ $\rm arcmin^2$ at a sampling of $0.13$ arc$\sec~\rm pix^{-1}$. The G141 grism covers the wavelength range $\lambda = (1.08 - 1.70) \micron$ with a throughput $> 10\%$ and was chosen as it samples H$\alpha$ at $0.65 < z < 1.59$, [\ion{O}{3}] at $1.16 < z < 2.40$, H$\beta$ at $1.22 < z < 2.50$, and [\ion{O}{2}] at $1.90 < z < 3.56$, enabling us to identify strong galaxy features at the redshifts of our cluster candidates ($1.4 < z < 2.8$) and potentially measure dust extinction via the Balmer decrement for $1.22 < z < 1.59$.\\

\begin{figure}
\centering
\includegraphics[scale=0.46]{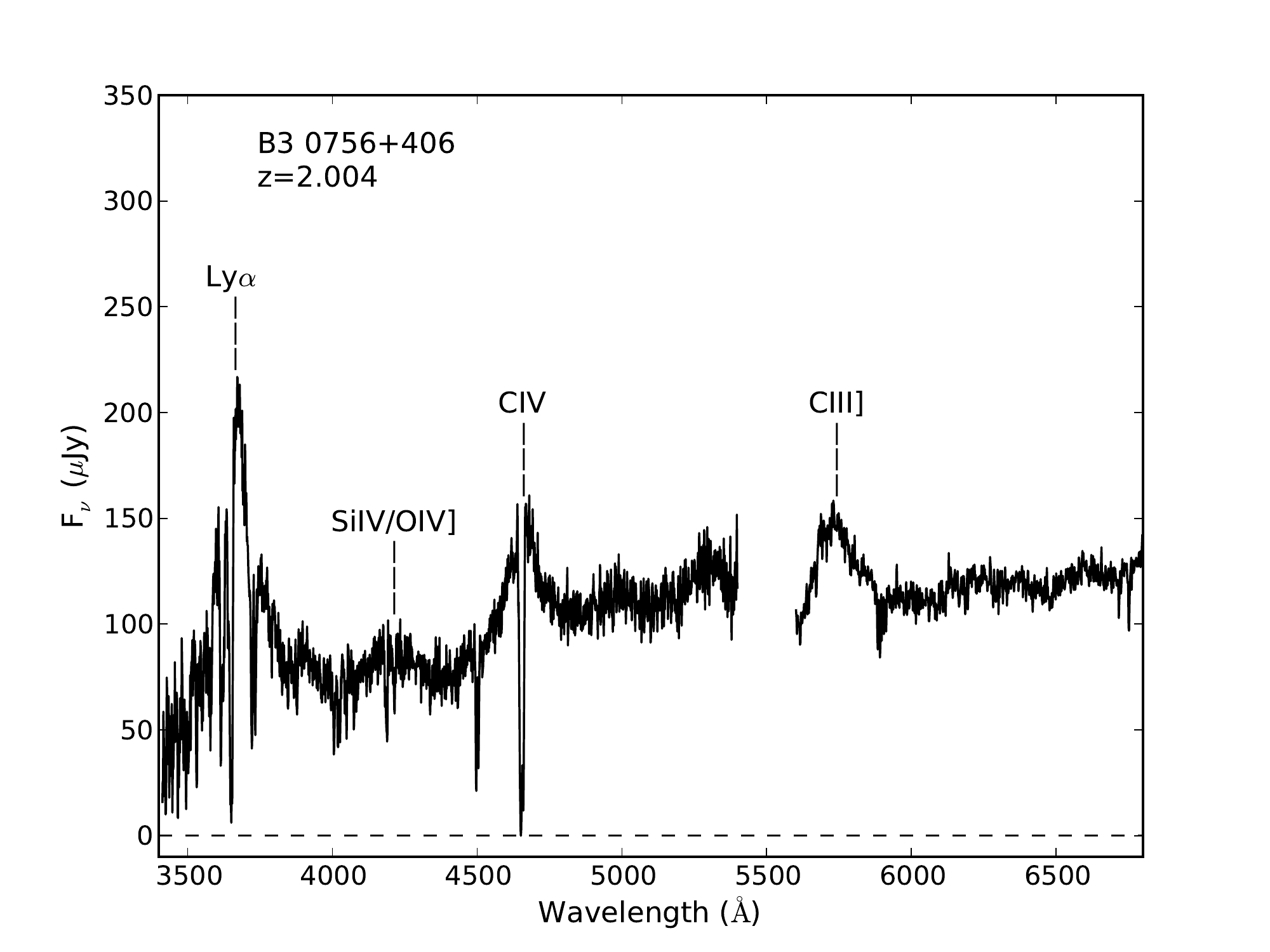}
\caption[Palomar spectrum]{Optical Palomar spectrum of B3~0756+406, the targeted RLQ of CARLA J0800+4029. We identify Ly$\alpha$, \ion{C}{4}, \ion{C}{3}], and \ion{Mg}{2} ($\lambda_{obs}=8411$ \AA, not shown), and measure a new redshift of $z = 2.004\pm0.002$.}
\label{fig:palomar}
\end{figure}

\subsection{Spectroscopy}

\subsubsection{HST Observations of CARLA J2039$-$2514 and CARLA J0800+4029}

The field around MRC~2036$-$254 (HzRG at $z = 1.997$) and the field around B3~0756+406\footnote{Note that \cite{Wylezalek13, Wylezalek14} refer to this source by its SDSS coordinates, SDSS~J080016.09+402955.6, rather than by its B3 radio catalog name.} (RLQ at $z = 2.004$) were observed in October and November 2014, respectively, partitioned in two single-orbit visits each, using two different orientations. Total exposure times on the field around MRC~2036$-$254 (B3~0756+406) reached $4023~\sec$ ($4123~\sec$) in grism mode and $1023~\sec$ ($1073~\sec$) in direct imaging mode. Table \ref{table:obs} lists the observation dates, exposure times, and orientation angles. We refer to the spectroscopically confirmed structures by their CARLA names, CARLA J2039$-$2514 and CARLA J0800+4029.

Each visit was divided into four dithered blocks of exposures, with the direct images taken just after the grism exposures to enable wavelength calibration of the spectra based on source position. We retrieved calibrated and flat-fielded individual exposures (FLT)\footnote{See details in Appendix \ref{ap:flt}.} from MAST\footnote{Mikulski Archive for Space Telescope: \url{https://archive.stsci.edu}.}, which uses the most up-to-date calibration files for WFC3. These were our primary data products before further reduction and extraction of the spectra.\\

\begin{figure}
\centering
{%
\setlength{\fboxsep}{0pt}%
\setlength{\fboxrule}{1pt}%
\fbox{\includegraphics[width=8.5cm]{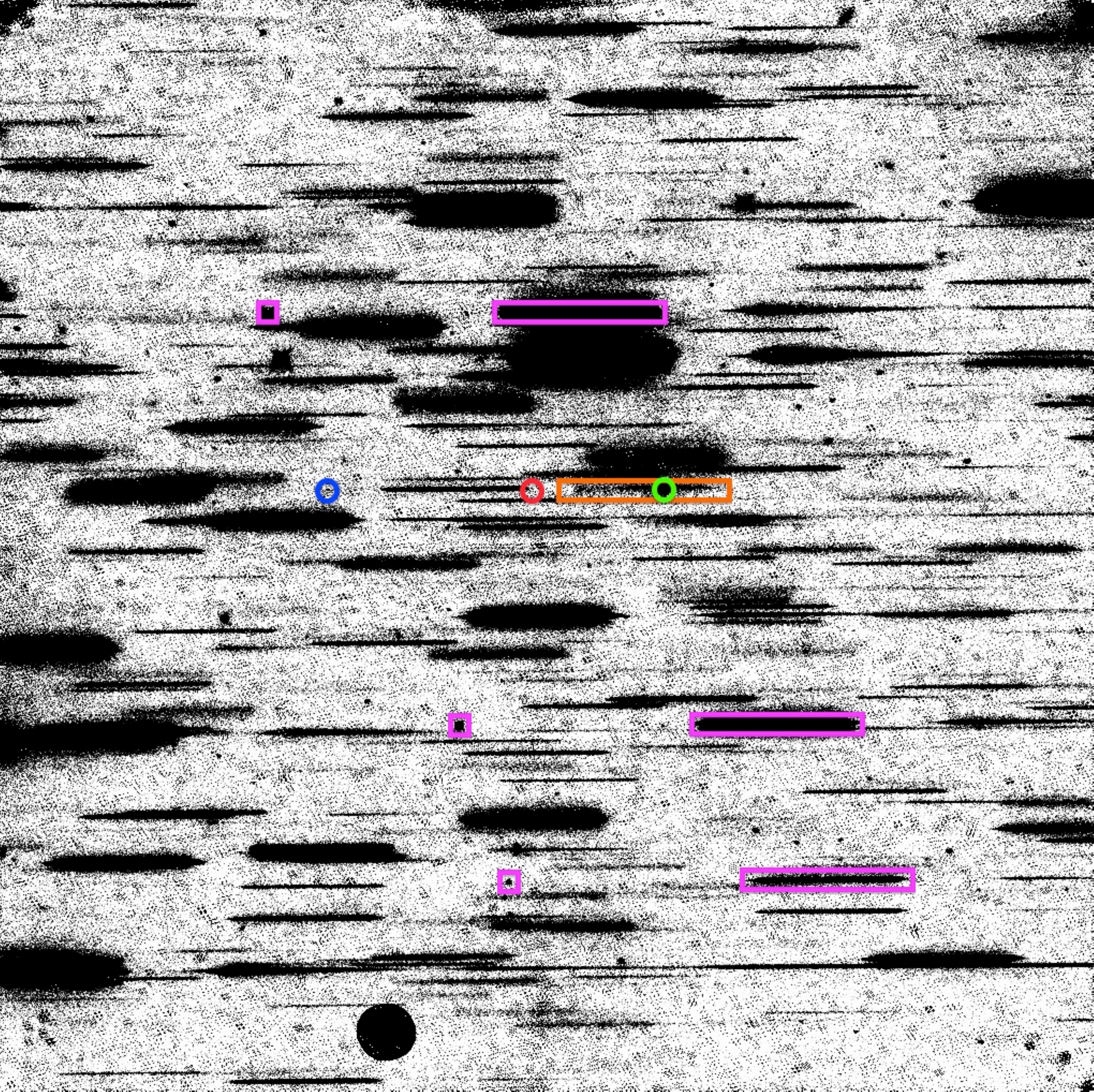}}%
}%
\caption[Grism]{Combined G141 grism exposures of the first visit to MRC 2036$-$254. The red circle shows the physical location of the HzRG. Its zeroth order is shown by the blue circle, and the trace of its first order by the orange rectangle. The green circle on top of the rectangle shows the location of the bright [\ion{O}{3}] emission line of the source. The other purple squares and rectangles highlight some zeroth and first orders of bright sources in the field.}
\label{fig:grism}
\end{figure}

\subsubsection{B3~0756+406 Palomar Spectrum}\label{p:palomar}

Because the very broad lines of B3~0756+406 in our {\it HST} spectroscopy prevent us from measuring a reliable redshift (see Appendix \ref{ap:clmemb}), and because of the relatively low signal-to-noise ratio of its SDSS
spectrum, we obtained an optical spectrum of the source on UT 2015 February 17
using the Double Spectrograph on the Hale~200" telescope at Palomar
Observatory.  The conditions were relatively clear, but not
photometric, with $\sim 1\farcs3$ seeing.  We observed the target
through a 1\farcs5 slit for two 900~s exposures using the 600 $\ell\,
{\rm mm}^{-1}$ grating on the blue arm of the spectrograph
($\lambda_{\rm blaze} = 4000$~\AA), the 316 $\ell\, {\rm mm}^{-1}$
grating on the red arm of the spectrograph ($\lambda_{\rm blaze} =
7500$~\AA), and the 5500~\AA\ dichroic.  The data were processed
using standard techniques within IRAF, and flux calibrated using
standard stars from \citet{Massey90} observed on the same night.
The optical spectrum (Figure \ref{fig:palomar}) shows strong and broad emission lines from
Ly$\alpha$, \ion{C}{4}, \ion{C}{3}], and \ion{Mg}{2} at a redshift
$z = 2.004\pm0.002$, slightly lower than the redshift $z = 2.021$ derived
from the lower quality SDSS spectrum. Narrow self-absorption is
seen in all but the \ion{C}{3}] line.\\

\section{Data Processing}\label{sec:red}

\subsection{Data Reduction}

Our grism data contain traces of several hundred objects (see Figure \ref{fig:grism}), including first and zeroth orders and even additional orders (-1, +2) in the case of bright objects. Therefore, in addition to standard reduction procedures (cosmic-ray rejection, sky subtraction, etc.) spectra need to be carefully extracted. We generally follow the steps presented in the WFC3 IR grism cookbook\footnote{Available at \url{http://axe-info.stsci.edu}.} (v1.3), applying identical methods for both fields.

\begin{table}
\caption{Noise levels}\smallskip
\label{table:noiselvl}
\centering
\begin{tabular}{ lccc }
\hline \hline
\multicolumn{1}{p{2.25cm}}{Target}& \multicolumn{2}{p{2.25cm}}{\centering F140W}&  \multicolumn{1}{p{2.25cm}}{\centering G141} \\
\cline{2-3}
\mbox{}& $\sigma \tablenotemark{a}$
 &  mag$_{\rm AB}$ ($5\sigma$)
 &  $\sigma \tablenotemark{b}$
     \\ \hline
MRC 2036$-$254&               	  			2.7&     26.5& - \\
~~~~Orient01&						- &        - & 	51\\
~~~~Orient02&						- &        - & 	60\\
\hline
B3~0756+406& 			2.4&        26.7& - \\
~~~~Orient01&						- &        - & 	44\\
~~~~Orient02&						- &        - &        	59\\
\hline
\end{tabular}
\tablenotetext{1}{Standard deviation in flux density ($10^{-21}$ $\rm erg~  \sec^{-1}~cm^{-2} ~ \AA^{-1}$).}
\tablenotetext{2}{Same as \tablenotemark{a} but at $15,000~\rm \AA$.
\smallskip}
\end{table}

We combine the individual F140W exposures using the aXe software (v2.2.4) to create deep drizzled direct images of the fields. From these images we perform source extraction using SExtractor (\citealp{BertinArnouts96}). In this step we create two catalogs: a deep catalog cleaned by hand from spurious detections (the master catalog), and a shallow catalog with a more conservative magnitude limit. The latter allows us to subtract the sky background from the grism images in a later step\footnote{See additional details on these steps in Appendix \ref{ap:combi}.}.
Since our targets were selected as CARLA overdense fields, we cross-correlate the color-selected IRAC $z > 1.3$ candidates with our {\it HST} master catalog, even though in practice we extract and analyze all {\it HST} sources (see details in Appendix \ref{ap:cross}).

The master catalog is then projected back onto each individual direct exposure and the background level is subtracted from each grism exposure using up-to-date master sky background (v1.0) and grism mode configuration files (v2.5)\footnote{Available at \url{http://www.stsci.edu/hst/wfc3/analysis/grism_obs/calibrations/wfc3_g141.html}\label{foot:calfiles}.}. Details on these two steps are presented in Appendix \ref{ap:back}.\\

\begin{figure*}
\centering
\includegraphics[width=9cm]{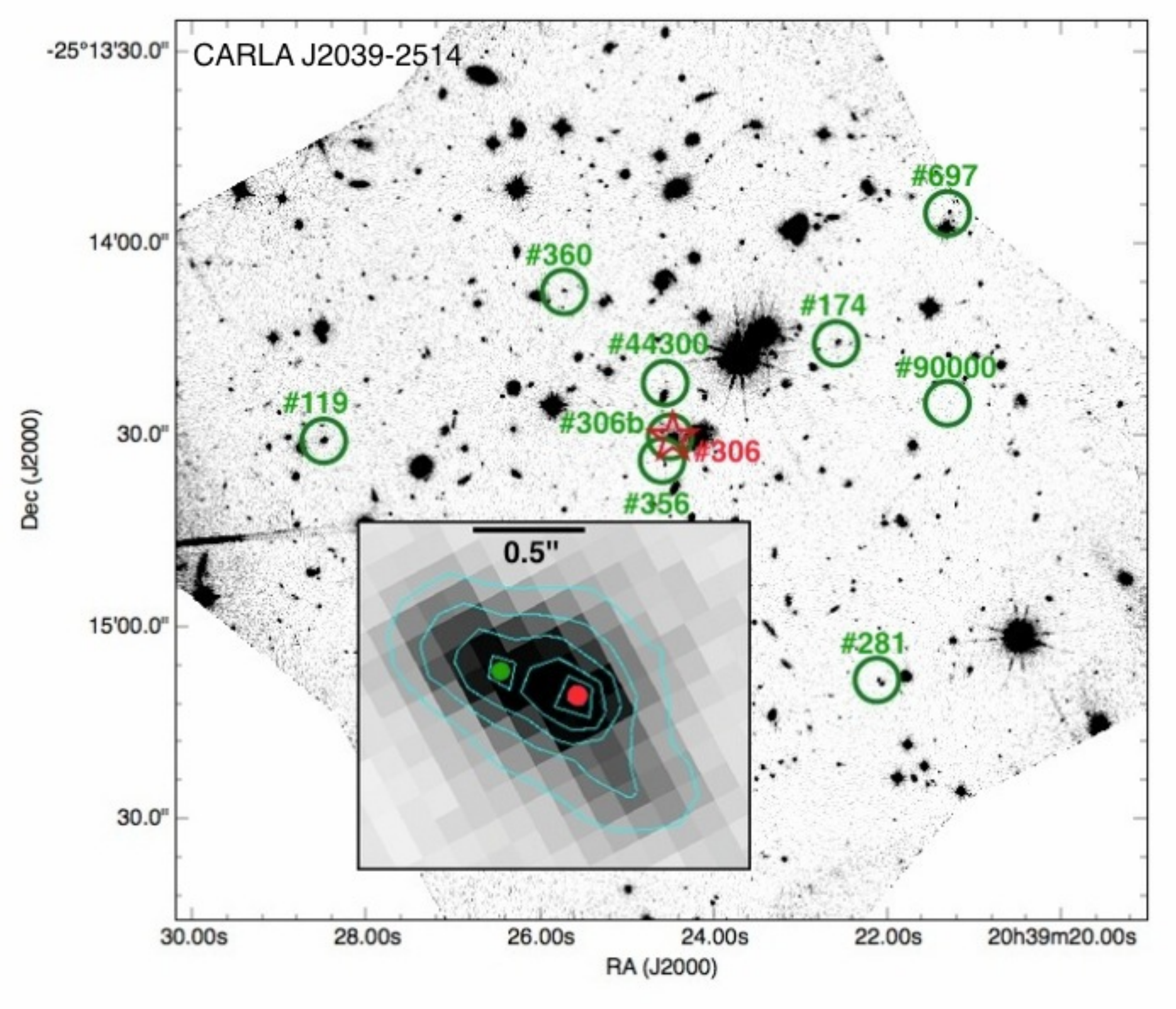}
\includegraphics[width=8.9cm]{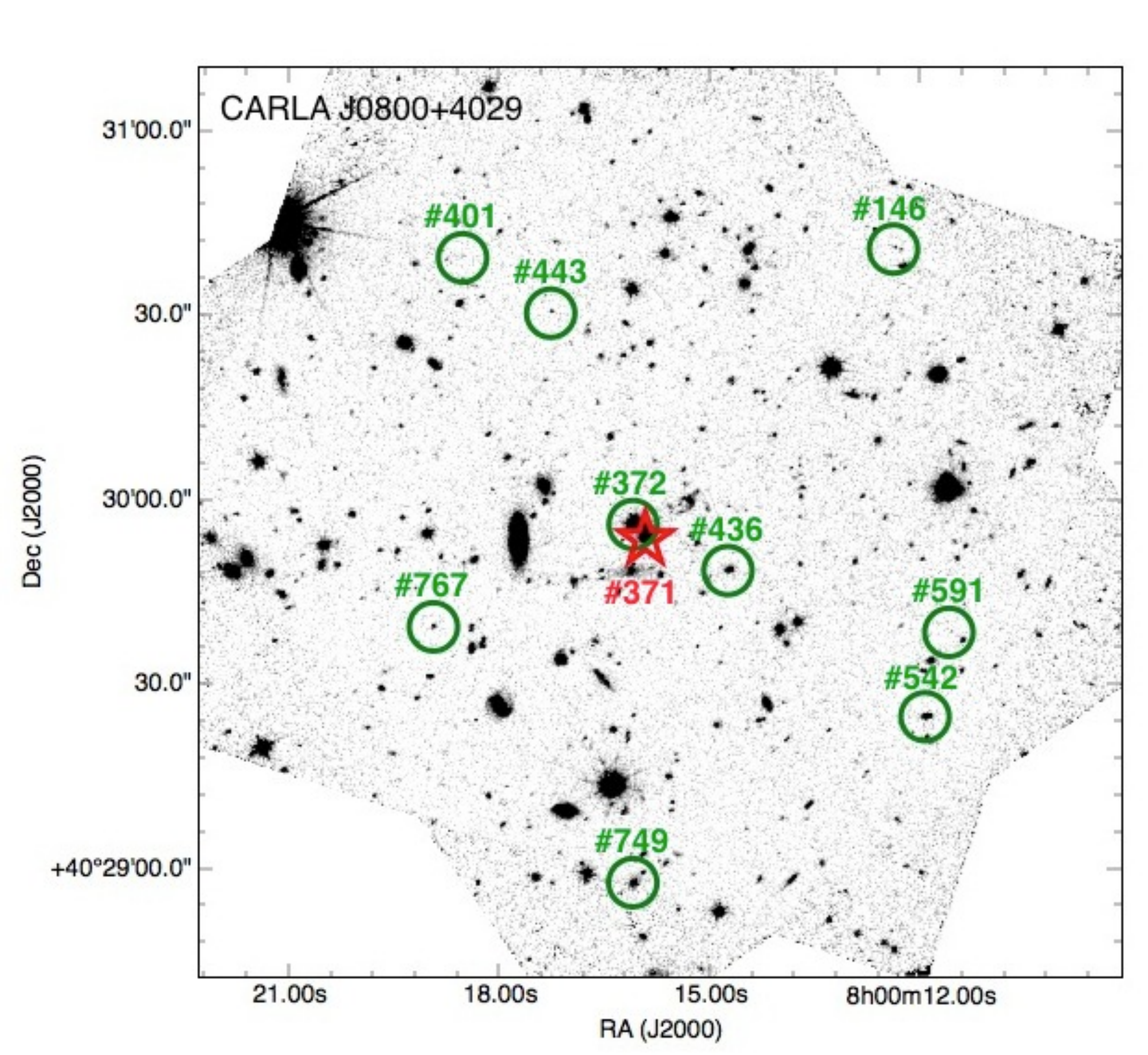}
\caption[F140W spatial distribution of CARLA J2039$-$2514 and CARLA J0800+4029 cluster members]{F140W images of our two RLAGN fields, showing the spatial distribution of CARLA J2039$-$2514 (left) and CARLA J0800+4029 (right) confirmed members. North is up and East is to the left. The red stars indicate the RLAGN, and the green circles indicate confirmed member galaxies. The inset in the left panel shows a close-up of the targeted HzRG, MRC~2036$-$254 ($\#306$). The red dot shows the position of the source, and the green dot the position of its companion $\#306$b; these two components are highlighted by contour lines.}
\label{fig:spadist}
\end{figure*}

\subsection{Extraction of Spectra and Contamination Models}\label{p:ExtrContam}

We then extract individual spectra from each scaled grism frame based on source positions and sizes stored in the individual master catalogs, and stack the spectra of same observation/orientation. The aXe task {\ttfamily axecore} determines contamination estimates from neighboring and/or overlapping objects. We use the ``Gaussian" method of the qualitative contamination model as a first approximation, adopting a width scale factor of $1$ instead of the standard factor of $3$ which we found makes the spectral traces in the spatial (cross-dispersion) direction too wide\footnote{See more details on the contamination models in Appendix \ref{ap:contam}.}.
To obtain deep 2D first order spectra of constant dispersion in wavelength and linear spatial sampling, we run the tasks {\ttfamily drzprep} and {\ttfamily axedrizzle}. Note that we keep each orientation separate in order to better handle contamination issues. We obtain for each available orientation a deep 2D cutout of the first spectral order of each source as well as a deep 2D cutout containing the traces of its possible contaminants estimated from the ``Gaussian" contamination model.

Finally, we manipulate the 2D cutouts to extract our own 1D spectra, flux-calibrated using the G141 first order sensitivity function.
The software identifies continuum and zeroth orders of possible contaminants on the 2D cutouts.\\

\subsection{Detection Limits and Noise Level}

\subsubsection{Imaging Noise Level}\label{p:Fnoise}

We measure the F140W noise level from object-free regions based on the SExtractor segmentation map corresponding to our deep catalog.
We use this map to mask all detected sources, and we measure the noise level by randomly placing 5000 non-overlapping $0.4 \arcsec$ diameter apertures\footnote{Typical size of our {\it HST} sources.} on the unmasked area covered by both orientations. For both fields the noise values follow a Gaussian distribution, with best-fit values given in Table \ref{table:noiselvl}. The images of both fields have $1\sigma$ noise levels of $\sim 2.5\times 10^{-21} \rm ~erg~ \sec^{-1}~cm^{-2}~ \AA^{-1}$, corresponding to $\sim 26.6$ mag (AB) at $5\sigma$.\\

\subsubsection{Spectroscopic Noise Level}\label{p:Gnoise}

We use the full contamination image produced by aXe as a mask on the G141 frames to compute the G141 noise distribution over spectrum-free regions. As aXe does not produce a co-added image for the G141 exposures, we use the IRAF task {\ttfamily imcombine} on the G141 sky-subtracted frames and corresponding contamination model frames to produce the grism combined (averaged) images of both the contamination model and the real exposures. For both fields the noise distribution closely follows a Gaussian distribution. For the spectroscopic analysis, we calibrate this baseline pixel noise level to the relevant wavelength under consideration using the G141 sensitivity function, and this provides the internal uncertainty for flux measurements. Table \ref{table:noiselvl} provides the values at $15,000~\rm \AA$, of the order of $5\times 10^{-20} \rm ~erg~ \sec^{-1}~cm^{-2}~ \AA^{-1}$. The contamination model contours shown on the spectral 2D cutouts (see Appendix \ref{ap:clmemb}) represent $1\sigma$, $2\sigma$, $5\sigma$, and $10\sigma$ deviations above the mean level.\\

\subsubsection{Line Detection Limit}

The grism sensitivity is maximal and roughly constant over the wavelength range $(14,500 - 15,500) \rm~ \AA$ where H$\beta$ and [\ion{O}{3}] fall at $z \sim 2$. We use cluster member spectra of undetected H$\beta$ emission lines to determine our detection limit (note that all cluster members have detected [\ion{O}{3}] lines). We generate $100$ Gaussians of evenly spaced height values in the range $(0-3)\times10^{-19}~ \rm erg~  \sec^{-1}~cm^{-2}~ \AA^{-1}$ that we add to the spectra at the expected locations of H$\beta$. Using the same fitting procedure as described in \S\ref{s:mpfit}, we define our detection limit as the minimum flux which has a signal-to-noise ratio $\geq 5$ and a measured-to-true flux ratio in the range $0.8 - 1.2$. Our detection limit differs slightly from spectrum to spectrum, but is generally in the range $(1.2 - 4.0) \times 10^{-17} ~\rm erg~ \sec^{-1}~cm^{-2}$. We adopt the mean value, $2.5\times 10^{-17}~ \rm erg~ \sec^{-1}~ cm^{-2}$, as our detection limit for both fields.\\ 

\begin{figure*}[!ht]
\includegraphics[width=9.4cm]{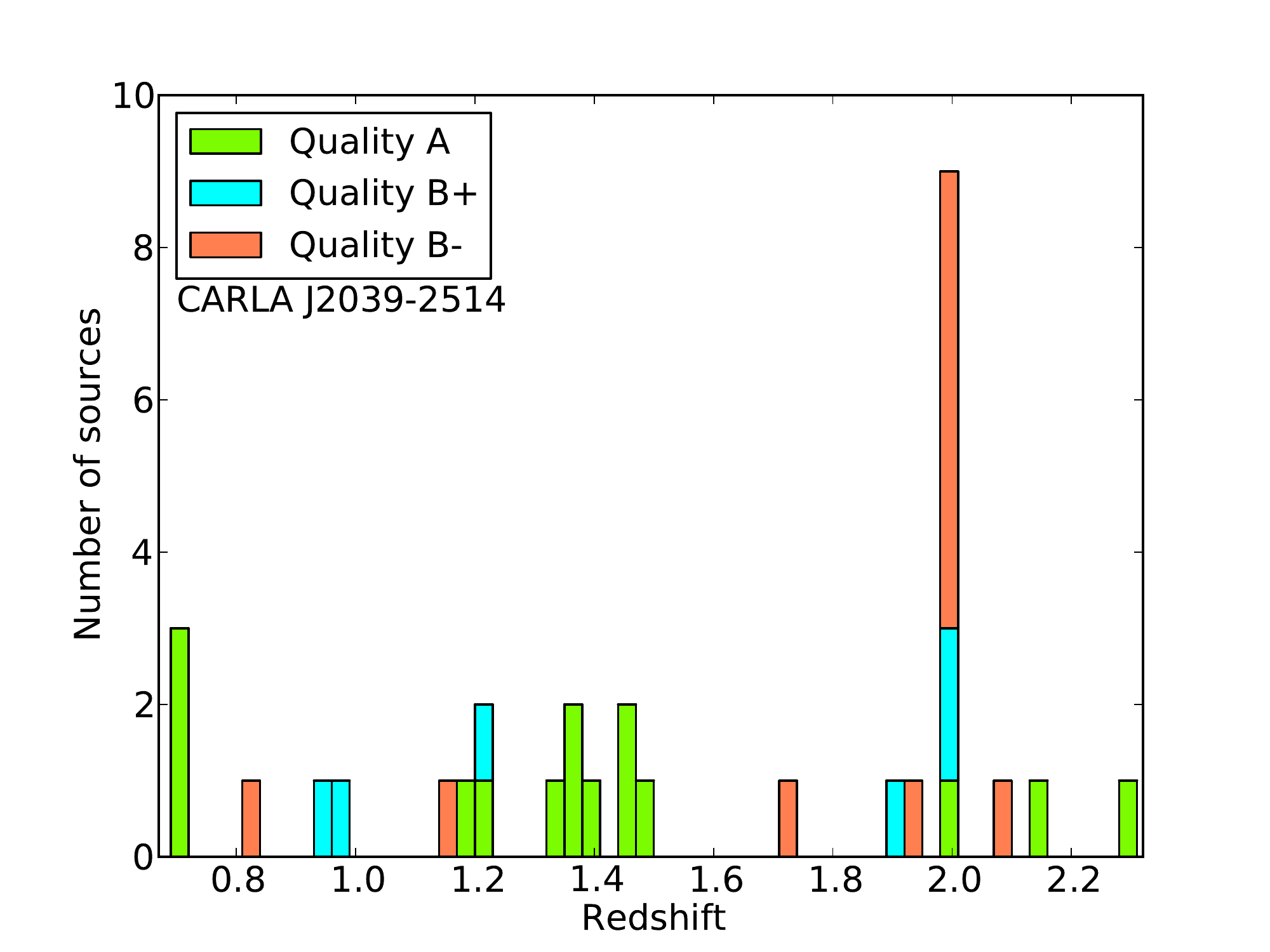}
\includegraphics[width=9.4cm]{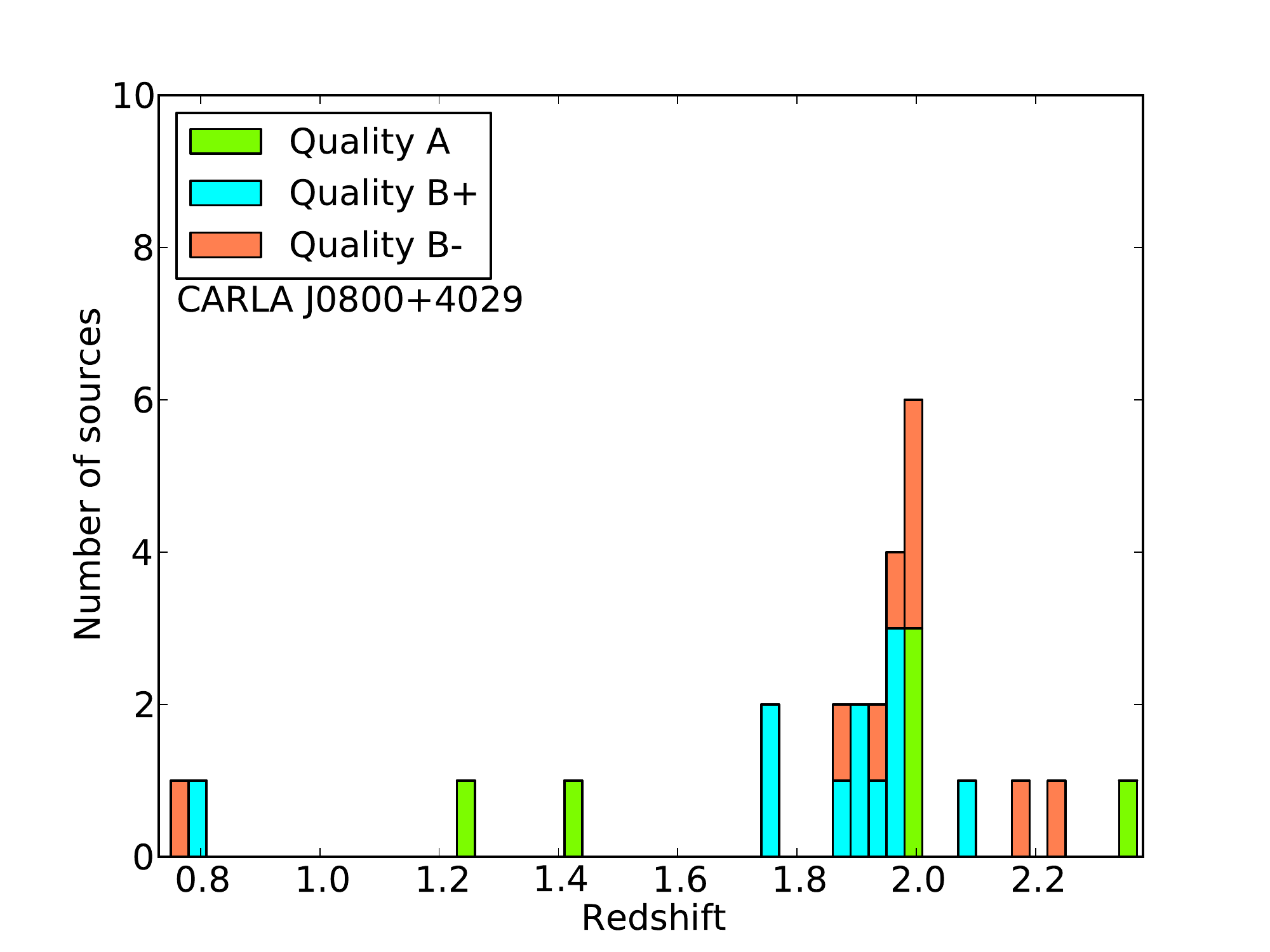}
\caption[Cluster redshift distributions]{Distribution of WFC3 grism redshifts in the fields around MRC 2036$-$254 (left) and B3~0756+406 (right). The bin size is 0.03 in redshift. The quality flags are described in Section \ref{sec:reddet}. Both fields have strong peaks at $z \simeq 2$, with a larger scatter for the latter field. Cluster member redshifts are listed in Table \ref{table:examplelargetable}, and redshifts of other sources in Table \ref{table:nonCl}.}
\label{fig:redtot}
\end{figure*}

\subsection{Measurements}

\subsubsection{Emission Line Fitting}\label{s:mpfit}
We determine the redshift of the sources and emission line fluxes using the python version of {\ttfamily mpfit}\footnote{Available here: \url{http://code.google.com/p/agpy/source/browse/trunk/mpfit/mpfit.py?r=399}}, a Levenberg-Marquardt least-squares minimization fitting procedure. When required (e.g., for all cluster members), we fit the [\ion{O}{3}] doublet ($\lambda 4959, \lambda 5007$) with two Gaussians constrained to have the same width, same redshift, and a 1:3.2 flux ratio (\citealp{Osterbrock89}). Simultaneously, we fit the continuum using a third order polynomial and include a third Gaussian corresponding to H$\beta$, constrained to have the same redshift as [\ion{O}{3}]. We substitute [\ion{O}{3}] with, or add to the model, other Gaussians to account for H$\alpha$ or [\ion{O}{2}] when required (e.g., for non-cluster members). Table \ref{table:examplelargetable} shows the redshifts and emission line fluxes of our cluster members, and Table \ref{table:nonCl} presents the redshifts of non-cluster members.\\

\subsubsection{Uncertainties}\label{p:zerr}
From the WFC3 Data Handbook, the internal accuracy of the G141 grism dispersion is 0.25 pixel for G141. In addition, the wavelength calibration of the spectroscopy depends on the accuracy of the object centroid in the direct image, and the accuracy of the grism wavelength zero point based on the direct image source positions. Small offsets between the orientations will also produce spectral offsets. The handbook recommends a calibration uncertainty of 0.3 pixel, corresponding to a wavelength uncertainty of $20\rm~\AA$ or a redshift uncertainty of $0.003 - 0.006$ at $z = 2$, depending on the identified spectral feature. The internal error on the emission line fluxes is simply the $1\sigma$ level of the grism detector noise distribution (see \S\ref{p:Gnoise}).

The fitting procedure described in \S\ref{s:mpfit} produces a formal $1\sigma$ error on all measured parameters, directly given by {\ttfamily mpfit}. To obtain the global parameter uncertainties (i.e., the measurement errors from the fitting procedure) we scale the formal errors with the reduced $\chi^2$ of the fit ($\chi^2$ divided by the number of degrees of freedom): $scaled\_error = formal\_error \times \sqrt{\chi^2 / DoF}$. The total redshift uncertainty is the quadratic sum of the internal and measurement errors. However, since we use the flux internal error as an additional input parameter to the fitting procedure, we only use {\ttfamily mpfit} scaled errors to calculate the total flux uncertainties (i.e., the scaled errors already take the flux internal error into account).\\

\section{Results}\label{sec:res}

We visually inspect all extracted spectra, and use our fitting procedure on those showing clear emission lines. The 2D cutouts, 1D spectra, and contamination contours of all cluster members are shown in Appendix \ref{ap:clmemb}, including their direct image stamps.\\

\subsection{Redshift Determination}\label{sec:reddet}
We have three proxies for identifying cluster members: emission lines, IRAC $[3.6]-[4.5]$ colors, and the redshift priors of the targeted RLAGN. We assign three different quality redshifts; A, B$^{+}$, and B$^{-}$, defined as follows:

\textbullet\ {\textbf{Quality A.}} When several lines are identified, we assign a quality A to the source redshift, and the redshift is considered very secure.

\textbullet\ {\textbf{Quality B$^{+}$.}} When only one line is detected but the source has a robust IRAC counterpart, we use the mid-infrared color to determine the most likely line identification (e.g., [\ion{O}{2}], [\ion{O}{3}], or H$\alpha$) and assign a quality B$^{+}$ to the redshift. Such sources are considered to have relatively secure redshifts.

\textbullet\ {\textbf{Quality B$^{-}$.}} When only one line is robustly detected but the source does not have an IRAC counterpart, the line identification is less secure and we assign a quality B$^{-}$ to the redshift. Such redshifts are considered likely correct, albeit with the potential for some mis-identifications.

\begin{figure*}[!ht]
\begin{center}
\includegraphics[width=14.5cm]{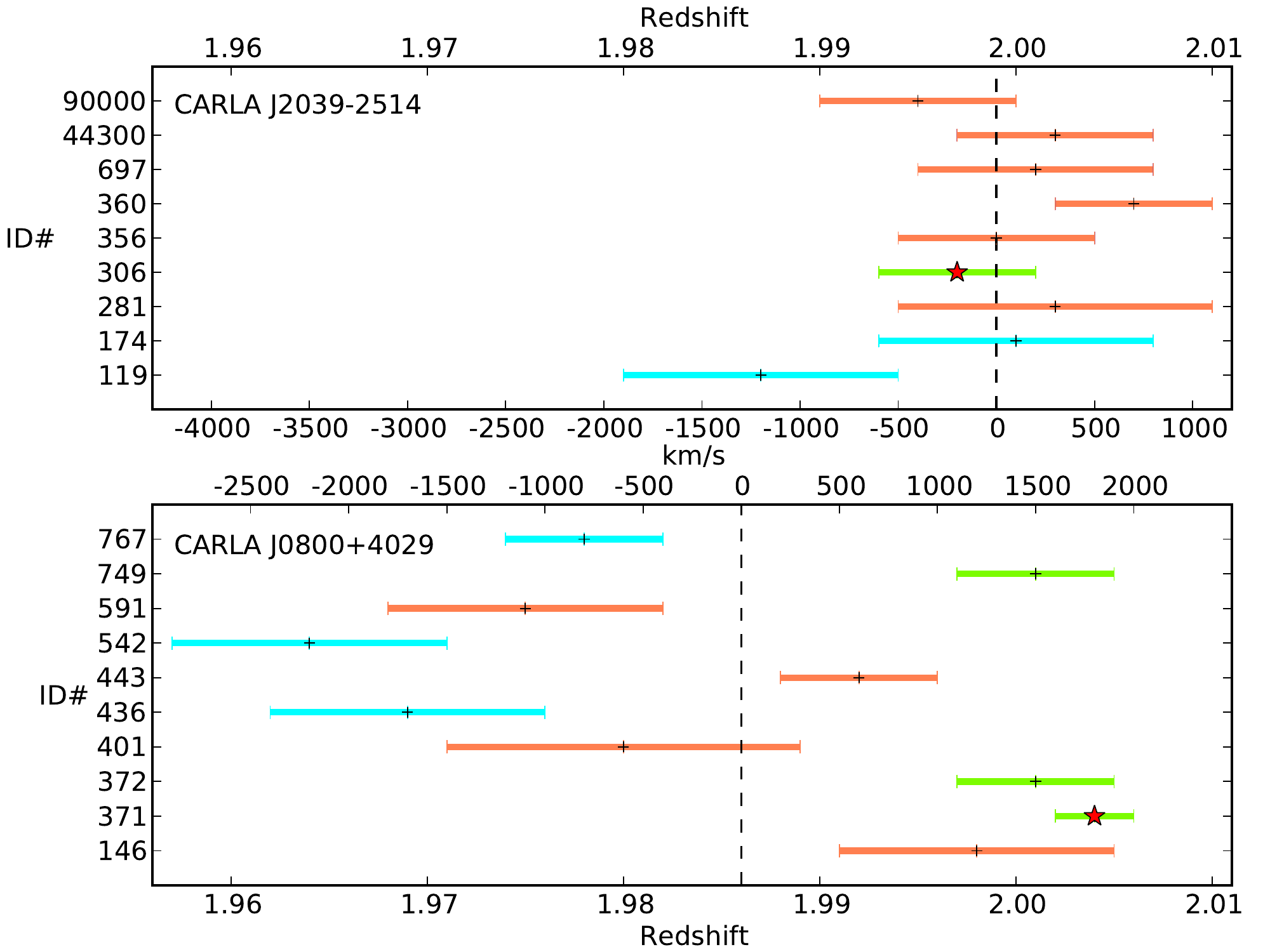}
\end{center}
\caption[Cluster redshift distributions]{Redshift distribution of the members of CARLA J2039$-$2514 (top) and CARLA J0800+4029 (bottom). The panels show the redshift of each source and its uncertainty. The vertical dashed lines indicate the mean redshift of each cluster. Both panels show the same redshift range (i.e., $1.956 < z < 2.011$). The velocity shifts relative to each cluster mean redshift are also indicated on the horizontal axes. Line colors correspond to the redshift qualities of Fig.\ref{fig:redtot}, and the red stars indicate the targeted RLAGN. The redshift distribution of  CARLA J2039$-$2514 members spans 0.03 in redshift, whereas the one of CARLA J0800+4029 members spans 0.05 in redshift, including $1\sigma$ uncertainties.}
\label{fig:reddist}
\end{figure*}

When only one line is detected, we assume it is a strong line characteristic of star-formation or AGN activity, given our observation depth. We discard the possibility of recovering a Ly-$\alpha$ line as it would imply a redshift $> 8$, and therefore we only consider [\ion{O}{2}]$\lambda3727$, [\ion{O}{3}]$\lambda\lambda4959,5007$, and H$\alpha$.
We also use [\ion{O}{3}] or H$\alpha$ non-detections to disentangle line identifications assuming a [\ion{O}{3}]:H$\alpha$ ratio of unity; empirically the line ratio has a dispersion of a factor of $\sim 2$ (e.g., \citealp{Mehta15}).
We do not use [\ion{O}{2}] non-detections as the line can be below our detection limit when the typically stronger [\ion{O}{3}] line is detected.
In cases where a source is detected in our {\it Spitzer} data, we use its IRAC mid-infrared color to segregate $z > 1.3$ galaxies from foreground sources since, empirically, galaxies at $z > 1.3$ tend to have $([3.6]-[4.5])_{\rm AB} > -0.1~ \rm mag$ (\citealp{Papovich08}; see \citealp{Galametz12} for a comprehensive discussion of possible contaminants). Based on the 11 quality A redshifts from our spectroscopic sample with {\it Spitzer} detections, we find that 8 out of 9 sources with $([3.6]-[4.5])_{\rm AB} > -0.1~ \rm mag$ are at $z > 1.3$, while 1 out of 2 sources with bluer mid-infrared colors is at $z < 1.3$.  This is consistent with Papovich (2008), who found that red mid-infrared colors effectively identify distant ($z > 1.3$) galaxies, but that distant galaxies do not exclusively have red mid-infrared colors.
Cluster member sources are identified based on the [\ion{O}{3}] line, which is observed at $\sim 15,000 \rm~\AA$ at $z \sim 2$, the redshift of the targeted RLAGN. Based on our identification scheme, a single emission line observed around $15,000 \rm~\AA$ can also correspond to [\ion{O}{2}] at $z\sim3$. However, based on the $79,609$ {\it HST} grism redshifts from the 3D-HST field survey (\citealp{Momcheva15}), there are $405$ [\ion{O}{3}] and $63$ [\ion{O}{2}] emitters above our detection limit in the wavelength range $(14,840-15,042) \rm~\AA$. With eight (CARLA J2039$-$2514) and seven (CARLA J0800+4029) quality B members (see Sec. \ref{sec:clmbship} and Fig. \ref{fig:reddist}), we therefore expect that no more than one spectroscopically confirmed source per structure is likely [\ion{O}{2}] at $z\sim3$ rather than [\ion{O}{3}] at the redshift of the RLAGN. We performed two independent redshift determinations (GN and DS) which yield consistent assessments.
Table \ref{table:examplelargetable} shows the cluster member properties, and Figure \ref{fig:spadist} shows their spatial distributions. Table \ref{table:nonCl} lists the measured redshifts of other sources in the {\it HST} fields of view, Figure \ref{fig:redtot} shows the redshift distribution of identified sources in each field, and Figure \ref{fig:reddist} highlights the cluster redshift distributions. We measure our member properties independently for both orientations when possible, and determine the final physical parameters as the average of the results from the two orientations, with uncertainties added in quadrature.\\

\subsection{Cluster Membership}\label{sec:clmbship}

\cite{Eisenhardt08} defined a spectroscopically confirmed $z > 1$ cluster as a structure containing at least five galaxies within a physical radius of 2 Mpc whose spectroscopic redshifts are confined to within $\pm 2000(1+\left<z_{\rm spec}\right>) \rm~km~\sec^{-1}$.
From emission lines and IRAC colors (when available), we identify 9 members in CARLA J2039$-$2514 at $\left<z\right> = 1.999 \pm 0.002$ (median $2.000$). This includes the RLAGN which is kinematically complex, showing two components, potentially a dual AGN, separated by $0.5 \arcsec$ (see Figure \ref{fig:spadist} inset). The standard deviation between the members is $0.005$ in redshift.
We identify 10 members in CARLA J0800+4029 at $\left<z\right> = 1.986 \pm 0.002$ (median 1.986), with a $0.014$ deviation between the members.
In both cases, all confirmed members are located within $1\arcmin$ radii, corresponding to projected radii of $500~ \rm kpc$ at the redshifts of the RLAGN ($1.5 ~\rm Mpc$ co-moving). As seen in Figure \ref{fig:reddist}, members are confined within $\rm [-1200; +700] ~km~\sec^{-1}$ of their cluster mean redshift for CARLA J2039$-$2514, spanning 0.02 in redshift space (removing one outlier, we have $\rm [-500; +600]~km~\sec^{-1}$ for eight of the nine confirmed members), and within $\rm [-2300; +1700] ~km~\sec^{-1}$ for CARLA J0800+4029, spanning 0.04 in redshift space.

While these two structures conform to the \cite{Eisenhardt08} confirmation criteria, these criteria were initially developed for ground-based spectroscopic programs, and have the potential to misidentify as clusters less massive structures such as protoclusters, groups, sheets, and filaments.
While keeping these concerns in mind, in the following we will sometimes
refer to the confirmed structures as clusters. In particular, as
discussed in Section \ref{sec:intro}, other hallmarks of a bona fide cluster
include a significant population of evolved massive galaxies and
a centrally concentrated distribution of galaxies. We show in
Section \ref{subsec:passivedendity} that 
CARLA J2039$-$2514 contains an overdensity of (spectroscopically unconfirmed) red galaxies with properties
consistent with being passive galaxies at the redshift of the
structure, and prior work by our team has demonstrated that the
color-selected (i.e., red) candidate cluster members are centrally concentrated
around the target RLAGN (\citealp{Wylezalek13}. The overdensities reach $9.0\sigma$ and $7.8\sigma$ above the field value for the fields around MRC 2036$-$254 and B3~0756+406, respectively. Therefore, the
balance of the evidence leans in favor of these structures being
galaxy clusters/forming clusters, though a conservative approach refers to them
simply as structures.

The overall redshift distribution of our two fields form a strong peak at $z\simeq 2$ (Figure \ref{fig:redtot}). However, the coarse resolution of the WFC3 grism and low number statistics prevent us from inferring reliable velocity dispersions.
The combined-quality A $\cup$ B$^{+}$ redshifts of CARLA J2039$-$2514 and CARLA J0800+4029 have median redshifts of $z=1.999$ and $z=1.990$, respectively; the three quality A redshifts of CARLA J0800+4029 have a median redshift of $z=2.001$.\\

\subsection{Star-Formation Rates and Dust Extinction}\label{p:sfr}
H$\alpha$, one of the most reliable SFR indicators, unfortunately falls outside of the grism wavelength range at the redshift of our clusters. We therefore assume [\ion{O}{3}]$\lambda5007$/H$\alpha = 1$ and use the \cite{Kennicutt83} relation, ${\rm SFR} = L{\rm (H\alpha)/(1.12 \times 10^{41} ~erg~s^{-1})} ~M_{\odot}~\rm yr^{-1}$, to convert our [\ion{O}{3}]$\lambda5007$ fluxes into SFRs. We note however that there is a large scatter in the [\ion{O}{3}]$\lambda5007$/H$\alpha$ ratio reported for $z\sim2$ star-forming galaxies. \cite{Mehta15} indicate a linear relation between [\ion{O}{3}]$\lambda5007$ and H$\alpha$ luminosities with [\ion{O}{3}]$\lambda 5007$/H$\alpha \sim 2$ albeit with a large scatter. On the other hand, \cite{Juneau14} suggests a [\ion{O}{3}]$\lambda 5007$/H$\alpha$ ratio in the range $0.5 - 1.5$ for the redshift and mass of our confirmed cluster members.
Using the mean [\ion{O}{3}]$\lambda 5007$ flux of our SF cluster members ($8 \times 10^{-17}~\rm  erg ~ s^{-1} ~cm^{-2}$), but excluding the RLAGN whose [\ion{O}{3}] fluxes are likely enhanced by AGN photoionization, we compute a mean SFR of $20~ M_{\odot} \rm ~ yr^{-1}$ for our SF cluster members, with no dust correction applied.
Given the spread in literature values for the
[\ion{O}{3}]$\lambda 5007$/H$\alpha$ ratio, this is considered a
very crude estimate of the SFR, only considered robust at the level
of a factor of 2.

To better constrain our SFR lower limit, we tentatively estimate the contribution of dust extinction to the SFR. One robust method is to use the ratio of the measured Balmer decrement (the flux ratio of H$\alpha$ over H$\beta$).
However, we deem it too uncertain to use [\ion{O}{3}]$\lambda5007$ as a proxy for H$\alpha$ to calculate Balmer decrements given {\it (i)} the scatter in the [\ion{O}{3}]$\lambda5007$ to H$\alpha$ ratio, {\it (ii)} the ten times shorter wavelength baseline between H$\beta$ and [\ion{O}{3}]$\lambda5007$ as compared to between H$\beta$ and H$\alpha$, and {\it (iii)} the low SNR for our measured fluxes, typically of the order of $5$.
We therefore assume a constant $A_V = 1$ mag, a typical dust attenuation in the $V$-band for SF galaxies (\citealp{Kewley04}). Other groups have applied similar approximations
(e.g., \citealp{Zeimann12}, \citealp{Newman14}). In particular,
the level of dust extinction is expected to correlate with stellar
mass, though the improvement taking that effect into account is
likely small compared to the uncertainties inherent to our oxygen-based
SFRs. 
Using the \cite{Calzetti00} extinction curves with $R_{V} = 4.05$, we have $A_{\rm[O~III]} = 0.96$ mag. Our SF cluster members therefore have dust-corrected SFRs in the range $\sim (20 - 140)~M_{\odot} \rm~yr^{-1}$, with a median (mean) cluster member SFR\footnote{Excluding the extrema, our mean dust-corrected SFR is $\sim 40~M_{\odot} \rm~yr^{-1}$. Using the \cite{Cardelli89} Galactic extinction law instead implies an average SFR of $\sim 60~M_{\odot} \rm~yr^{-1}$ for our SF members (including the extrema).} of $\sim 35~M_{\odot} \rm~yr^{-1}$ ($\sim 50~M_{\odot} \rm~yr^{-1}$). This leads to a lower limit on the total dust-corrected SFR of $\geq 400 ~M_{\odot} \rm~yr^{-1}$ for both of our clusters.\\

\subsection{Stellar Masses}
To estimate stellar masses, we scaled {\it Spitzer}/IRAC [3.6] and [4.5] magnitudes to \cite{BruzualCharlot03} stellar population synthesis models using a \cite{Chabrier03} initial mass function (IMF), single stellar population (SSP, i.e., single delta-burst population), and a $z_f = 4.5$ formation redshift.
We estimate stellar masses in the range $(0.1 - 1.9) \times 10^{11} ~M_{\odot}$ for our CARLA cluster members (median $1.1 \times 10^{11} ~M_{\odot}$). For cluster members without CARLA counterparts, we assign upper limits of $<10^{10} ~M_{\odot}$ based on the IRAC depths. The RLAGN contributions to the SEDs likely contaminate the stellar component (e.g., \citealp{Drouart12}) and overestimate the derived stellar masses. This is particularly true for type$-1$ AGN like B3~0756+406, and typically less so for type$-2$ AGN like MRC~2036$-$254. Without additional longer wavelength photometry to disentangle the stellar and AGN components in the SEDs (e.g., \citealp{Seymour07}, \citealp{DeBreuck10}), we simply determine upper limits on the stellar masses of the RLAGN MRC 2036$-$254 and B3 0756+406, respectively $<3 \times 10^{11}$ and $<3 \times 10^{12} ~M_{\odot}$, and we determine a limiting stellar mass $<2 \times 10^{11} ~M_{\odot}$ for the CARLA J0800+4029 object $\#749$ identified as a QSO.\\

\subsection{Cluster Masses}
We next make an estimate of the total masses of the
two newly confirmed structures. The low resolution of the grism
spectroscopy precludes a measurement of a velocity dispersion, and
we currently lack the data to attempt other standard cluster mass
measurements, such as weak lensing, diffuse X-ray emission, and SZ
decrements. Therefore, 
we use the
{\it Spitzer} imaging to estimate the total stellar mass in the
systems, and then compare it to other, better studied clusters at
comparable redshift.

Following the methodology discussed in detail in \cite{Wylezalek14},
we fit the CARLA J2039$-$2514 and CARLA J0800+4029 background-subtracted
luminosity functions with a \cite{Schechter76} function, integrating
over the range $(m^{\star}-2.5)$ to $(m^{\star}+10)$. This allows
us to estimate the total luminosity density of the two structures,
which we convert to total stellar masses using \cite{BruzualCharlot03}
solar metallicity SSP models with formation redshift $z_f = 4.5$.
We assume a constant stellar mass to light ratio, determined for
both the \cite{Chabrier03} and the \cite{Salpeter55} IMF. The
\cite{Chabrier03} IMF provides total stellar masses of $M_{\star}
= 1.9 \times10^{12} \, M_{\odot}$ for CARLA J2039$-$2514, and
$M_{\star} = 2.9 \times10^{12} \, M_{\odot}$ for CARLA J0800+4029.
We apply the same methodology to the well studied distant galaxy
clusters ClG 0218.3$-$0510 ($z=1.62$; \citealp{Papovich10}, \citealp{Tanaka10})
and IDCS J1426.5+3508 ($z=1.75$; \citealp{Stanford12}, \citealp{Brodwin16}),
finding total stellar masses of $M_{\star} = 1.2 \times10^{12} \,
M_{\odot}$ for ClG 0218.3$-$0510 and $M_{\star} = 1.9\times10^{12}
\, M_{\odot}$ for IDCS J1426.5+3508. 
Our derived total stellar masses for the two new systems at $z \sim
2$ are therefore similar to IDCS J1426.5+3508 and slightly higher
than ClG 0218.3$-$0510. 
These total masses are all
$\sim 50\%$ larger if we instead adopt the \cite{Salpeter55}
IMF.

\cite{Tanaka10} report an X-ray derived mass $M_{200} = (5.7 \pm 1.4)\times10^{13} \, M_{\odot}$ for ClG 0218.3$-$0510. \cite{Brodwin12} and \cite{Brodwin16} report a total
halo mass in the range $M_{500} = (1.9-3.3)\times10^{14} \, M_{\odot}$
($M_{200} \sim 4\times10^{14} \, M_{\odot}$) for IDCS J1426.5+3508 from a variety of halo mass tracers, including
SZ, X-rays, and strong gravitational lensing (see also
\citealp{Gonzalez12}; and \citealp{Mo16} for a weak lensing analysis), making IDCS J1426.5+3508 the most massive
galaxy cluster currently known at $z>1.5$. Galaxy clusters in this
mass range typically have stellar to halo mass ratios of a few
percent (e.g., \citealp{Andreon12}, \citealp{Kravtsov14}), which
is consistent with our {\it Spitzer}-derived stellar mass estimate
above. We note, however, that these relations were derived using
lower redshift clusters. Based on their estimated stellar masses,
CARLA J2039$-$2514 and CARLA J0800+4029 have comparable stellar
mass to IDCS J1426.5+3508, implying their total halo masses are
likely $> 10^{14}\, M_{\odot}$. If confirmed by additional data,
this would place both structures among the most massive galaxy
clusters known at $z>1.5$.\\

\subsection{Individual Sources}\label{s:sourcesinterest}
This section discusses in detail cluster members of particular interest. The full sample of cluster members is discussed in Appendix \ref{ap:clmemb}. All spectra are shown in the same Appendix.\\

\subsubsection{CARLA J2039$-$2514}

\textbullet\  {\textbf{MRC~2036$-$254 (\#306):}} This source is the targeted HzRG. It was first identified in the Molongo Reference catalog of Radio Sources (\citealp{Large81}), and \cite{McCarthy96} reported a redshift of $z = 2.0$. Both \cite{Carilli97} and \cite{Kapahi98} presented radio observations of the source. The latter identified two lobes ``without an unambiguously identified core", whereas the former identified three lobes, including a pair of hot spots parallel to the radio axis in the north (\citealp{Overzier05}). \cite{Overzier05} also detected an X-ray component coincident with the southwestern lobe identified in \cite{Carilli97}. In both orientations we detect several emission lines ([\ion{O}{2}], H$\beta$, [\ion{O}{3}]), providing a redshift quality A. However, we only use the first observation to infer the source properties due to spectral overlap in the second observation.
We measure a new redshift of $z = 1.997 \pm 0.004$, consistent with the previous measurement. The source IRAC color is also consistent with this redshift, as expected for an AGN (e.g., \citealp{Stern05}).

\textbullet\  {\textbf{\#306b:}} The source is a previously 
unknown companion to the targeted HzRG, located $0.35 \arcsec$ to the 
northeast, with distinct continuum and emission lines seen offset
from MRC~2036$-$254.  The sources overlap in the second orientation,
so all measurements are based on the first visit (i.e., orientation).  Several lines
are detected, providing a redshift of $z = 1.999 \pm 0.004$ (quality
A) consistent with the HzRG. We find a slightly lower [\ion{O}{3}]
flux but a higher H$\beta$ flux as compared to MRC~2036$-$254, but
still with a sufficiently high line ratio to be suggestive (though
not conclusive) of an AGN ($\log_{10}(\rm[O~III]/H\beta) = 0.55$).
MRC~2036$-$254 and $\#306$b are likely a merging system\footnote{Note that type-2 RLAGN are often found to be associated with close merging systems (e.g., \citealp{Chiaberge15}).}, perhaps a
dual AGN, whose components are separated by $ 3 \rm~kpc$. With
this close separation, we take the conservative approach and do not
list $\#306$b as an isolated additional galaxy in the cluster, but
instead consider the system as a single galaxy at the redshift
measured for MRC~2036$-$254.\\

\begin{figure*}[!htb]
\begin{center}
\includegraphics[width=13cm]{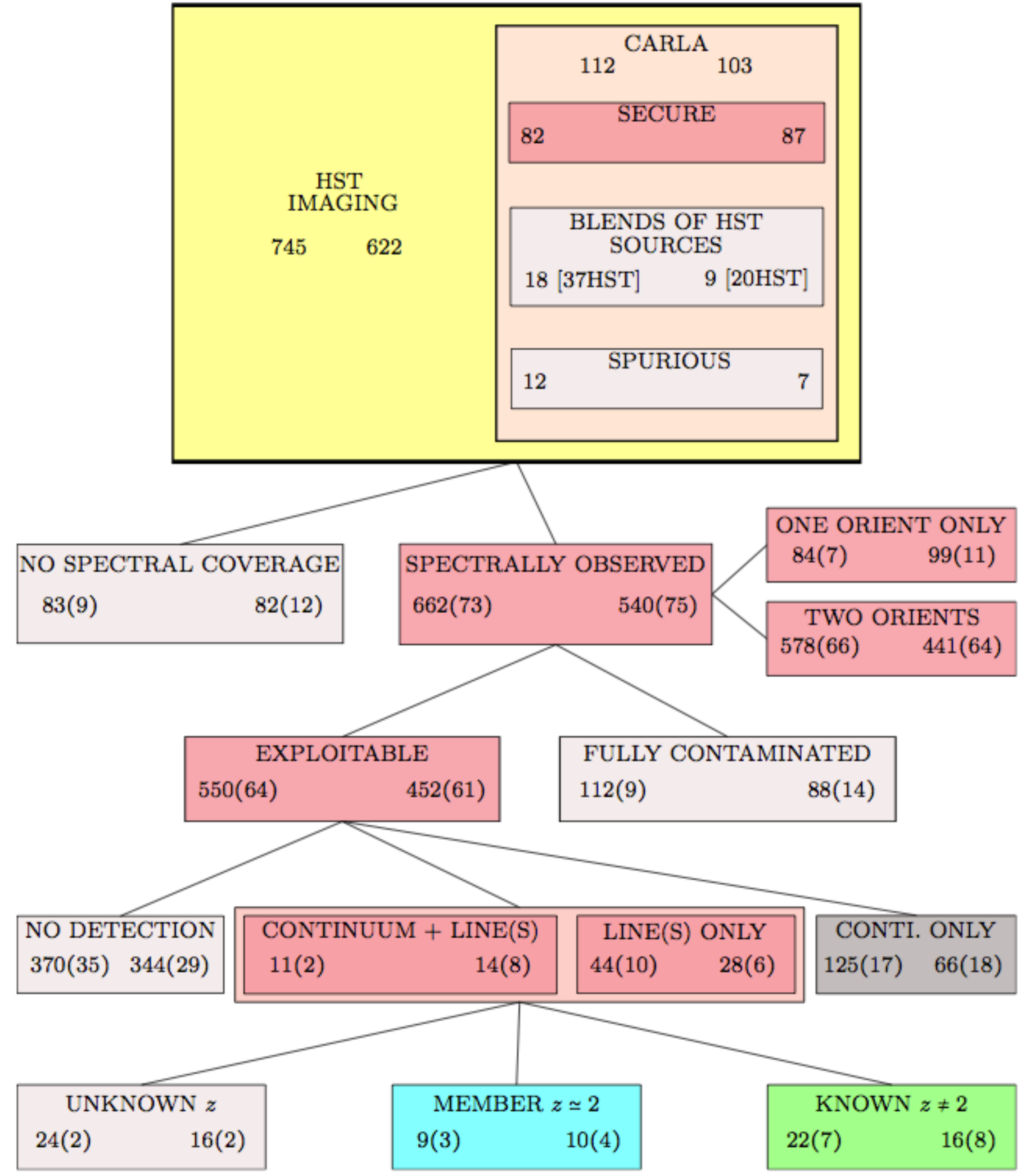}
\end{center}
\caption[Global Flowchart]{Flowchart of the classification of all sources from our master catalog. Inside each box, left side numbers correspond to the MRC 2036$-$254 field whereas the right side numbers correspond to the B3~0756+406 field. Numbers in parentheses refer to secure CARLA sources -- i.e., non-spurious {\it Spitzer} color-selected sources with a single {\it HST} counterpart.
	In total, we confirmed 9 and 10 cluster members for CARLA J2039$-$2514 and CARLA J0800+4029, respectively. Three and four of the members were selected as CARLA sources, respectively.
	}
\label{fig:Globflowchart}
\end{figure*}

\subsubsection{CARLA J0800+4029}

\textbullet\  {\textbf{B3~0756+406 (\#371):}} This source is the targeted QSO. It was first identified in the new Bologna sky survey (\citealp{Ficarra85}), and its redshift determined at $z = 2.021$ in SDSS Data Release 7 (DR7, \citealp{Schneider10}), albeit with some scatter across various SDSS analyses:  \cite{Hewett10} reported $z=2.014$ based on their improved redshifts for $91,000$ quasar spectra from SDSS DR6, while SDSS DR2 reported $z=2.026$ (\citealp{Abazajian04}). We did not reliably identify broad lines in the G141 spectra. Therefore, we re-observed the target at Palomar Observatory (see details in \S\ref{p:palomar}). From several emission lines (Ly$\alpha$, \ion{C}{4}, \ion{C}{3}], and \ion{Mg}{2}) we measure a new redshift of $z = 2.004 \pm 0.002$ (quality A), slightly lower than its previous lower-quality SDSS redshift. The source has an IRAC color consistent with this redshift, as expected for an AGN.

\textbullet\  {\textbf{\#372:}} The first observation of this galaxy is contaminated by the target quasar, B3~0756+406, but we obtain $z = 2.001 \pm 0.004$ from the second observation. We detect a break around $12,000\rm ~\AA$ consistent with a D4000 break at the redshift measured from [\ion{O}{3}]\footnote{Even though we extracted the 1D spectrum of the second observation from a contaminant-free region, we cannot totally exclude that the bright nearby continuum seen on the 2D cutout might be partly causing the D4000 break-like feature in the source continuum.}. H$\beta$ is tentatively detected slightly below our detection limit. Therefore, we assign a quality A to the redshift based on the detections of [\ion{O}{3}]$\lambda5007$ and the D4000 break. The detections indicate that this source possesses both old and young stellar population, and an active star-forming region. This source is located $\sim 3''$ to the northeast of B3~0756+406, implying a separation of $\sim 25\rm~kpc$ from the QSO (see Appendix \ref{ap:clmemb}).

\textbullet\  {\textbf{\#749:}} The second observation is contaminated over the full wavelength range, save for some pixels around $15,000\rm ~ \AA$ which only allow us to confirm the presence of emission lines seen in the first observation. Our flux measurements come from the first observation, and we measure both strong [\ion{O}{3}] and H$\beta$ emission. We measure a redshift of $z = 2.001 \pm 0.004$ (quality A). The source IRAC color is consistent with the measured redshift. This source shows broad H$\beta$, narrow [\ion{O}{3}], and a tentative [\ion{Ne}{3}] emission line, as seen in Appendix \ref{ap:clmemb}, which likely identifies this source as a QSO, even though we do not totally exclude that the poor G141 resolution might possibly bias its identification as an AGN.\\

\begin{figure*}
\centering
\includegraphics[width=8.97cm]{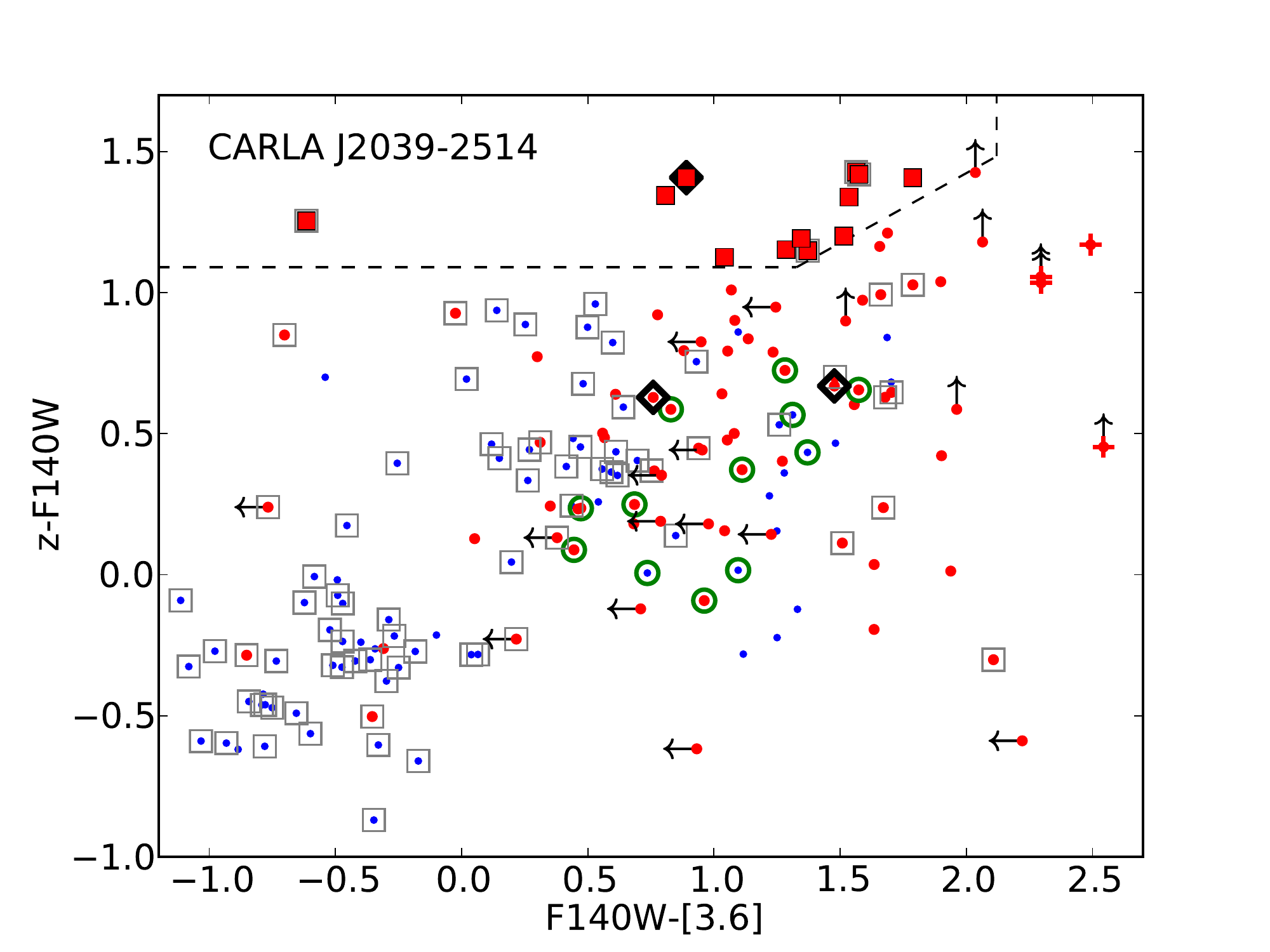}
\includegraphics[width=8.97cm]{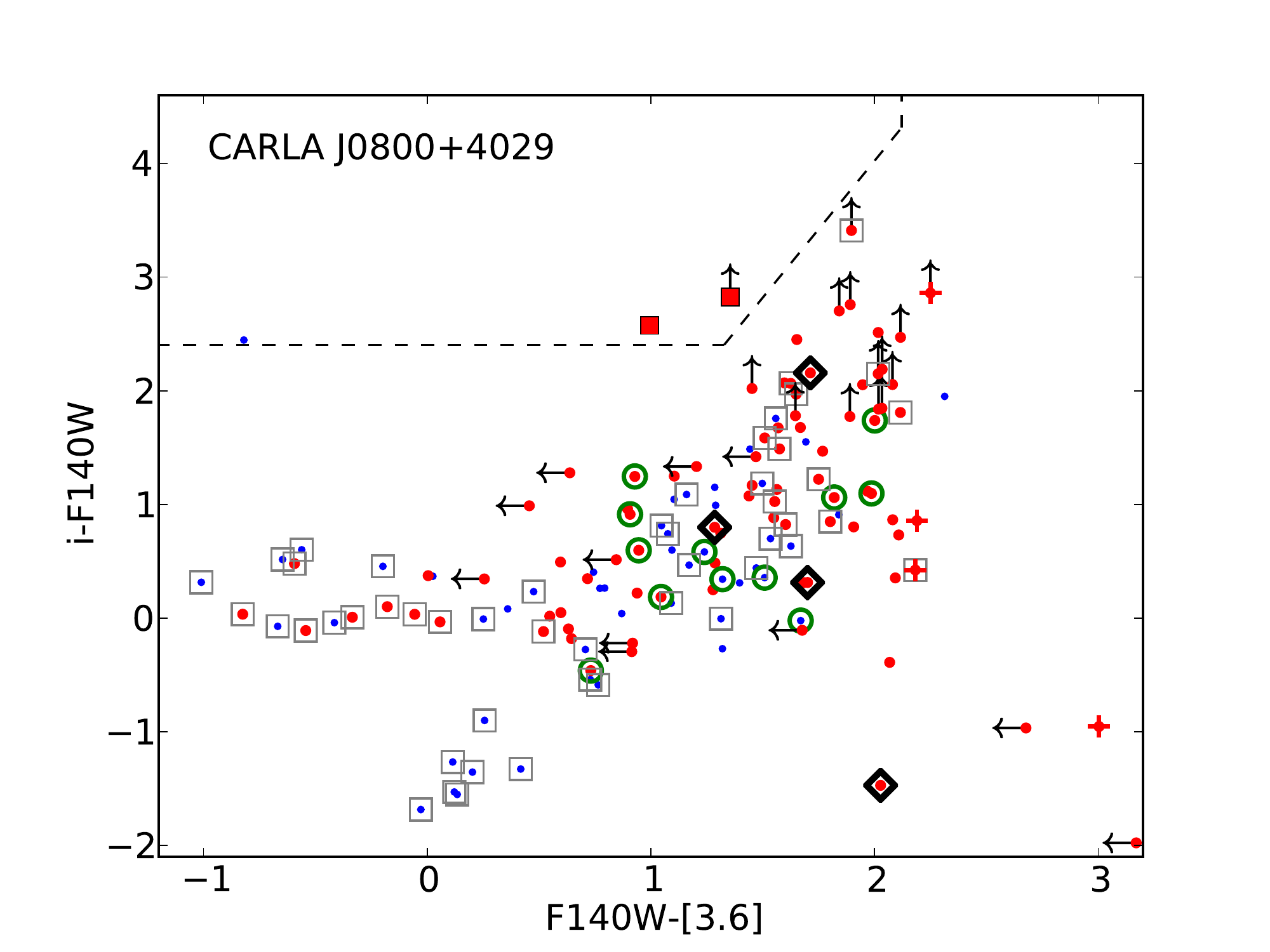}
\caption[Color-color diagrams of M2036 and J0800]{Color-color diagrams of CARLA J2039$-$2514 (left) and CARLA J0800+4029 (right). The dashed lines, adapted from \cite{Williams09} as described in Section \ref{sec:passivef}, separate passive, dusty SF, and SF galaxies. Passive galaxies are located inside the upper left quadrant and dusty SF galaxies on the right side of the quadrant vertical boundary. All CARLA candidates are shown by red markers, with the passive CARLA candidates highlighted by red squares and the dusty SF CARLA candidates by red plus ``+" signs; bluer sources with IRAC colors $<-0.1$ mag (AB) are shown by solid blue circles. Black diamonds highlight confirmed cluster members, and green rings are spectroscopically confirmed non-members. Open gray squares indicate continuum-only sources as identified in the grism data. Leftward arrows denote sources below our detection limit at $3.6\micron$, and upward arrows in the optical ($2\sigma$); their positions in the color-color plane are set using these limits.}
\label{fig:ccd}
\end{figure*}

\section{Discussion}\label{s:discussion}

\subsection{Method Efficiency}\label{sec:dismethod}
The two $z=2$ structures reported in this paper likely represent some of the highest redshift clusters currently spectroscopically confirmed. They illustrate the high efficiency of our approach to target RLAGN with IRAC mapping and {\it HST} grism follow-up. As the two fields are only a pilot study for our full sample of 20 cluster candidates, we next explore the efficiency of our grism spectroscopy and $[3.6] - [4.5]$ selection.\\

\subsubsection{Grism Efficiency}
We assess in this section the outcome of our {\it HST} observations.
In Figure \ref{fig:Globflowchart} we show the flowchart of the classification of {\it HST} sources from our master catalog. The classification is shown for both fields, with the numbers on the left side of each box corresponding to the MRC 2036$-$254 field and the numbers on the right side corresponding to the B3~0756+406 field. We also display in parentheses numbers corresponding to the classification for secure {\it Spitzer} color-selected candidates which have a single {\it HST} counterpart. Overall, we determine redshifts for $6\%$ of our exploitable {\it HST} sources (31/550 and 26/452, respectively, for the fields around MRC 2036$-$254 and B3~0756+406); where `exploitable' is defined to mean that $>75\%$ of the source continuum falls on the detector and contamination is less than $60\%$ of the cutout length. We find that $2\%$ of sources (9/550 and 10/452) are confirmed cluster members -- i.e., for every three redshifts that we measure, we find one cluster member. This is consistent with probing a biased environment, and is not an instrumental bias. For example, the redshift distribution of the 3D-HST field survey (\citealp{Momcheva15}) is roughly flat in the same redshift window ($0.7<z<2.3$), based on their sample of $46,256$ grism redshifts in this window obtained over $626~ \rm arcmin^{2}$ with a similar two-orbit per field $HST$ grism program. Of the exploitable {\it HST} sources, $71\%$ (370/550 and 344/452) are not detected in the dispersed grism data. Moreover, of the sources with spectral detections, $66\%$ (125/180 and 66/108) show continuum only, for which we could not measure a redshift because no emission lines were detected. These sources are likely a mixture of:  {\it (i)} stars, {\it (ii)} old and passive (quiescent) galaxies with little star formation, {\it (iii)} galaxies at redshifts for which no strong features are covered by our grism observations, and {\it (iv)} SF galaxies for which we do not detect emission lines at the depth of our data (SFR $\la 20~M_{\odot} \rm~ yr^{-1}$, unless highly dust-obscured). Also, we have not been able to determine a redshift for $40\%$ (24/55 and 16/42) of sources with detected emission line(s) (with or without continuum) because of ambiguity on the nature of the line(s). This number improves by a factor of 3.5 when considering {\it Spitzer} color-selected sources since we have a color information which helps identify ambiguous emission lines. Overall, we also lose $26\%$ (195/745 and 170/622) of the source spectra because of full contamination of their spectral first orders ($15\%$, 112/745 and 88/622) or because their traces fell outside of the G141 detector ($12\%$, 83/745 and 82/622).\\

\subsubsection{IRAC Color Selection Efficiency}
The primary aim of our {\it HST} program is to confirm galaxy clusters which were selected as overdense fields of mid-infrared color-selected galaxies. Hence, we also evaluate the success of the CARLA selection method given our {\it HST} observations. Focusing on the numbers in parentheses in the flowchart, we also find that a large fraction, $51\%$ (35/64 and 29/61, respectively for the fields around MRC 2036$-$254 and B3~0756+406) of the exploitable CARLA sources do not have any spectroscopic detection at the depth of our grism observations. This is better than the $71\%$ of all {\it HST} sources, implying that CARLA sources are brighter, on average, than our {\it HST} sources.
This is unsurprising since rest-frame near-infrared luminosity strongly correlates with stellar mass (e.g., \citealp{Gavazzi96}), and our IRAC $\rm 4.5 \micron$ flux cut imposes a limiting stellar mass around $1 \times 10^{10} ~M_{\odot}$ on the CARLA sources, whereas the {\it HST} F140W imaging also detects sources below the IRAC flux limit.

We determine the redshift of $18\%$ (10/64 and 12/61) of the exploitable {\it Spitzer} color-selected sources, and we find that $6\%$ (3/64 and 4/61) are cluster members. Again, we identify one cluster member for every three sources with redshift determinations. Among the CARLA sources with spectral detections, $57\%$ (17/29 and 18/32) show continuum without detectable emission lines. This suggests a large fraction of quiescent galaxies and/or galaxies with low or dust-extincted SFRs as potential cluster members (see Section \ref{sec:passivef}).
Note the similarity with \cite{Brodwin13}, albeit at lower redshift, who found that $\sim 40\%$ of cluster members at $1.37<z<1.50$ are SF galaxies based on the {\it Spitzer} 24$\micron$ emission, using an infrared-selected sample with stellar masses $>10^{10.1}~M_{\odot}$, and a SFR lower limit of $50~M_{\odot}\rm ~yr^{-1}$ for the SF members.
This is also consistent with the work of \cite{Cooke16}, who found that $76\%$ of CARLA galaxies ($M>10^{10}~M_{\odot}$) in the field of CARLA J1753+6311 ($z=1.58$) are quiescent. The CARLA selection method therefore finds both passive galaxies and SF galaxies, though the shallow grism data presented here are only capable of confirming the latter.\\

\subsection{Stellar Populations}\label{sec:passivef}
\cite{Williams09}  (see also \citealp{Labbe05}, \citealp{Wuyts07}, \citealp{Whitaker11}) have shown that it is possible to use rest-frame $UVJ$  colors to separate the SF, dusty SF and passive populations. To build our color-color diagrams (Fig. \ref{fig:ccd}), we use observed $z/i - \rm F140W$ vs. $\rm F140W - [3.6]$. Following \cite{Mei09} and using Mei's codes and the python version of {\ttfamily EZGAL} (\citealp{ManconeGonzalez12}), we transform \cite{Williams09} color limits into our observed apparent colors. We use a \cite{BruzualCharlot03} SSP model with galaxy formation redshifts averaged between $z_f=3$ and $8$, and metallicities equal to $40\%$ solar, solar and $2.5$ times solar; and we age the templates to $z=2$. We average the color conversions using a set of formation redshift ranges from $z_f=3-8$ to $z_f=5-8$ increasing by steps of $0.1$. This is a conservative simple model and assumes that passive galaxies are mostly located close to the passive boundary defined in  \cite{Williams09}. A more detailed analysis of the CARLA cluster stellar population will be performed in future work.

\begin{figure*}
\centering
\includegraphics[width=8.97cm]{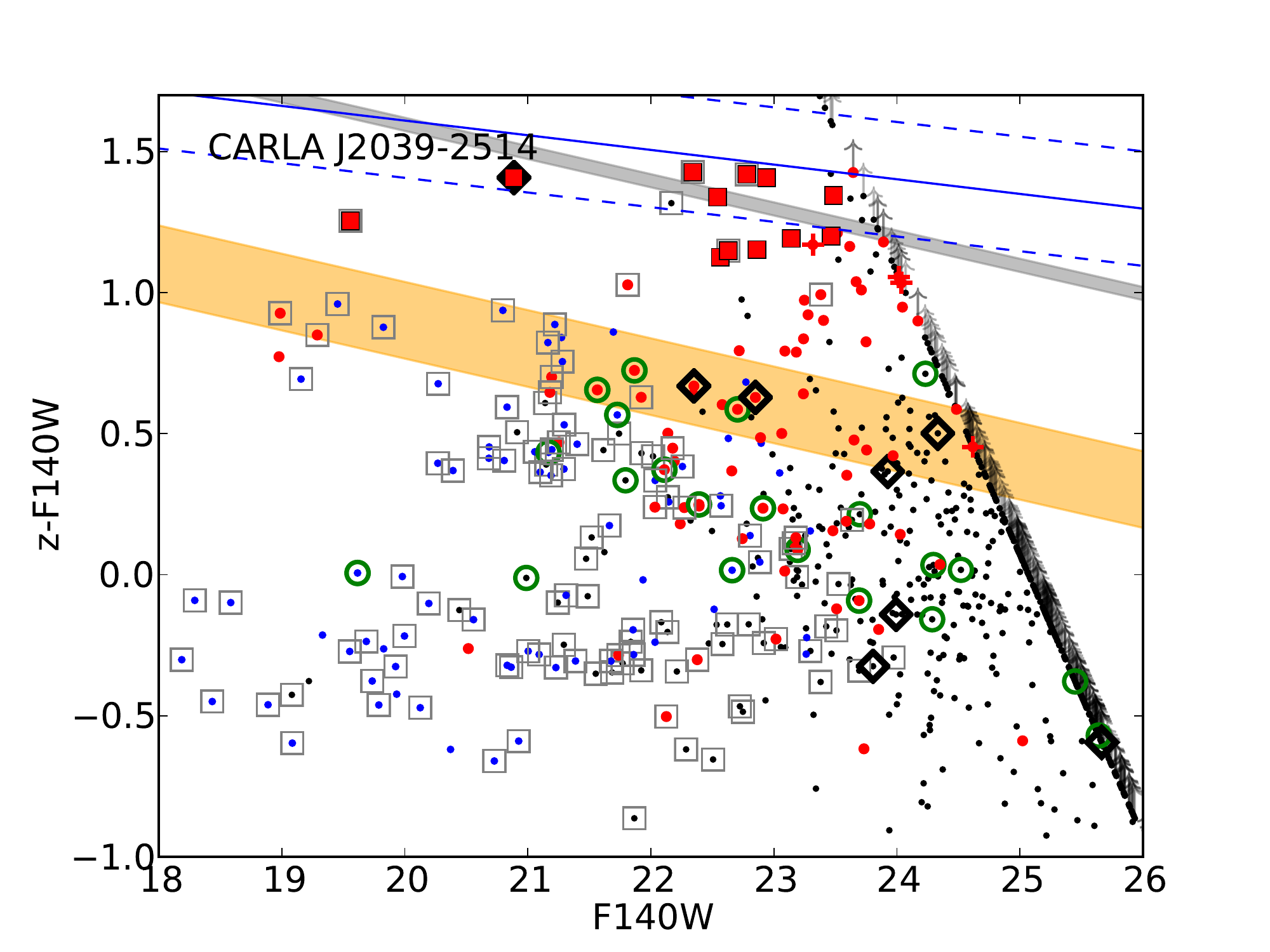}
\includegraphics[width=8.97cm]{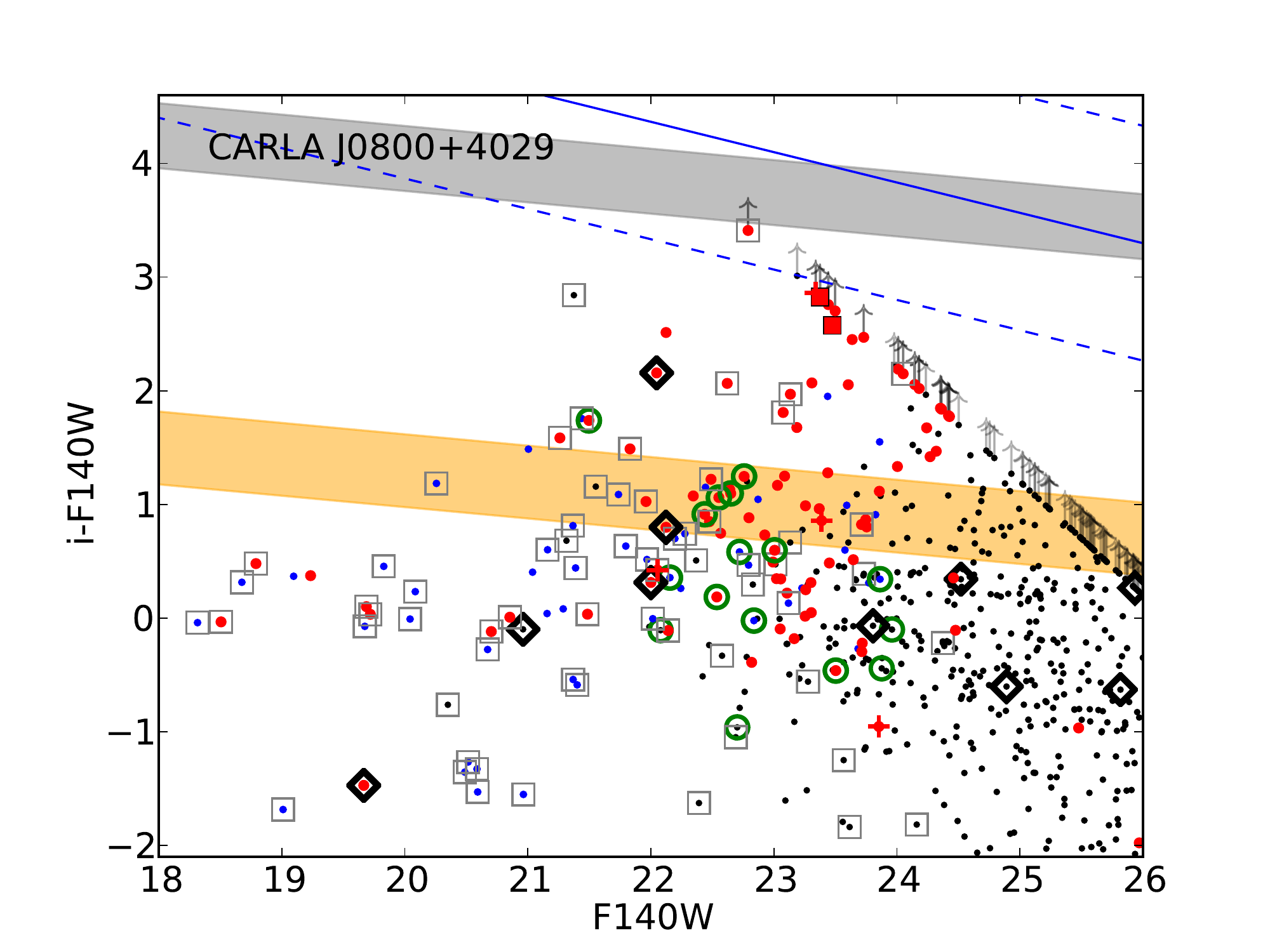}
\caption[Color-magnitude diagrams of M2036 and J0800]{Color-magnitude diagrams of CARLA J2039$-$2514 (left) and CARLA J0800+4029 (right). Same markers as in Figure \ref{fig:ccd}, with the addition of sources detected in the optical and F140W but not detected in IRAC shown by small black dots. Sources below our optical detection limits ($2\sigma$) are set to these limits and shown with upward arrows. The gray and orange shaded areas represent estimates of a $z=2.0$ red sequence for delta-burst and exponentially decaying stellar populations respectively, described in Section \ref{subsec:redseq}. The thickness of these regions correspond to formation redshifts in the range $z_{f} =  3.0 - 8.0$. The solid blue lines represent the $z=1$ \cite{Mei09} color-magnitude relation passively evolved to $z=2.0$, as described in Section \ref{subsec:redseq} (the dashed blue lines represent the $3\sigma$ range).}
\label{fig:cmd}
\end{figure*}

In Figure \ref{fig:ccd} we plot color-color diagrams for the two fields to investigate the presence of passive cluster members associated with CARLA J2039$-$2514 and CARLA J0800+4029, and in Figure \ref{fig:cmd} we plot their color-magnitude diagrams (CMDs). We use $i$-band data from \cite{Cooke15a} for CARLA J0800+4039 and a $4800~\sec$ $z$-band image obtained with VLT/ISAAC on UT 2002 July 17 for CARLA J2039$-$2514\footnote{Archival data from run ID 69.A-0234.}. For both CMDs, we use {\it HST}/F140W data to bracket the D4000 break at $z=2$.
In the color-color diagrams, the passive candidates are located inside the upper left quadrant. The SF population is located below the horizontal line, whereas dusty SF galaxies lie on the right of the vertical boundary.
We identify fourteen passive galaxies for CARLA J2039$-$2514 (two of them outside the plot) and two for CARLA J0800+4029.\\

\subsubsection{Density of Passive Candidates}\label{subsec:passivedendity}
We compare the density of passive, red ($([3.6]-[4.5])_{\rm AB} > -0.1$), sources in our RLAGN fields with densities of sources 
similarly selected in wide-field surveys. We make use of the 3D-HST multi-wavelength catalogs 
(\citealp{Skelton14}) in the five CANDELS fields (GOODS-North, GOODS-South, AEGIS, COSMOS and 
UDS; \citealp{Grogin11}). All CANDELS fields were at least partially covered by $i$-band, $z$-band, 
F140W, $3.6\micron$ and $4.5\micron$ observations. The F140W images have the most limited 
coverage; we therefore use the F140W image field of view to derive source densities (\citealp{Brammer12}). 
We isolate passive, red (i.e., CARLA), sources using the same color-selections as described above. At the depth of our RLAGN field data, we expect 
$\sim 9$ sources ($1.69 \pm 0.04$ arcmin$^{-2}$) selected by the $z$/F140W/[3.6] criterion in the $5.38$ arcmin$^{2}$ {\it HST} field of view of CARLA J2039$-$2514, and 
$\sim 2$ sources ($0.38 \pm 0.02 $ arcmin$^{-2}$) selected by the $i$/F140W/[3.6] criterion in the $5.54$ arcmin$^{2}$ {\it HST} field of view of CARLA J0800+4029. 
We identify 
$14$ ($\sim 2.6$ arcmin$^{-2}$) and 
$2$ ($\sim 0.4$ arcmin$^{-2}$) passive CARLA candidates for CARLA J2039$-$2514 and CARLA J0800+4029, respectively. 
CARLA J2039$-$2514 is therefore consistent with possessing an overdensity of
$z \sim 2$ passive red galaxies relative to the field, while our comparison suggests that CARLA J0800+4029 has similar values than the field. However, these are likely underestimates of the true (over)densities given the limiting magnitudes in the CARLA imaging of those fields.\\

\subsubsection{Red Sequences and SF Populations}\label{subsec:redseq}
In the CMDs (Fig. \ref{fig:cmd}), the orange and gray shaded areas show estimates of the expected color of $L^{*}$ early-type galaxies computed with a \cite{BruzualCharlot03} passive evolution model of solar metallicity and a \cite{Salpeter55} IMF\footnote{Using a \cite{Chabrier03} IMF instead does not significantly change the results.}. These were calculated using the python version of {\ttfamily EZGAL} (\citealp{ManconeGonzalez12}) with the appropriate filter transmission curves. We evolve a SSP (i.e., delta-burst), shown by the gray shaded areas, and an exponentially decaying stellar population of characteristic time-scale $\tau = 1 \rm ~Gyr$, shown by the orange shaded areas. The thickness of these regions correspond to formation redshifts over the range $3 < z_{f} < 8$. Both model evolutions are normalized to match the typical CARLA $L^{*}$ at $z=2.0$ (\citealp{Wylezalek14}), and assume slopes of $-0.1$, similar to what is observed in the Coma cluster at similar rest-frame wavelengths (\citealp{Eisenhardt07}).
To compare with the color-magnitude relation (CMR) observed in confirmed X-ray and infrared detected clusters at redshift $z\sim1$, we plot the \cite{Mei09} early-type CMR (solid blue lines; the blue dashed lines show the $3\sigma$ dispersion around the mean) evolved passively to $z=2$. 
Following \cite{Mei09}, we use a \cite{BruzualCharlot03} SSP model with galaxy formation redshifts between $3$ and $8$, and metallicities equal to $40\%$ solar, solar and $2.5$ times solar, to convert rest-frame $(U-B)$ vs. $M_{B}$ derived at $0.8<z<1.3$ to our observed bandpasses at $z=2$. We again average the color conversions using a set of ranges from $z_f=3-8$ to $z_f=5-8$. 
The \cite{Mei09} early-type galaxy CMR parameters are derived using {\it HST}/ACS filters that correspond to rest-frame $(U-B)$ and $M_{B}$ in the range $0.8<z<1.3$. While the $z$-band and F140W-band correspond to the same rest-frame for CARLA J2039$-$2514, the $i$-band probes bluer stellar populations than the $U$-band rest-frame. Within the uncertainties, these relations are consistent with the predictions from \cite{BruzualCharlot03}.

The majority of CARLA J2039$-$2514 passive candidates identified in Figure \ref{fig:ccd}, with $z-\rm F140W \ga 1$ mag and $\rm F140W > 20$ mag, agree well with the SSP models shown on the CMD (Fig. \ref{fig:cmd}), suggesting CARLA J2039$-$2514 hosts a population of quiescent galaxies consistent with a cluster red sequence at $z=2.00$. On the other hand, CARLA J0800+4029, unlike CARLA J2039$-$2514, does not exhibit $\rm F140W < 23$ mag passive candidates consistent with a cluster red sequence, and the depth of our $i$-band data prevents us from confirming $\rm F140W > 23$ mag sources for this field. 
Specifically, thirteen sources (marked with upward arrows in Figures \ref{fig:ccd} and \ref{fig:cmd}) do not have $i$-band detections and therefore require deeper data to determine whether they are passive candidates populating what could constitute a nascent cluster red sequence. 
In each color-color diagram, we only identify a handful of dusty SF CARLA candidates, none of which are spectroscopically confirmed as members or non-members.
All spectroscopically confirmed members and non-members are identified as SF galaxies in the color-color diagrams, as expected, with the exception of the target RLAGN in CARLA J2039$-$2514 which falls in the passive region as expected for a type-2 RLAGN.  
Four of the confirmed members of CARLA J2039$-$2514 seem to be well described by the exponentially decaying model of star formation. The remaining two members that have both F140W and $z$-band detections, and most of the confirmed members of CARLA J0800+4029 do not seem to agree with this model, suggesting that the build-up of their stellar content followed a diversity of star-formation histories. This supports \cite{Cooke15a} who showed that the star formation histories (SFHs) of CARLA cluster galaxies are best described by multiple bursts of star formation normally distributed over a few Gyrs, or distributed following the cosmic SFH. Exploration of a range of SFHs is however beyond the scope of this paper and will be addressed in future work. Finally, we note that 
only one blue ($([3.6] - [4.5])_{\rm AB} < -0.1$ mag) source falls in the passive region of the color-color diagrams, and that the putative red sequence of CARLA J2039$-$2514 almost exclusively comprises a population of quiescent galaxies up to $\rm F140W<24$ mag. The spectroscopic confirmation of these passive CARLA candidates, however, will require significantly deeper data.\\

\subsection{Comparison to Other High-Redshift Clusters}\label{p:comparison}
We compare our results to other high-redshift clusters spectroscopically confirmed using {\it HST}/WFC3 grism data. \cite{Gobat13}, hereafter G13, reported on a rich cluster at $z \simeq 2$. They first reported evidence of a fully established galaxy cluster at $z=2.07$ (\citealp{Gobat11}) from X-ray emission, ground-based spectroscopy, and ground- and space-based photometry, later revised by G13 to a slightly lower redshift using deep {\it HST} grism observations. The system consists of two unrelated aligned structures, with the background overdensity being ``sparse and sheet-like". The G13 median redshift of the {\it HST}/WFC3 confirmed members of the foreground overdensity is $z = 1.993$, with a standard deviation of $0.012$. With median redshifts of $z = 2.000$ (standard deviation 0.005) and $z = 1.986$ (standard deviation 0.014), CARLA J2039$-$2514 and CARLA J0800+4029, respectively, are among the most distant confirmed clusters currently known. G13 confirmed 22 members with 18 orbits and three orientations (12.5 hrs on source) including five quiescent sources, and found no evidence for an already formed red sequence within $20\arcsec$ of the cluster core (\citealp{Gobat11}). More recently, \cite{Newman14}, hereafter N14, confirmed another rich cluster at high redshift using {\it HST}/WFC3 slitless spectroscopy. With 14 orbits, they confirmed 19 members at $z = 1.80$, of which more than $75\%$ are quiescent. They found a clear red sequence of observed mean color $\left<z - J\right> = 1.98 \pm 0.02$, which includes 13 of their 15 quiescent cluster members. \cite{Zeimann12}, hereafter Z12, also confirmed both emission-line and quiescent sources in a cluster at $z  = 1.89$ using six orbits of {\it HST}/WFC3 grism spectroscopy. Z12 showed that a significant fraction of early-type galaxies in the cluster field were consistent with forming a red sequence. These deep observations show that clusters can host a substantial fraction of quiescent galaxies even at early epochs. By design, our shallow two-orbit per field strategy only confirms SF members. In addition to the confirmed emission-line members, we find a large fraction ($79\%$, 52/64 and 47/61) of {\it Spitzer} color-selected cluster candidates below our detection limit or showing continuum only. 
CARLA J2039$-$2514 and CARLA J0800+4029 are therefore relatively robust confirmations, and have a high potential for being richer structures than what our shallow {\it HST} observations allowed us to unveil.
Furthermore, we note that our two-orbit strategy spectroscopically confirms $\sim 5$ cluster members per orbit, which is approximately four times more efficient than the $>10$-orbit programs reported in G13 and N14.

From X-ray emission and richness, G13 estimated a mass of $M_{200} \sim 5\times 10^{13} ~M_{\odot}$ for their cluster. N14 reported a massive cluster with $M_{200} = (2-3) \times 10^{14}~M_{\odot}$, including five very massive members whose stellar masses are in the range $(4-10) \times 10^{11}~M_{\odot}$. We do not find such high masses for our confirmed cluster members (with the exception of the RLAGN); our seven IRAC-detected sources have a median stellar mass of $1.1 \times 10^{11}~M_{\odot}$, and the remaining twelve non-IRAC detected members have masses $< 10^{10}~M_{\odot}$. Z12 derived SFRs from the [\ion{O}{2}] and H$\beta$ lines and find SFRs in the range $(20-40) ~M_{\odot} \rm~yr^{-1}$. Based on the admittedly crude [\ion{O}{3}] SFR indicator, we find similar SFRs for 9/16 of our SF cluster members. We also find higher SFRs, in the range $(40-140) ~M_{\odot} \rm~yr^{-1}$, for a significant number of SF cluster members (7/16). Note that our line detection limit imposes a SFR lower limit of $>20 ~M_{\odot}\rm ~yr^{-1}$ (similar to Z12 but based on a different line). According to the SFR/stellar-mass relation in \cite{Rodighiero11} for $1.5 < z < 2.5$ galaxies, our low-mass cluster members (typically all the non-IRAC detected members, with masses $\la 10^{10}~M_{\odot}$) are above the star-forming main-sequence towards starburst galaxies. Our detection limits prevent us from confirming main-sequence galaxies at these masses and redshifts.

In Section \ref{p:sfr} we showed that confirmed emission line members of
both structures imply total cluster SFRs of at least $\sim 400\,
M_\odot\, {\rm yr}^{-1}$ within $\sim 500$~kpc of the cluster
centers, which are assumed coincident with the target RLAGN. Based
on {\it Spitzer} $24 \micron$ imaging of a sample of clusters from the
{\it Spitzer}/IRAC Shallow Cluster Survey, \cite{Brodwin13}
showed a steeply increasing SFR in cluster cores out to $z = 1.50$,
with an average SFR of several hundred $M_\odot\, {\rm yr}^{-1}$
found in the cores of the highest redshift clusters in that study.
\cite{Alberts14, Alberts16} find similar results based on longer
wavelength {\it Herschel} data of a similar cluster sample. The
results found here at $z \sim 2$ for CARLA~J2039$-$2514 and
CARLA~J0800+4029 are consistent with those studies, though higher
fidelity SFR indicators for these newly confirmed structures would
be highly preferable to the current [\ion{O}{3}]-based values.

Although G13 found no evidence for an already formed red sequence within $20\arcsec$ of the cluster core, they determined, from their best subsample of ($96$) cluster candidates comprising $14$ spectroscopic members and $82$ photo-$z$ candidates (expected to include $50$ interlopers), that $\sim(60-80)\%$ of candidates within $\sim 20\arcsec$ of the cluster core are passive regardless of mass, compared to $\sim 20\%$, $\sim 40\%$ and $\sim 60\%$ in the field for $\log M/M_{\odot} > 10, 10.5,$ and $11$, respectively (\citealp{Strazzullo13}).
We additionally note that the structure reported in \cite{Spitler12} and \cite{Yuan14}, respectively discovered at $z=2.2$ from photometric redshifts and later spectroscopically confirmed at $\left<z\right>=2.095$ with $57$ members, comprises several $1\arcmin$ radius overdense groups covering a $12\arcmin \times 12\arcmin$ area in COSMOS (\citealp{Scoville07}). This structure has a slightly enhanced number of red galaxies for two groups compared to the field, with $N_{red}=0.5\pm0.2 \times N_{tot}$ compared to $0.2\pm0.03$ in the field.
Our results are also indicative of the presence of passive CARLA candidates associated with CARLA J2039$-$2514, 
while CARLA J0800+4029 exhibits a number of passive candidates similar to the field in our comparison.
A careful analysis accurately evaluating the passive fraction of galaxies in our structures will be addressed in Cooke et al. (in prep).\\

\section{Summary}\label{sec:con}

We conclude the following from our {\it HST}/WFC3 F140W and G141 follow-up observations on two overdense CARLA fields, which are the first two of a sample of 20 cluster candidates at $1.4 < z < 2.8$.

\begin{itemize}
\item[1.] We spectroscopically confirm two {\it Spitzer} color-selected
overdensities as high-redshift structures. 
Adopting the \cite{Eisenhardt08} criteria defining a $z>1$ galaxy
cluster through spectroscopic confirmation, these two structures are among the most distant clusters
currently known. Furthermore, though we note that the \cite{Eisenhardt08} criteria may also identify sheets, filaments, groups and
protoclusters when applied to grism data, the structures reported here possess additional
attributes typically used to identify galaxy clusters:  CARLA
structures are, on average, centrally concentrated, and CARLA J2039$-$2514 has an overdensity of red galaxies
consistent with being passive cluster members. Our results suggest that CARLA J2039$-$2514 is a bona fide galaxy cluster. While CARLA J0800+4029 conforms to the \cite{Eisenhardt08} criteria, has a centrally concentrated $7.8\sigma$ overdensity of {\it Spitzer} color-selected (i.e., red) candidate cluster members, and has comparable stellar mass to the most massive cluster known to date at $z>1.5$, it lacks a clear overdensity of passive candidates and a red sequence population in the current data, suggestive of a younger forming cluster.

\item[2.] We identify CARLA J2039$-$2514 at $\left< z \right> = 1.999$ (median $z=2.000$), consisting of 9 confirmed members, including a potential dual AGN. We also identify CARLA J0800+4029 at $\left< z \right> = 1.986$ (median $z=1.986$), consisting of 10 confirmed members including two quasars. We estimate median (mean) SFRs of $\sim 35~ M_{\odot} \rm ~yr^{-1}$ ($\sim 50~ M_{\odot} \rm ~yr^{-1}$) and average stellar masses of $\la 1 \times 10^{11}~M_{\odot}$ for the confirmed star-forming members of both CARLA J2039$-$2514 and CARLA J0800+4029.

\item[3.] Considering the total background-subtracted mid-infrared light
from the two structures, we find that the inferred total stellar
masses are comparable to the most massive clusters known at slightly
lower redshift ($z \sim 1.7$). This analysis crudely implies these
structures have total masses of $M_{500} \sim 10^{14}\, M_\odot$.

\item[4.] With just two orbits, we only confirm emission-line sources with SFRs $> 20~ M_{\odot} \rm ~yr^{-1}$.
We show that {\it HST} grism spectroscopy is efficient at confirming galaxy clusters at high-redshift even with shallow observations, but that deeper spectroscopic data is required to confirm the clusters' passive population.

\item[5.] We study the two cluster color-magnitude relations. 
While CARLA J2039$-$2514 shows a population of red and quiescent galaxies where we would expect a red sequence at these redshifts, CARLA J0800+4029 does not exhibit such a population up to $F140W < 23$ mag. However, a promising thirteen red CARLA J0800+4029 sources with $F140W > 23$ mag require deeper optical imaging to determine whether they are passive candidates and populating a nascent cluster red sequence. We conclude that CARLA J2039$-$2514 already hosts a population of quiescent galaxies, with little contamination from obscured star-forming galaxies.

\item[6.] We show that our CARLA selection is robust. It efficiently selects overdense fields at high redshifts while potentially selecting at the same time the most likely detectable emission-line sources. Our {\it Spitzer} mid-infrared color-selection increased the source likelihood of being a cluster member by a factor of three compared to no color-selection in these biased environments.

\item[7.] Our low-resolution observations have redshift uncertainties up to $1000~\rm km~\sec^{-1}$. Higher resolution spectroscopy will be required to reliably measure velocity dispersions and cluster masses. Future observations could also provide accurate SFRs for our cluster members using more robust estimators such as H$\alpha$ luminosities and far-infrared observations, and better determine stellar masses using additional photometric data.\\
\end{itemize}

\acknowledgments
We thank our anonymous referee for her/his comments and suggestions that improved the quality of this paper.
This work is based on observations made with the NASA/ESA {\it Hubble Space Telescope}, obtained at the Space Telescope Science Institute, which is operated by the Association of Universities for Research in Astronomy, Inc., under NASA contract NAS 5-26555. This work is also based in part on observations made with the {\it Spitzer Space Telescope}, which is operated by the Jet Propulsion Laboratory, California Institute of Technology, under a contract with NASA. This work is also based in part on observations made with the $200$-inch Hale Telescope, Palomar Observatory, operated by the California Institute of Technology. D.W. acknowledges support by Akbari-Mack Postdoctoral Fellowship. S.M. acknowledges financial support from the Institut Universitaire de France (IUF), of which she is senior member. E.A.C acknowledges the support of the STFC. N.A.H acknowledges support from STFC through an Ernest Rutherford Fellowship.\\

{\it Facilities:} \facility{HST (WFC3; STScI)}, \facility{Spitzer (IRAC; JPL/Caltech)}, \facility{Palomar (DBSP; Caltech)}.\\

\appendix

\section{Notes on data processing}\label{ap:notesdataprep}

\subsection{FLT Files}\label{ap:flt}
Three different kinds of calibrated data can be retrieved from MAST: ``calibrated, flat-fielded individual exposures" (FLT), ``calibrated, cosmic-ray-rejected, combined images" (CRJ), and ``calibrated, geometrically corrected, dither-combined images" (DRZ; WFC3 Data Handbook, \citealp{Rajan11}). According to the handbook, the {\ttfamily Astrodrizzle} package of the aXe software for slitless spectroscopy data extraction (\citealp{Kümmel09}) supersedes the CRJ and DRZ data preparations of the standard WFC3 calibration program ({\ttfamily calwf3}). Therefore, we only retrieved FLT files, which are used as input by aXe.\\

\subsection{Image Combination and Source Extraction}\label{ap:combi}
We first used the aXe (v2.4.4) {\ttfamily astrodrizzle} task to combine the eight F140W dithered direct exposures of each field (four from each orientation). This step creates a deep ($1023-1073 \sec$) drizzled direct image with bad pixels and cosmic rays rejected. The combined direct images of each field are shown in Figure \ref{fig:spadist}, with the positions of confirmed cluster members indicated.

In order to associate sources and spectra and to wavelength calibrate the spectra, we first create a catalog of sources in each field using SExtractor (\citealp{BertinArnouts96}) on the drizzled image. We create a deep catalog using SExtractor detection parameters selected to identify even the faintest sources ($\rm DETECT\_MINAREA = 4$, $\rm DETECT\_THRESH = 2$, and $\rm ANALYSIS\_THRESH = 2$). Lower values only add spurious detections (e.g., noise clumps, stellar diffraction spikes, dismemberment of extended sources) while higher values miss a significant number of faint sources. This deep catalog, gathering information on source positions, magnitudes, sizes, and orientations, is then cleaned by hand from spurious detections.
We also generate a more conservative catalog using $\rm DETECT\_MINAREA = 10$, $\rm DETECT\_THRESH = 3$, and $\rm ANALYSIS\_THRESH = 3$ to restrict the number of detections for the grism sky background subtraction (see Appendix \ref{ap:back}).\\

\subsection{{\it HST--CARLA} Cross-Correlation}\label{ap:cross}
We cross-correlate our {\it HST} and CARLA/{\it Spitzer} catalogs for two reasons. First, we update the nature of the CARLA sources themselves. With nearly ten times better spatial sampling than {\it Spitzer}, {\it HST} provides morphological information and identifies some IRAC sources as spurious detections or blends, potentially leading to misfigured IRAC colors and erroneous CARLA selection. Second, we wish to investigate the spectra of all CARLA sources in our grism data. To the cleaned deep catalog we add {\it Spitzer} CARLA candidates too red to be detected in the {\it HST} direct imaging. Even though these undetected sources may not show any continuum, they can potentially show emission lines in the grism data. This step required a slight astrometric correction ($\sim 0.2\arcsec$) to the IRAC positions. We based our correction on secure CARLA sources identified in the first step. At this point, we now have our cleaned, complete and final catalog of sources for spectroscopic analysis, referred to as the master catalog.\\

\subsection{Back Projection and Background Subtraction}\label{ap:back}
Exposure blocks (pairs of direct and grism observations) are slightly offset from one another. Therefore, we cannot directly associate the master catalog sources to their spectra. Using the {\ttfamily Astrodrizzle} aXe task {\ttfamily iolprep}, we then project back the master catalog to each individual direct image and obtain individual catalogs corresponding to each grism exposure, required for stacking the grism data of each observation.

Prior to stacking the grism data, we remove the overall background level of the G141 detector, which varies significantly along its surface ($\sim (0.9 - 2.4)~e^{-}~\sec^{-1}$).
The aXe task {\ttfamily axeprep} scales a master sky background on spectral-free regions. We use the shallow catalog, projected back to individual frames, as our master catalog is too deep for aXe to produce enough sky-free regions.\\

\subsection{Contamination Models}\label{ap:contam}
Two kinds of contamination models can be generated using the aXe task {\ttfamily axecore}: quantitative or qualitative. For each source, the quantitative contamination model produces a 2D cutout of the first order grism spectrum, and shows how many potential contaminating spectral traces from other sources fall in the cutout. However, this model is purely geometrical and includes all spectral orders from all sources regardless of their likelihood of actually being detected. This leads to large over-estimation of the contamination and typically no contaminant-free regions. On the other hand, the qualitative contamination model computes the shapes and flux intensities of the spectral traces, based on position, magnitude, and size of the sources. One can choose between the ``Gaussian" and the ``fluxcube" methods (described in the aXe User Manual (v2.3), \citealp{Kümmel11}). Briefly, the ``Gaussian" method approximates the shapes of sources as Gaussians and requires magnitudes from at least one band to estimate flux intensities (more bands provide better estimates). The ``fluxcube" method creates more realistic spectral trace shapes as it models the morphologies of the sources from at least one {\it HST} image and a segmentation map (again, more images from multiple bands provide better estimates). \cite{Gobat13} report that the ``fluxcube" method is often a poor approximation of the spectrum, even with several multi-band {\it HST} images available. With only F140W available to us, we use the ``Gaussian" method to locate traces and zeroth orders of possible contaminants.\\

\section{Cluster member properties}\label{ap:membprop}

We present in Table \ref{table:examplelargetable} the properties of the members of our two clusters, CARLA J2039$-$2514 and CARLA J0800+4029.\\

\begin{sidewaystable*}[htbp]
\caption{Confirmed cluster members}\smallskip
\label{table:examplelargetable}
\centering
\begin{tabular}{ lccccccccccc }
\hline \hline
Field & Object & \multicolumn{1}{p{1.8cm}}{\centering RA (hh:mm:ss.ss)} & \multicolumn{1}{p{1.8cm}}{\centering Dec (dd:mm:ss.s)} & \multicolumn{1}{p{1.6cm}}{\centering F140W (AB)} & \multicolumn{1}{p{1.6cm}}{\centering [3.6] \\ (AB)} & \multicolumn{1}{p{1.6cm}}{\centering [4.5] \\ (AB)} & $f_{\rm[OIII]}$\tablenotemark{a} & $f_{\rm H\beta}$\tablenotemark{a} & $z$ & Quality\tablenotemark{b} & Remarks\tablenotemark{c}\\ \hline
CARLA J2039$-$2514&	119&  	20:39:28.51&	-25:14:31.2&   $22.35 \pm 0.02$&  20.87 $\pm$ 0.02& 	20.88 $\pm$ 0.02& 	6.5 $\pm$ 2.9& -& 1.987 $\pm$ 0.007& B$^{+}$& SF\\
\mbox{}& 			174&  	20:39:22.61&  	-25:14:16.0&   $22.87 \pm 0.03$&  22.09 $\pm$ 0.06& 	22.13 $\pm$ 0.05&   5.1 $\pm$ 1.5& -& 2.000 $\pm$ 0.007& B$^{+}$& SF\\
\mbox{}& 			281&  	20:39:22.14 &  	-25:15:08.6&   $23.98 \pm 0.06$&   	-& 				-& 				8.4 $\pm$ 3.2& -& 2.002 $\pm$ 0.008& B$^{-}$& SF\\
\mbox{}& 			306&  	20:39:24.49 &  	-25:14:30.7&   $20.94 \pm 0.01$&  19.99 $\pm$ 0.01& 	19.79 $\pm$ 0.01& 	74.7 $\pm$ 3.8& 7.3 $\pm$ 3.1& 1.997 $\pm$ 0.004& A& HzRG\\
\mbox{}& 			306b\tablenotemark{d}&  	20:39:24.52 &  	-25:14:30.6&   -&   		-& 				-& 				59.5 $\pm$ 2.7& 17.7 $\pm$ 4.0& 1.999 $\pm$ 0.004& A& AGN\\
\mbox{}& 			356&  	20:39:24.61 &  	-25:14:34.2&   $24.41 \pm 0.07$&   	-& 				-& 				8.0 $\pm$ 1.9& -& 1.999 $\pm$ 0.005& B$^{-}$& SF\\
\mbox{}& 			360&  	20:39:25.74 &  	-25:14:07.9&   $23.92 \pm 0.06$&   	-& 				-& 				10.1 $\pm$ 2.8& -& 2.006 $\pm$ 0.004& B$^{-}$& SF\\
\mbox{}& 			697&  	20:39:21.32 &  	-25:13:55.6&   $23.80 \pm 0.05$&   	-& 				-& 				5.1 $\pm$ 1.8& -& 2.001 $\pm$ 0.006& B$^{-}$& SF\\
\mbox{}& 			44300&  	20:39:24.57 &  	-25:14:22.2&   $25.66 \pm 0.16$&   	-& 				-& 				3.5 $\pm$ 1.3& -& 2.002 $\pm$ 0.005& B$^{-}$& SF\\
\mbox{}& 			90000&  	20:39:21.33 &	-25:14:25.3&   $26.35 \pm 0.24$&   	-& 				-& 				4.0 $\pm$ 1.5& -& 1.995 $\pm$ 0.005& B$^{-}$& SF\\
\hline
CARLA J0800+4029& 	146&  08:00:12.40&   +40:30:40.4&   $24.89 \pm 0.09$&  -& 				-& 				   4.4 $\pm$ 1.8&   	-& 1.998 $\pm$ 0.007& B$^{-}$& SF\\
\mbox{}& 					371&  08:00:16.11&   +40:29:55.6&   $19.66 \pm 0.01$&  17.64 $\pm$ 0.01& 17.22 $\pm$ 0.01& -&		         	      	-& 2.004 $\pm$ 0.002\tablenotemark{e}& A& QSO\\
\mbox{}& 					372&  08:00:15.94 &  +40:29:53.2&   $20.96 \pm 0.01$&   Blend& 				Blend& 				   22.8 $\pm$ 1.7& 	-& 2.001 $\pm$ 0.004& A& SF\\
\mbox{}& 					401&  08:00:18.54 &  +40:30:39.0&   $25.94 \pm 0.24$&   -& 				-& 				   5.4 $\pm$ 3.4&    	-& 1.980 $\pm$ 0.009& B$^{-}$& SF\\
\mbox{}& 					436&  08:00:14.76 &  +40:29:48.1&   $22.05 \pm 0.01$&   20.33 $\pm$ 0.01& 	20.19 $\pm$ 0.01& 	   5.6 $\pm$ 2.1&    	-& 1.969 $\pm$ 0.007& B$^{+}$& SF\\
\mbox{}& 					443&  08:00:17.29 &  +40:30:30.0&   $24.52 \pm 0.07$&   -& 				-& 				   7.6 $\pm$ 1.3&    	-& 1.992 $\pm$ 0.004& B$^{-}$& SF\\
\mbox{}& 					542&  08:00:11.95 &  +40:29:24.3&   $22.12 \pm 0.01$&   20.83 $\pm$ 0.02& 	20.78 $\pm$ 0.02& 	   18.4 $\pm$ 5.1&	-& 1.964 $\pm$ 0.007& B$^{+}$& SF\\
\mbox{}& 					591&  08:00:11.61 &  +40:29:38.0&   $25.82 \pm 0.18$&   -& 				-& 	   			   3.4 $\pm$ 1.6 &	-& 1.975 $\pm$ 0.007& B$^{-}$& SF\\
\mbox{}& 					749&  08:00:16.12 &  +40:28:57.3&   $22.00 \pm 0.02$&   20.30 $\pm$ 0.01& 	20.10 $\pm$ 0.01& 	   27.1 $\pm$ 2.2& 	27.8 $\pm$ 3.5& 2.001 $\pm$ 0.004& A& QSO\\
\mbox{}& 					767&  08:00:18.96 &  +40:29:38.9&   $23.81 \pm 0.05$&   23.84 $\pm$ 0.24& 	22.97 $\pm$ 0.09& 	   4.2  $\pm$ 1.2&	-& 1.978 $\pm$ 0.004& B$^{+}$& SF\\
\hline
\end{tabular}
\tablenotetext{1}{Flux in units of $\rm 10^{-17} \rm erg ~\sec^{-1} ~cm^{-2}$.}
\tablenotetext{2}{See Sec.\ref{sec:reddet}.}
\tablenotetext{3}{SF: star-forming galaxy, HzRG: high-redshift radio-galaxy, AGN: active galactic nucleus, QSO: quasi-stellar object.}
\tablenotetext{4}{This source is the companion to MRC~2036$-$254 ($\#306$).}
\tablenotetext{5}{Determined from the Palomar spectrum described in Sec.\ref{p:palomar}.}
\end{sidewaystable*}

\section{Non cluster member emission line sources}\label{ap:nonmembprop}

We present in Table \ref{table:nonCl} the measured redshifts of non-cluster member sources in the fields of view of our clusters.\\

\begin{table}
\caption{Confirmed non-cluster members}\smallskip
\label{table:nonCl}
\centering
\begin{tabular}{ lccccccc }
\hline \hline
\multicolumn{1}{p{1.2cm}}{Field}& \multicolumn{1}{p{1.2cm}}{\centering Object} & \multicolumn{1}{p{1.8cm}}{\centering RA (hh:mm:ss.ss)} & \multicolumn{1}{p{1.8cm}}{\centering Dec (dd:mm:ss.s)} & \multicolumn{1}{p{1.8cm}}{\centering F140W \\ (AB)} & \multicolumn{1}{p{1.8cm}}{\centering $z$} & \multicolumn{1}{p{1.2cm}}{\centering Quality\tablenotemark{a}} & \multicolumn{1}{p{2.0cm}}{\centering Line(s) fitted\tablenotemark{b}}\\ \hline
CARLA J2039$-$2514&               	  			54&     20:39:28.65& -25:14:51.9& $21.79 \pm 0.01$& $1.219 \pm 0.005$ & A & H$\alpha$\\
\mbox{}&							83&     20:39:28.83& -25:14:36.9& $21.72 \pm 0.01$& $0.984 \pm 0.005$ & B$^{+}$ & H$\alpha$\\
\mbox{}&							133&   20:39:21.92& -25:14:19.3& $24.42 \pm 0.08$& $2.289 \pm 0.006$ & A & [\ion{O}{3}], [\ion{O}{2}]\\
\mbox{}&							261&     20:39:25.48& -25:13:35.9& $24.34 \pm 0.06$& $1.379 \pm 0.004$ & A &  [\ion{O}{3}], H$\alpha$\\
\mbox{}&							265&     20:39:25.05& -25:13:49.7& $21.16 \pm 0.01$& $0.813 \pm 0.003$ & B$^{-}$ &  H$\alpha$\\
\mbox{}&							298&     20:39:24.13& -25:14:24.5& $22.68 \pm 0.02$& $0.944 \pm 0.004$ & B$^{+}$ &  H$\alpha$\\
\mbox{}&							382&     20:39:25.40& -25:14:25.3& $24.01 \pm 0.09$& $1.452 \pm 0.006$ & A &  [\ion{O}{3}], H$\alpha$\\
\mbox{}&							422&     20:39:26.81& -25:14:06.4& $22.71 \pm 0.02$& $2.130 \pm 0.004$ & A &  [\ion{O}{3}], [\ion{O}{2}]\\
\mbox{}&							429&     20:39:25.05& -25:14:55.5& $21.86 \pm 0.02$& $1.908 \pm 0.008$ & B$^{+}$ &  [\ion{O}{3}]\\
\mbox{}&							496&     20:39:24.63& -25:15:29.5& $25.62 \pm 0.17$& $1.934 \pm 0.006$ & B$^{-}$ &  [\ion{O}{3}]\\
\mbox{}&							564&     20:39:28.53& -25:14:13.6& $19.62 \pm 0.01$& $0.714 \pm 0.004$ & A &  H$\alpha$\\
\mbox{}&							564b\tablenotemark{c}&   20:39:28.53& -25:14:13.6& -& $0.7$ & A &  H$\alpha$ (visual)\\
\mbox{}&							565&     20:39:28.52& -25:14:15.3& $20.98 \pm 0.01$& $0.709 \pm 0.005$ & A &  H$\alpha$\\
\mbox{}&							570&     20:39:26.13& -25:15:17.6& $23.70 \pm 0.05$& $1.330 \pm 0.006$ & A &  [\ion{O}{3}], H$\alpha$\\
\mbox{}&							579&     20:39:27.13& -25:14:57.8& $23.70 \pm 0.05$& $1.444 \pm 0.004$ & A &  [\ion{O}{3}], H$\alpha$\\
\mbox{}&							581&     20:39:25.77& -25:15:33.6& $23.19 \pm 0.03$& $1.381 \pm 0.006$ & A &  [\ion{O}{3}], H$\alpha$\\
\mbox{}&							582&     20:39:27.74& -25:14:42.8& $24.52 \pm 0.06$& $1.360 \pm 0.004$ & A &  [\ion{O}{3}], H$\alpha$\\
\mbox{}&							588&     20:39:27.93& -25:14:40.1& $22.10 \pm 0.02$& $1.472 \pm 0.006$ & A &  [\ion{O}{3}], H$\alpha$\\
\mbox{}&							619&     20:39:20.18& -25:14:56.0& $22.91 \pm 0.03$& $1.216 \pm 0.004$ & B$^{+}$ &  H$\alpha$\\
\mbox{}&							639&     20:39:21.03& -25:14:18.1& $22.39 \pm 0.02$& $1.171 \pm 0.004$ & A &  H$\alpha$\\
\mbox{}&							667&     20:39:20.73& -25:14:19.5& $21.57 \pm 0.01$& $1.151 \pm 0.003$ & B$^{-}$ &  H$\alpha$\\
\mbox{}&							20700&     20:39:29.21& -25:14:18.0& $25.45 \pm 0.15$& $1.732 \pm 0.007$ & B$^{-}$ &  [\ion{O}{3}]\\
\mbox{}&							55200&     20:39:26.54& -25:14:04.2& $25.64 \pm 0.13$& $2.094 \pm 0.004$ & B$^{-}$ &  [\ion{O}{3}]\\
\hline
CARLA J0800+4029& 			166&     08:00:13.74& +40:30:53.6& $23.88 \pm 0.05$& $2.345 \pm 0.005$ & A & [\ion{O}{3}], [\ion{O}{2}]\\
\mbox{}&							198&     08:00:14.20& +40:30:49.9& $22.72 \pm 0.02$& $1.927 \pm 0.007$ & B$^{-}$ &  [\ion{O}{3}]\\
\mbox{}&							204&     08:00:10.62& +40:29:47.4& $22.55 \pm 0.02$& $1.745 \pm 0.006$ & B$^{+}$ &  [\ion{O}{3}]\\
\mbox{}&							294&     08:00:18.39& +40:29:49.7& $22.16 \pm 0.01$& $0.753 \pm 0.005$ & B$^{-}$ &  H$\alpha$\\
\mbox{}&							313&     08:00:17.38& +40:29:59.1& $22.70 \pm 0.02$& $2.171 \pm 0.005$ & B$^{-}$ &   [\ion{O}{3}]\\
\mbox{}&							349&     08:00:12.42& +40:28:46.5& $22.54 \pm 0.02$& $2.234 \pm 0.005$ & B$^{-}$ &   [\ion{O}{3}]\\
\mbox{}&							355&     08:00:13.63& +40:29:10.0& $22.76 \pm 0.02$& $1.916 \pm 0.009$ & B$^{+}$ &   [\ion{O}{3}]\\
\mbox{}&							357&     08:00:18.59& +40:30:31.3& $22.44 \pm 0.02$& $1.905 \pm 0.005$ & B$^{+}$ &   [\ion{O}{3}]\\
\mbox{}&							387&     08:00:14.23& +40:29:26.4& $21.50 \pm 0.01$& $1.923 \pm 0.005$ & B$^{+}$ &   [\ion{O}{3}]\\
\mbox{}&							554&     08:00:13.72& +40:30:01.4& $23.96 \pm 0.07$& $1.876 \pm 0.006$ & B$^{-}$ &   [\ion{O}{3}]\\
\mbox{}&							584&     08:00:15.62& +40:29:14.6& $22.84 \pm 0.02$& $0.799 \pm 0.003$ & B$^{+}$ &   H$\alpha$\\
\mbox{}&							587&     08:00:16.78& +40:31:01.1& $22.08 \pm 0.01$& $1.258 \pm 0.003$ & A &   H$\alpha$\\
\mbox{}&							673&     08:00:16.41& +40:29:22.2& $22.65 \pm 0.02$& $1.743 \pm 0.007$ & B$^{+}$ &   [\ion{O}{3}]\\
\mbox{}&							697&     08:00:19.04& +40:29:57.6& $23.50 \pm 0.04$& $2.092 \pm 0.004$ & B$^{+}$ &   [\ion{O}{3}]\\
\mbox{}&							746&     08:00:19.47& +40:29:51.2& $23.01 \pm 0.03$& $1.873 \pm 0.006$ & B$^{+}$ &   [\ion{O}{3}]\\
\mbox{}&							776&     08:00:19.20& +40:29:27.4& $23.86 \pm 0.06$& $1.439 \pm 0.007$ & A &  [\ion{O}{3}], H$\alpha$\\
\hline
\end{tabular}\\
\tablenotetext{1}{See Sec.\ref{sec:reddet}.}
\tablenotetext{2}{Note that in some cases additional lines have been identified but not fitted.}
\tablenotetext{3}{This source is a second component to $\#564$, blended in the F140W imaging.}
\end{table}

\section{Individual sources}\label{ap:clmemb}

\begin{figure*}[!ht]
{%
\setlength{\fboxsep}{0pt}%
\setlength{\fboxrule}{1pt}%
\fbox{\includegraphics[page=2, scale=0.45]{M2036_119.pdf} \hfill \includegraphics[page=1, scale=0.40]{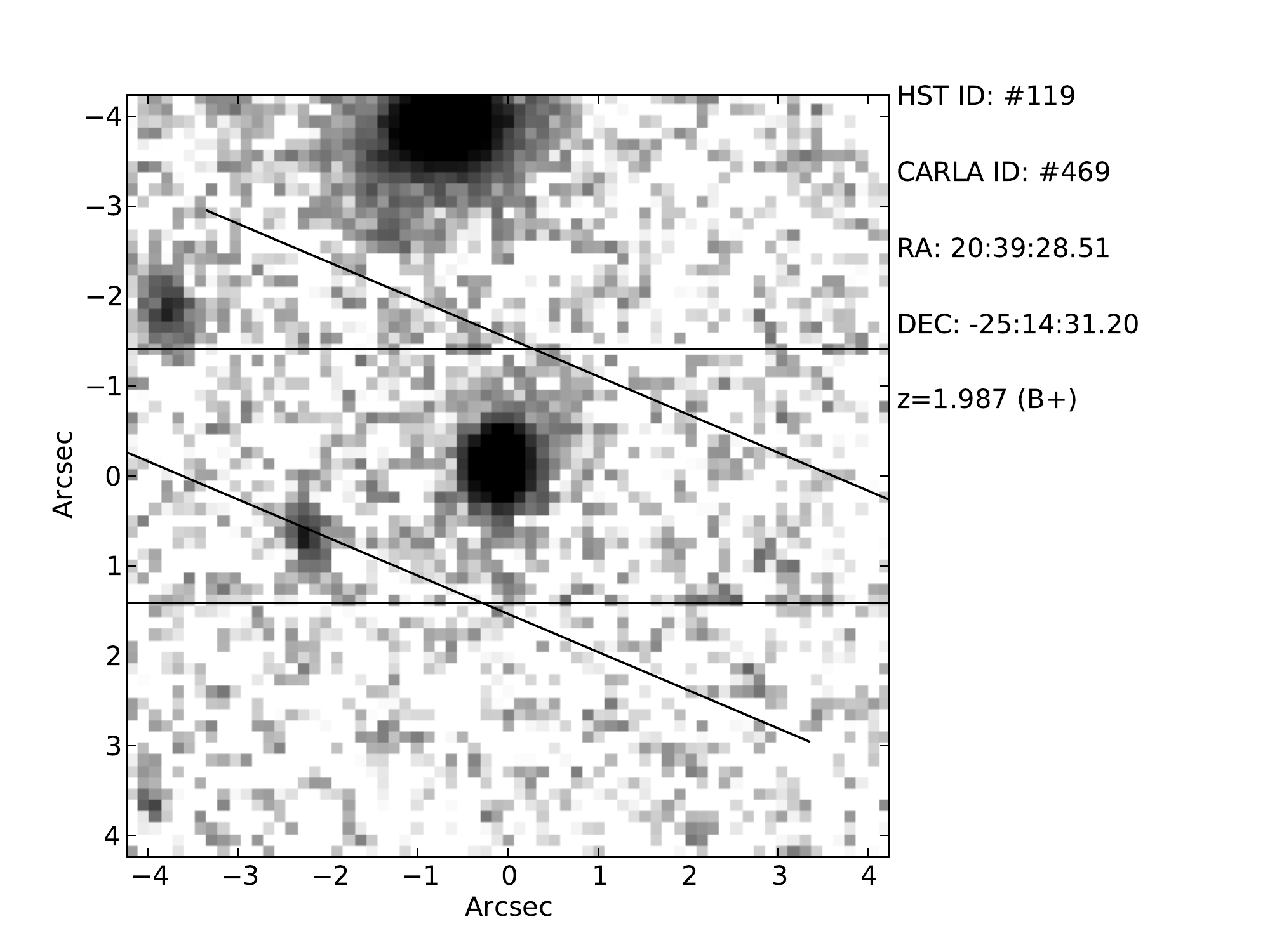} \mbox{(a)}}%
}\\%
{%
\setlength{\fboxsep}{0pt}%
\setlength{\fboxrule}{1pt}%
\fbox{\includegraphics[page=2, scale=0.45]{M2036_174.pdf} \hfill \includegraphics[page=1, scale=0.40]{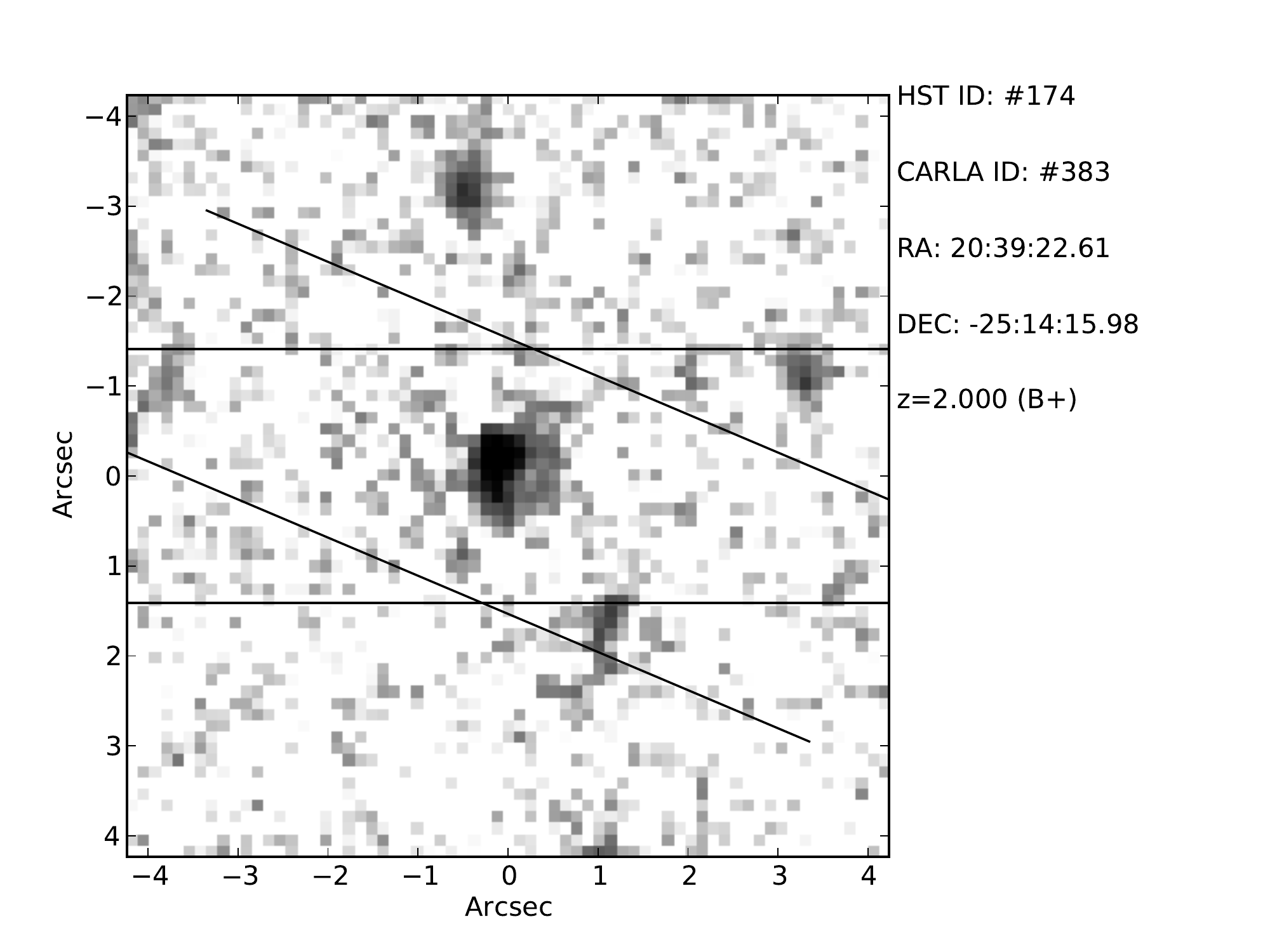} \mbox{(b)}}%
}\\%
{%
\setlength{\fboxsep}{0pt}%
\setlength{\fboxrule}{1pt}%
\fbox{\includegraphics[page=2, scale=0.45]{M2036_281.pdf} \hfill \includegraphics[page=1, scale=0.40]{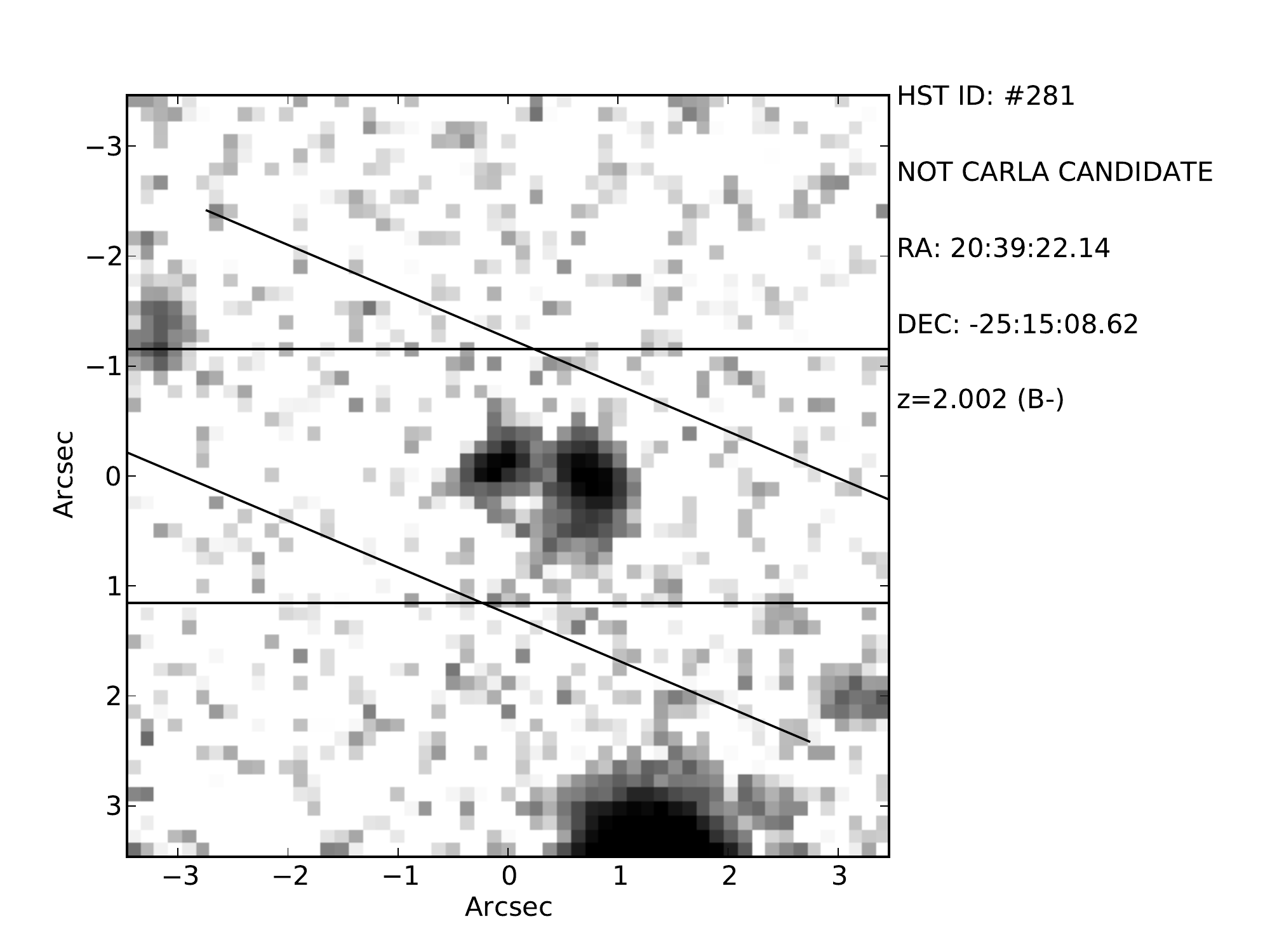} \mbox{(c)}}%
}\\%
\caption[Cluster member spectra]{Cutout spectra in 2D and 1D and direct imaging of all cluster members, for both fields. The slits in the direct imaging represent the dispersion direction and width of the 2D spectral cutouts, where the horizontal slit correspond to the first observation, and the inclined one to the second. The red slits in the 2D images, however, represent the width from which we extract the 1D spectra. Grey scales are arbitrary set to visually facilitate the identification of the source spectral features. The vertical dotted lines are visual aids locating possible emission lines at the measured redshifts of each source. The green contours represent the contamination model contours as described in Sections \ref{p:ExtrContam} and \ref{p:Gnoise}. The grey error bars on top of the 1D spectra represent the grism $1\sigma$ wavelength-depend background noise. Panels (a) to (j) correspond to CARLA J2039$-$2514, panels (k) to (t) to CARLA J0800+4029.\\ \mbox{}\\ \mbox{}\\}
\label{fig:spectra}
\end{figure*}

We discuss here the spectra and properties of all cluster members, except the ones already discussed in section \ref{s:sourcesinterest}. We also show the 2D cutouts, 1D spectra, fitting and contamination contours of all cluster members, as well as their direct image stamps (Figure \ref{fig:spectra}).

\subsection{CARLA J2039$-$2514}

\textbullet\  {\textbf{\#119:}} We identify contamination above $15,200$ \AA\ and below $11,500$ \AA\ in the first observation, and up to $15,200$ \AA\ in the second observation, albeit slightly offset from the center of the 2D extraction. This {\it Spitzer}-selected CARLA source has a mid-infrared color consistent with the emission line seen in both orientations as being [\ion{O}{3}]. We measure a redshift of $z = 1.987 \pm 0.007$, and we do not detect H$\beta$ or other emission lines. Hence, we assign a quality B$^{+}$ to the redshift based on the emission line and {\it Spitzer} color.

\textbullet\  {\textbf{\#174:}} The contamination model for this {\it Spitzer}-selected CARLA source does not identify strong spectroscopic contamination. However, the two emission-like features seen at $14,100$ and $14,600$ \AA\ in the first observation are not observed in the second observation. This suggests unidentified contamination at the level of our contamination contours. This could be due to pixel noise, a non-extracted source in the field, a source outside the field of view, or a shallow emission-line contaminant. However, the emission line at $15,000$ \AA\ is consistent between both orientations and the source's {\it Spitzer} color is consistent with [\ion{O}{3}] at $z \sim 2$. We measure a redshift of $z = 2.000 \pm 0.007$ from the [\ion{O}{3}] emission line.
Because of the contamination concerns, the line flux listed in Table \ref{table:examplelargetable} is from the second observation. We do not detect other lines, and therefore, assign a quality B$^{+}$ to this source.

\textbullet\  {\textbf{\#281:}} Another source ($\#280$) is located $1\arcsec$ from $\#281$. We detect the same emission lines in the $\#280$ spectra as in $\#281$, however, spatially offset above and below the continuum for the former (depending on the observation). Therefore, we associate the emission lines to $\#281$, and we do not exclude $\#280$ as a quiescent companion to $\#281$, even though it might be a foreground or background source. The contamination model indicates slight contamination around $15,000$ \AA. The contamination is likely the origin of the broad shape of the line in the second observation, not seen in the first one, and might enhance the true flux of the line, but we still use both orientations to determine the source properties. We measure $z = 2.002 \pm 0.008$ (quality B$^{-}$).

\textbullet\  {\textbf{\#356:}} Another source ($\#355$) is located $1\arcsec$ from $\#356$. For similar reasons as for $\#281$, we associate the emission lines to $\#356$, and do not exclude $\#355$ as a quiescent companion to $\#356$. We measure $z = 1.999 \pm 0.005$ (quality B$^{-}$) from the second observation, as the first one is strongly contaminated by a bright source.

\textbullet\  {\textbf{\#360:}} We detect contamination above $15,300$ \AA~in the first observation, and below $14,000$ \AA~in the second. However, our model does not indicate contamination around the well-detected line at $15,050$ \AA, which we identify as [\ion{O}{3}]. We do not identify any other emission lines, and we consider the emission-like feature seen in the second observation at the expected position of H$\beta$ to be due to contamination since it is not present in the first observation. We measure a redshift of $z = 2.006 \pm 0.004$ (quality B$^{-}$).

\textbullet\  {\textbf{\#697:}} The contamination model indicates shallow contamination in the first observation, and spatially offset contamination in the second. We use both observations and determine a quality B$^{-}$ redshift of $z = 2.001 \pm 0.006$, based on the detection of a single emission line in both orientations. 

\textbullet\  {\textbf{\#44300:}} From the second observation, free from contamination, we detect a single emission line which we associate to [\ion{O}{3}] at a redshift of $z = 2.002 \pm 0.005$ (quality B$^{-}$).

\textbullet\  {\textbf{\#90000:}} The first orientation is free from contamination within our 1D extraction region. The second observation is contaminated longward of the emission line, which affects our spectral modeling. Therefore, we only use the first observation to measure the line parameters in Table \ref{table:examplelargetable}, providing a quality B$^{-}$ redshift of $z = 1.995 \pm 0.005$.\\

\subsection{CARLA J0800+4029}

\textbullet\  {\textbf{\#146:}} The second observation is free from contaminant. The first observation is contaminated up to $15,700$ \AA, but the contamination is spatially offset by $\sim 0.5\arcsec$ from the source. Therefore, we carefully extract the 1D spectrum from the 2D cutout to avoid the contaminant, allowing us to use both observations. We measure a redshift of $z = 1.998 \pm 0.007$ (quality B$^{-}$).

\textbullet\  {\textbf{\#401:}} The second orientation of this source is visually contaminated below $15,000\rm\AA$, but we do not have a contamination model associated with this source as it was extracted alone after the main source extraction. We measure a redshift of $z = 1.980 \pm 0.009$ (quality B$^{-}$).

\textbullet\  {\textbf{\#436:}} This source has a {\it Spitzer} color consistent with $z> 1.3$. Contamination is detected offset from the source continuum in the 2D cutouts, without overlap. A single line is detected in both observations, providing a quality B$^{+}$ redshift of $z = 1.969 \pm 0.007$.

\textbullet\  {\textbf{\#443:}} Above $15,500$ \AA, the second observation is contaminated by a zeroth order image of a bright source. We identify an emission line at $\sim 14,980\rm~\AA$ in both observations, but the line is potentially contaminated in the second observation by the mentioned zeroth order. Therefore, we only use the first observation, and measure a redshift of $1.992 \pm 0.004$ (quality B$^{-}$). The fitting procedure tentatively detects H$\beta$ below our detection limit, and therefore is not taken into account.

\textbullet\  {\textbf{\#542:}} The second observation shows the same emission line as the first observation, but broadened in the dispersion direction (and slightly in the spatial direction). This could be due to the spatial extent of the source which is more elongated in the dispersion direction of the second observation, or due to an undetected contaminant in the second observation. The mid-infrared {\it Spitzer} color is consistent with [\ion{O}{3}] at $z\sim2$, but no other emission lines are detected. We use the two observations to measure its redshift and line flux despite the elongated spectral shape of the second observation (note that only using the first observation does not significantly change the results). We measure a redshift of $z = 1.964 \pm 0.007$ (quality B$^{+}$).

\textbullet\ {\textbf{\#591:}} Given our contamination model, both observations are free from contaminants. However, we identify in the direct imaging a source located $\sim 2\arcsec$ from $\#591$ and likely contaminating part of the spectral 2D cutouts. Therefore, we carefully extract the 1D spectrum from the 2D cutout to avoid potentially contaminated regions, allowing us to use both observations. We measure a redshift of $z = 1.975 \pm 0.007$ (quality B$^{-}$).

\textbullet\  {\textbf{\#767:}} This source was detected in our IRAC images and has a {\it Spitzer} color consistent with [\ion{O}{3}] at $z\sim2$. However, it was not selected as a CARLA candidate as it did not pass our CARLA flux cut. The second observation is contaminated below $14,900\rm~\AA$, with the contamination also somewhat affecting the emission line. The first observation is not contaminated around the emission line at $14,910\rm~ \AA$, but is likely contaminated at shorter wavelengths. Using the first observation, we determine a redshift of $z = 1.978 \pm 0.004$, of quality B$^{+}$. The observed feature at the expected location of H$\beta$ is likely due to contamination, just at our flux detection limit.\\

\mbox{}\\

\begin{figure*}[!ht]
{%
\setlength{\fboxsep}{0pt}%
\setlength{\fboxrule}{1pt}%
\fbox{\includegraphics[page=2, scale=0.45]{M2036_306.pdf} \hfill \includegraphics[page=1, scale=0.40]{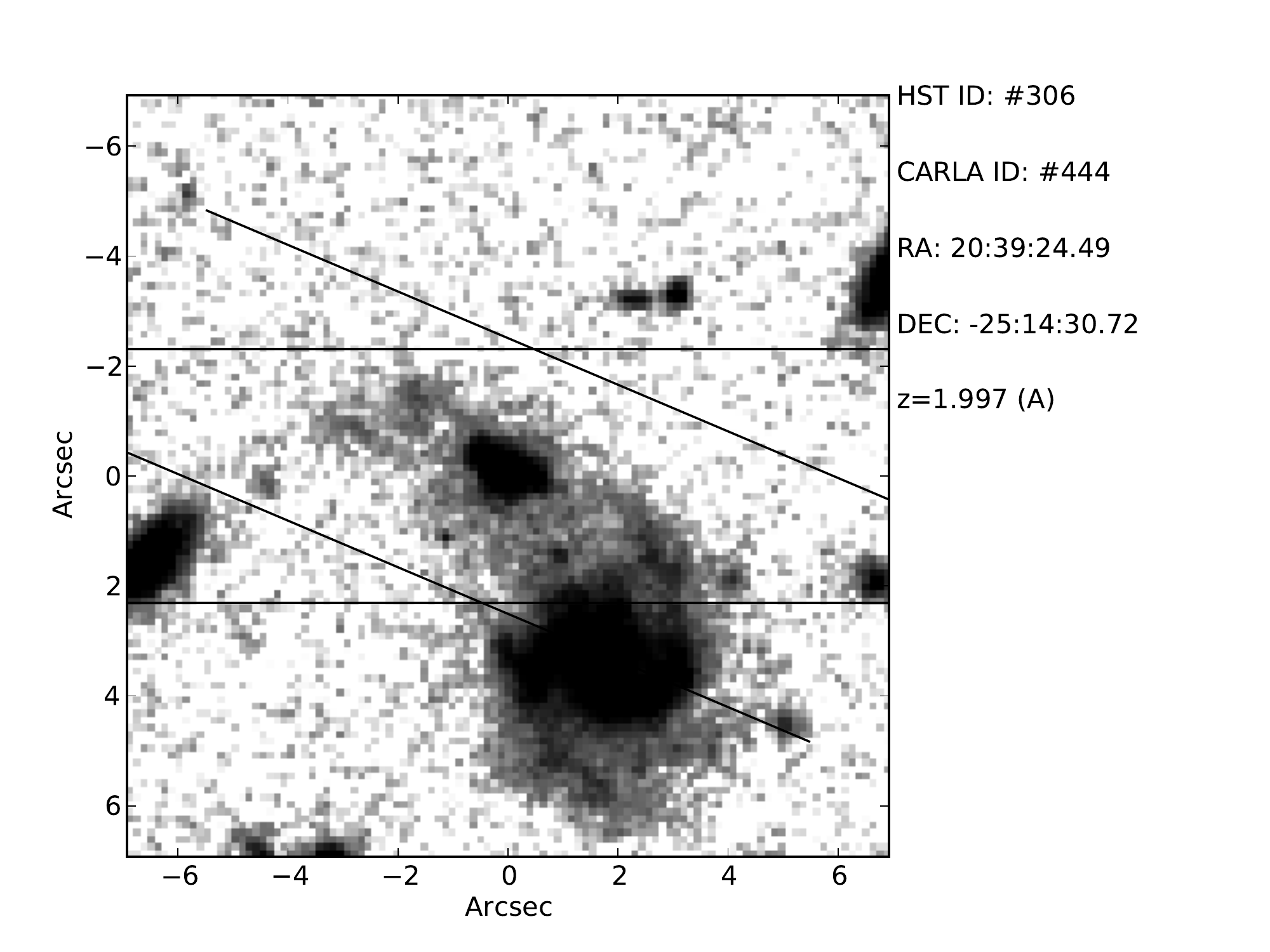} \mbox{(d)}}%
}\\%
{%
\setlength{\fboxsep}{0pt}%
\setlength{\fboxrule}{1pt}%
\fbox{\includegraphics[page=2, scale=0.45]{M2036_306b.pdf} \hfill \includegraphics[page=1, scale=0.4]{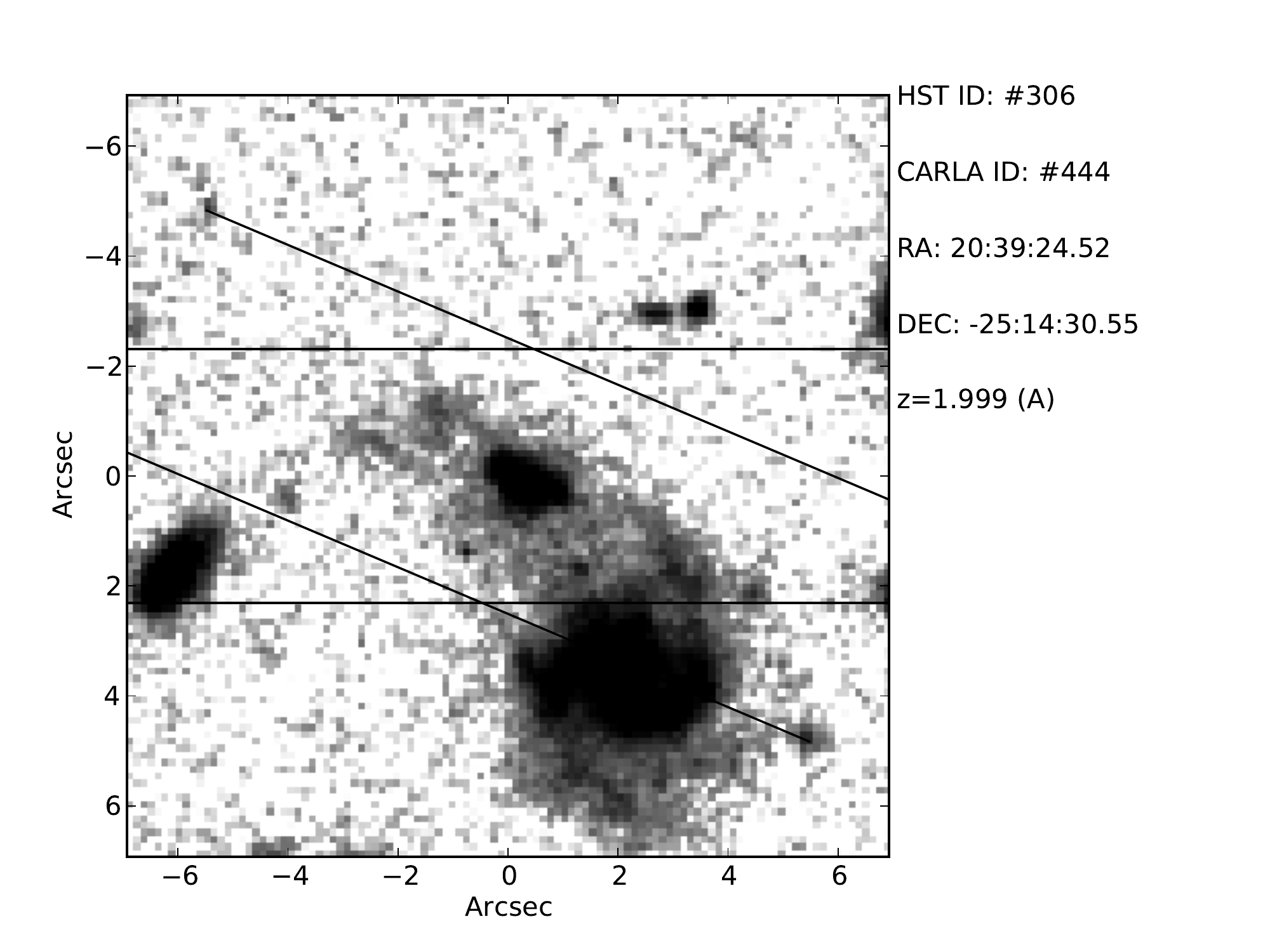} \mbox{(e)}}%
}\\
{%
\setlength{\fboxsep}{0pt}%
\setlength{\fboxrule}{1pt}%
\fbox{\includegraphics[page=2, scale=0.45]{M2036_356.pdf} \hfill \includegraphics[page=1, scale=0.4]{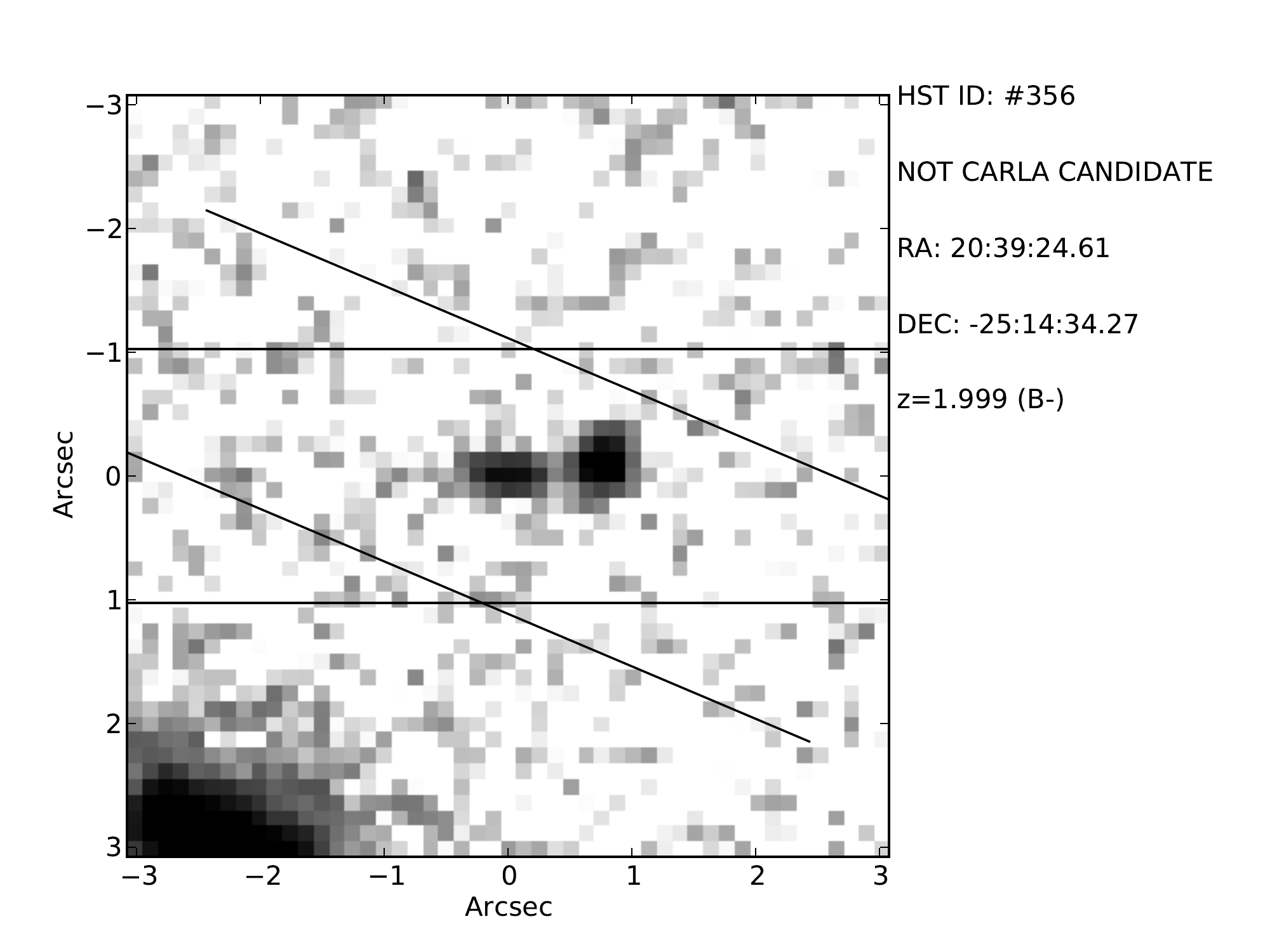} \mbox{(f)}}%
}\\
\textbf{\mbox{}\\ Figure \ref{fig:spectra}} --- Continued. Note that the 2D cutouts of $\#306$b do not show any contamination model contours since we re-extracted this source alone, however, we invite the reader to have a look at $\#306$ (Panel (d)) since the two sources are present on the same cutout.\\ \mbox{}\\ \mbox{}\\ \mbox{}\\ \mbox{}\\ \mbox{}\\ \mbox{}\\
\end{figure*}

\begin{figure*}[!ht]
{%
\setlength{\fboxsep}{0pt}%
\setlength{\fboxrule}{1pt}%
\fbox{\includegraphics[page=2, scale=0.45]{M2036_360.pdf} \hfill \includegraphics[page=1, scale=0.4]{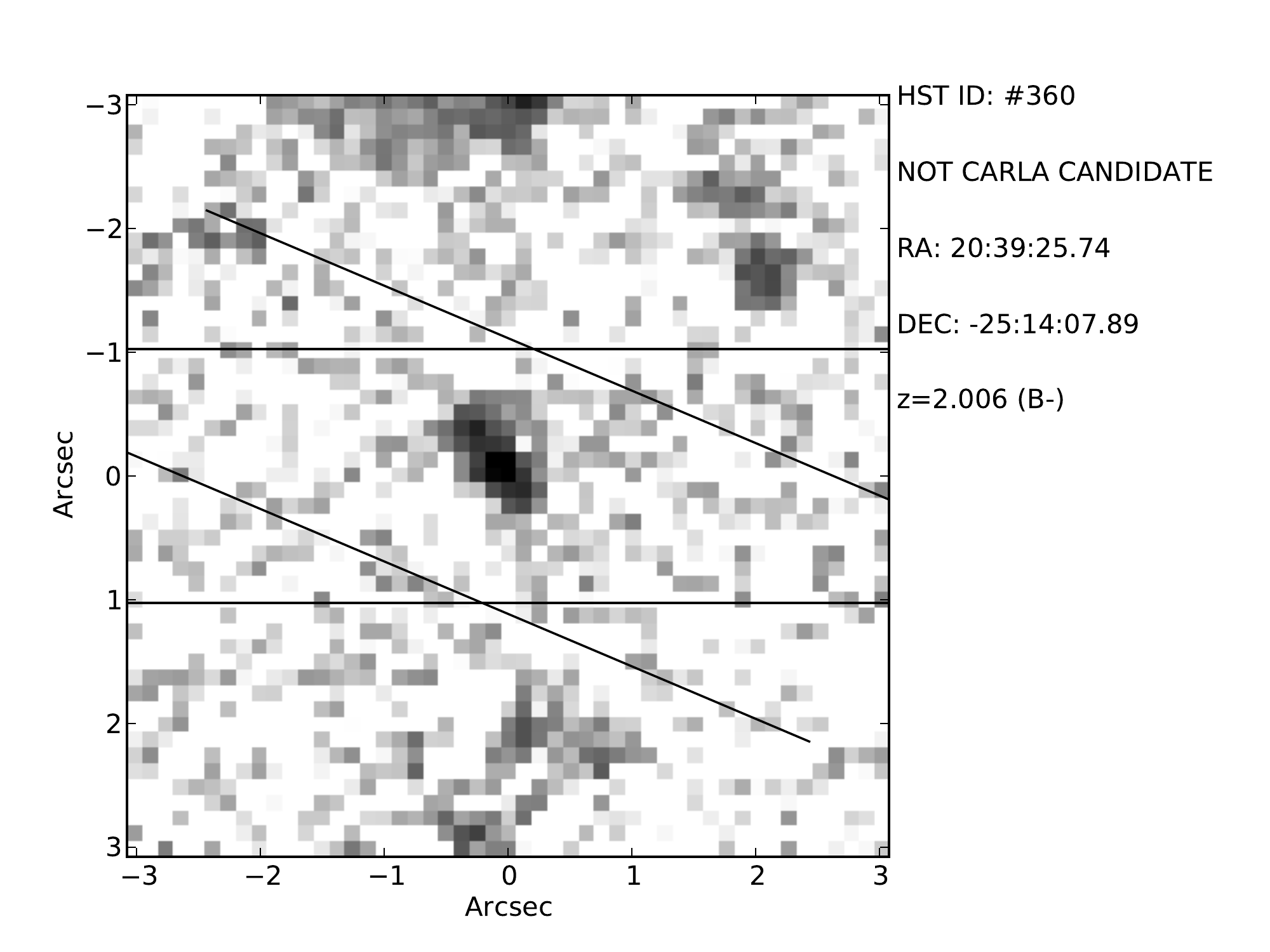} \mbox{(g)}}%
}\\
{%
\setlength{\fboxsep}{0pt}%
\setlength{\fboxrule}{1pt}%
\fbox{\includegraphics[page=2, scale=0.45]{M2036_697.pdf} \hfill \includegraphics[page=1, scale=0.4]{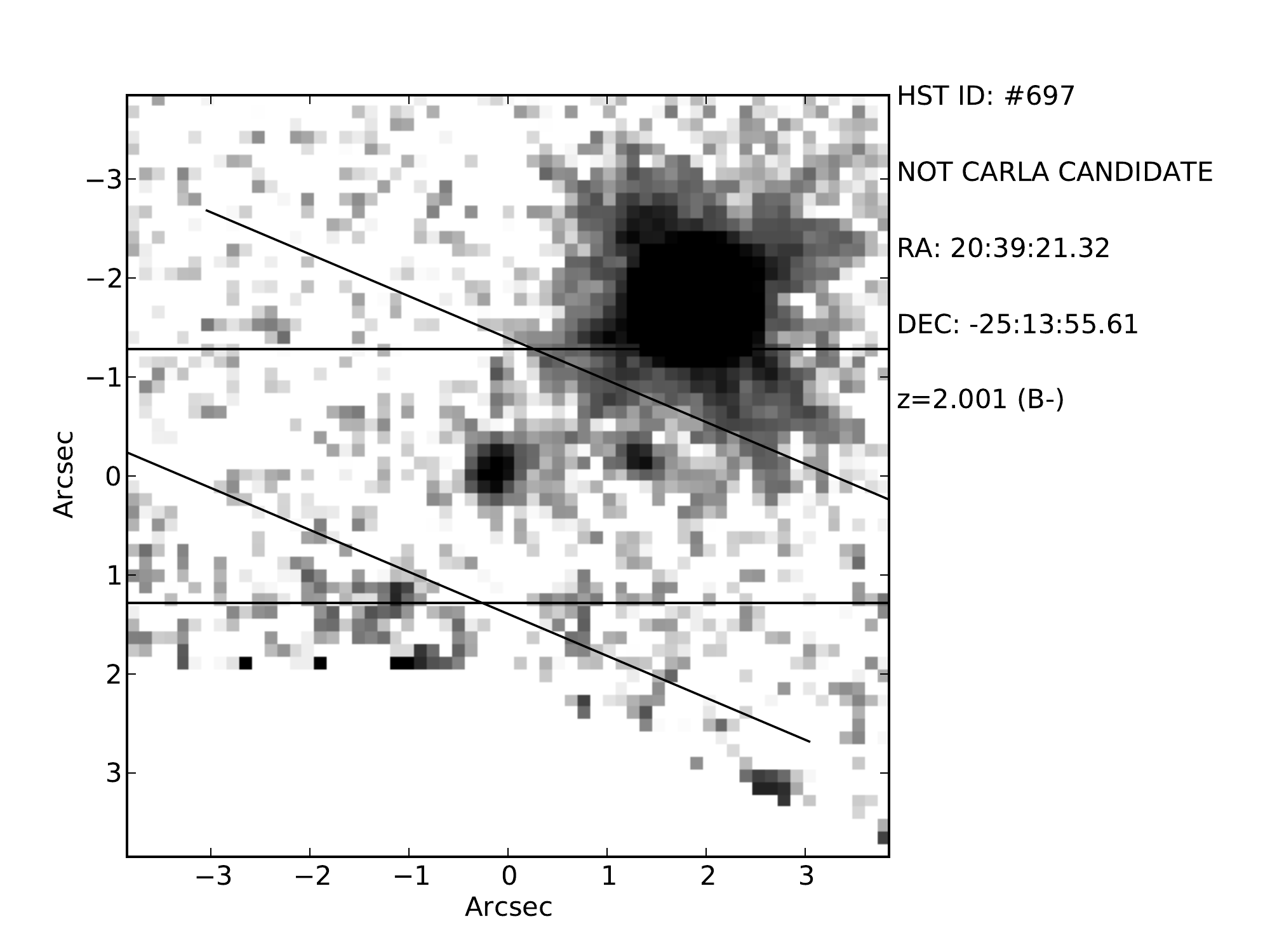}  \mbox{(h)}}%
}\\
{%
\setlength{\fboxsep}{0pt}%
\setlength{\fboxrule}{1pt}%
\fbox{\includegraphics[page=2, scale=0.45]{M2036_44300.pdf} \hfill \includegraphics[page=1, scale=0.4]{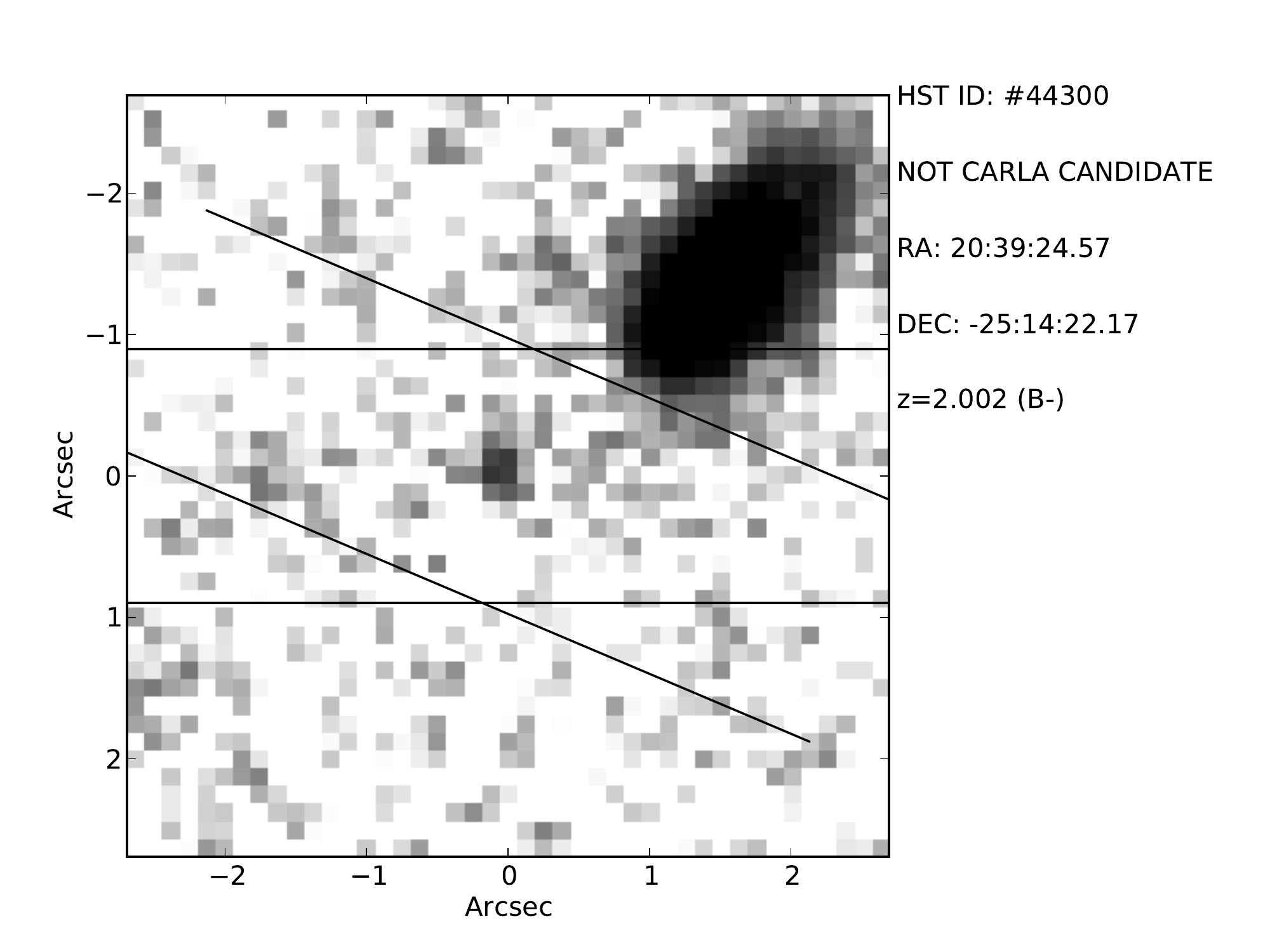}  \mbox{(i)}}%
}\\
\textbf{\mbox{}\\ Figure \ref{fig:spectra}} --- Continued.\\
\mbox{}\\ \mbox{}\\ \mbox{}\\ \mbox{}\\ \mbox{}\\ \mbox{}\\
\end{figure*}

\begin{figure*}[!ht]
{%
\setlength{\fboxsep}{0pt}%
\setlength{\fboxrule}{1pt}%
\fbox{\includegraphics[page=2, scale=0.45]{M2036_90000.pdf} \hfill \includegraphics[page=1, scale=0.4]{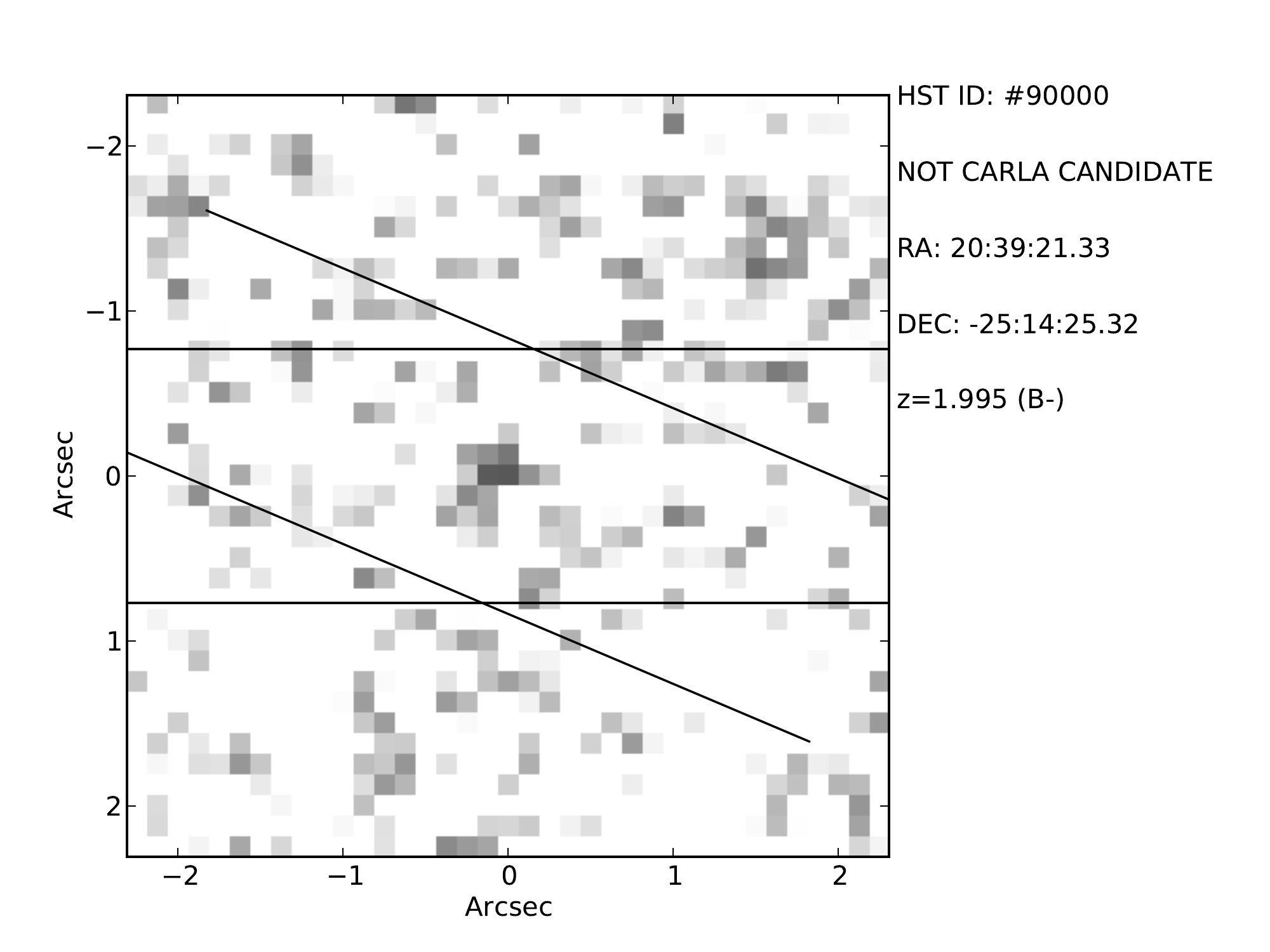}  \mbox{(j)}}%
}\\
{%
\setlength{\fboxsep}{0pt}%
\setlength{\fboxrule}{1pt}%
\fbox{\includegraphics[page=2, scale=0.45]{J0800_146.pdf} \hfill \includegraphics[page=1, scale=0.4]{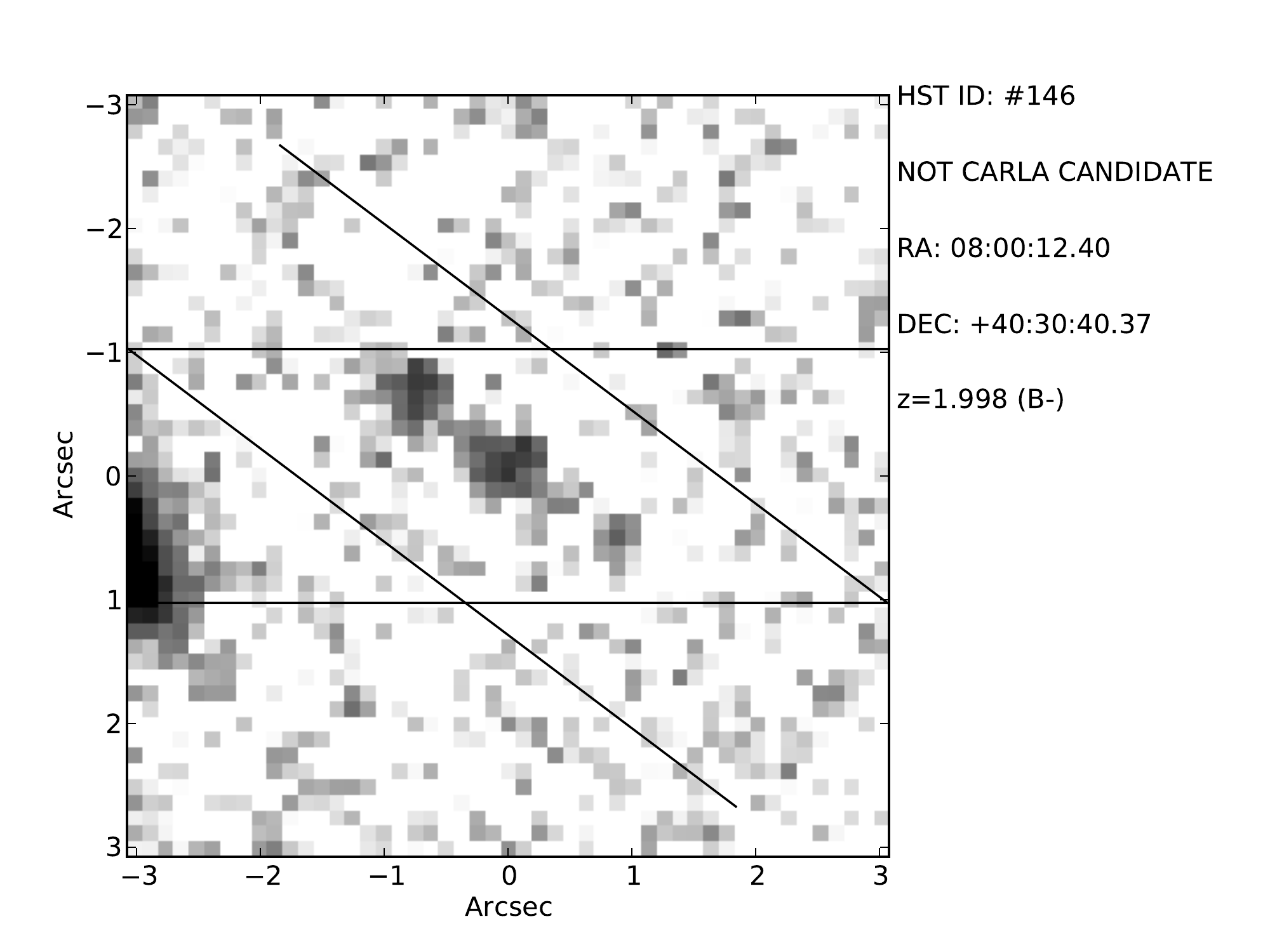}  \mbox{(k)}}%
}\\
{%
\setlength{\fboxsep}{0pt}%
\setlength{\fboxrule}{1pt}%
\fbox{\includegraphics[page=2, scale=0.45]{J0800_371.pdf} \hfill \includegraphics[page=1, scale=0.4]{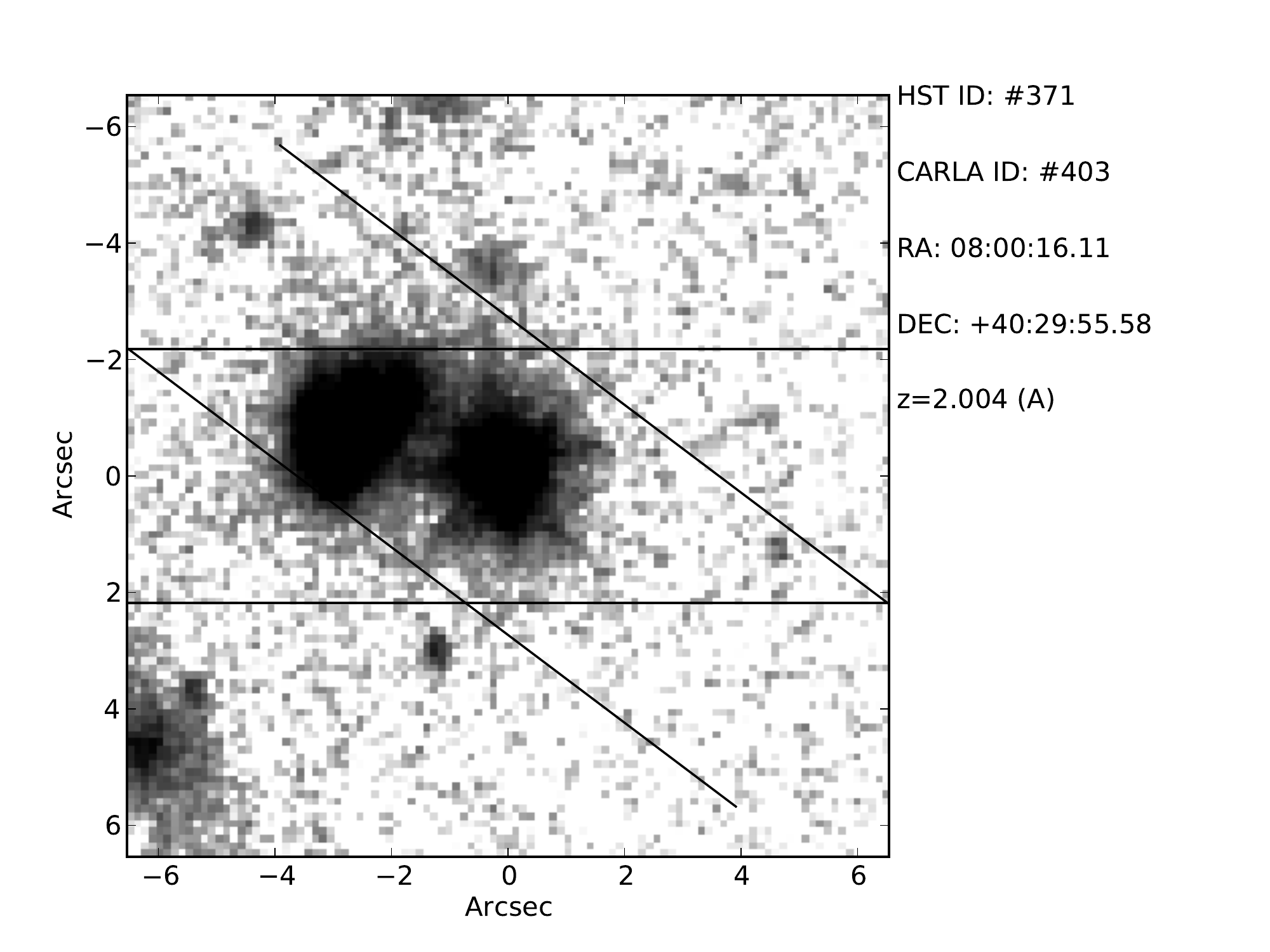}  \mbox{(l)}}%
}\\
\textbf{\mbox{}\\ Figure \ref{fig:spectra}} --- Continued.\\
\mbox{}\\ \mbox{}\\ \mbox{}\\ \mbox{}\\ \mbox{}\\ \mbox{}\\
\end{figure*}

\begin{figure*}[!ht]
{%
\setlength{\fboxsep}{0pt}%
\setlength{\fboxrule}{1pt}%
\fbox{\includegraphics[page=2, scale=0.45]{J0800_372.pdf} \hfill \includegraphics[page=1, scale=0.4]{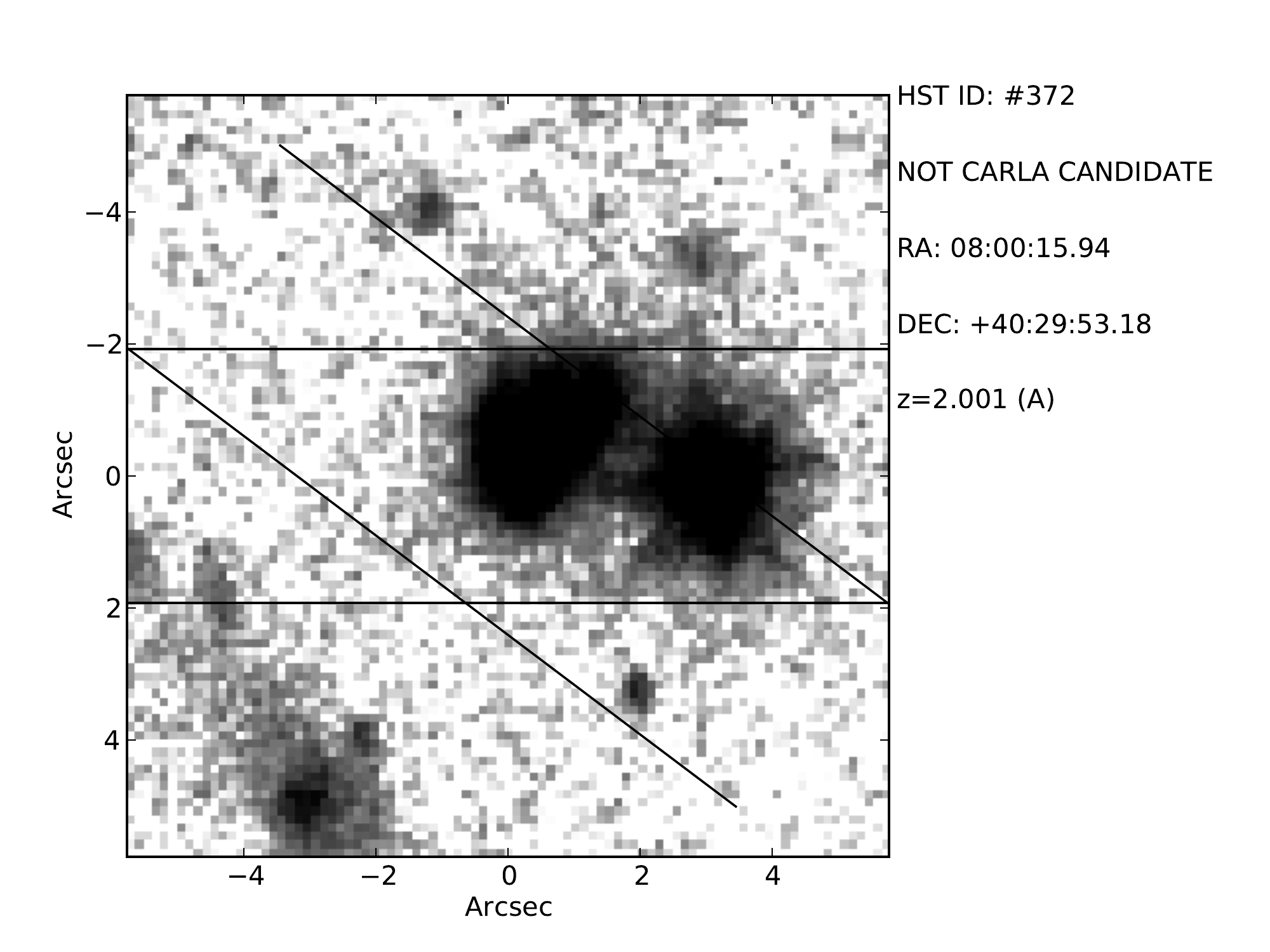}  \mbox{(m)}}%
}\\
{%
\setlength{\fboxsep}{0pt}%
\setlength{\fboxrule}{1pt}%
\fbox{\includegraphics[page=2, scale=0.45]{J0800_401.pdf} \hfill \includegraphics[page=1, scale=0.4]{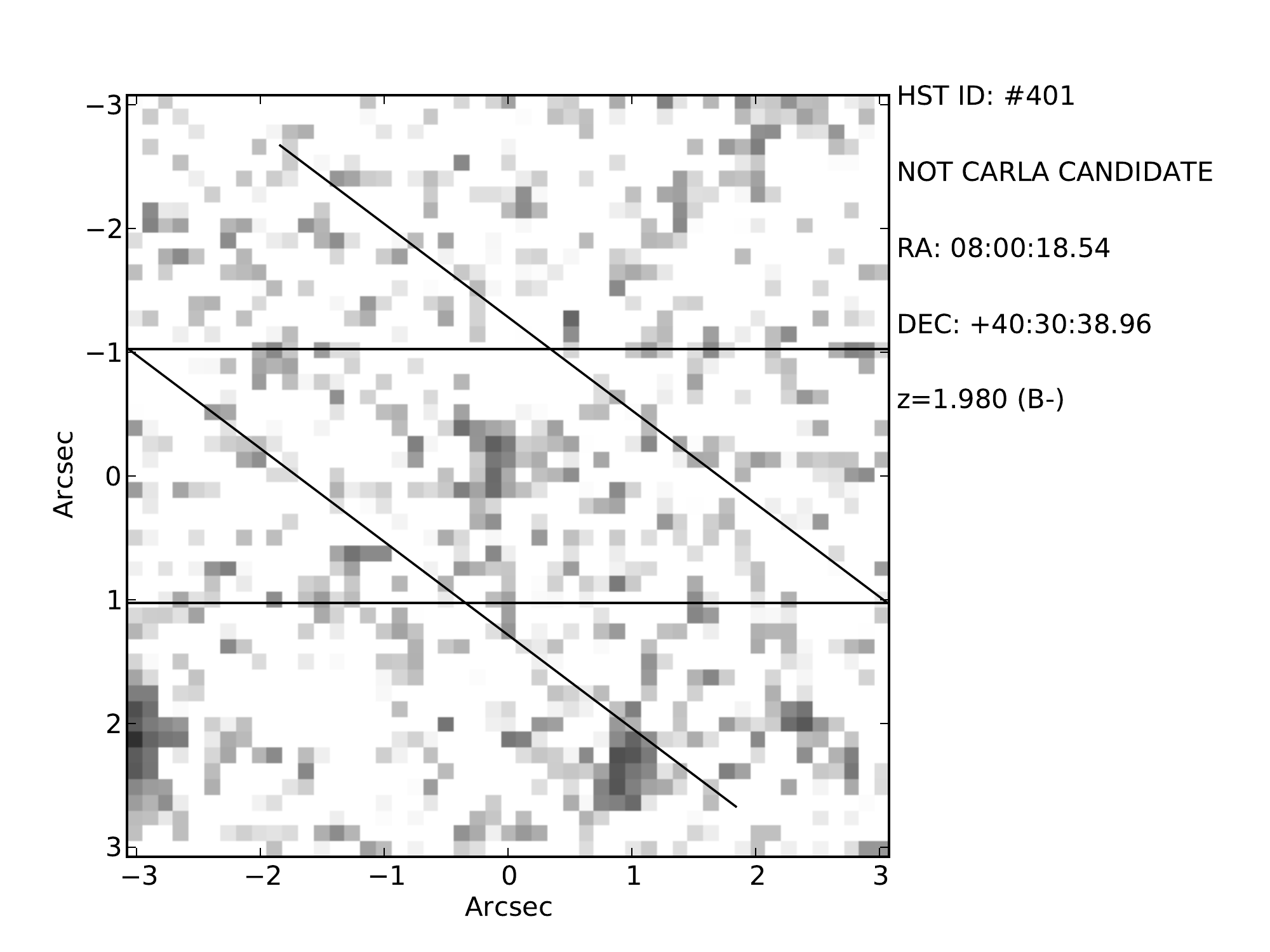}  \mbox{ (n)}}%
}\\
{%
\setlength{\fboxsep}{0pt}%
\setlength{\fboxrule}{1pt}%
\fbox{\includegraphics[page=2, scale=0.45]{J0800_436.pdf} \hfill \includegraphics[page=1, scale=0.4]{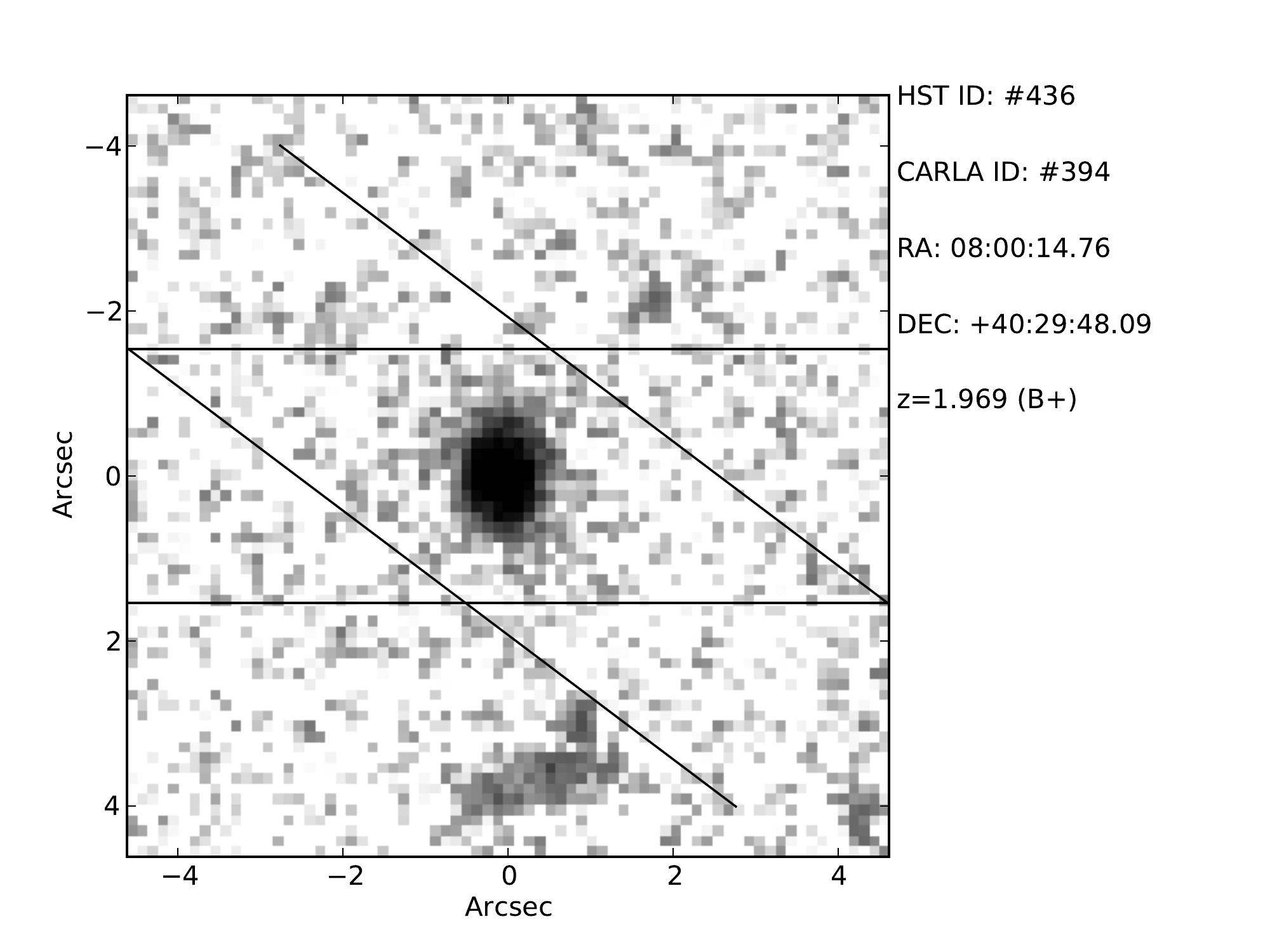}  \mbox{ (o)}}%
}\\
\textbf{\mbox{}\\ Figure \ref{fig:spectra}} --- Continued.\\
\mbox{}\\ \mbox{}\\ \mbox{}\\ \mbox{}\\ \mbox{}\\ \mbox{}\\
\end{figure*}

\begin{figure*}[!ht]
{%
\setlength{\fboxsep}{0pt}%
\setlength{\fboxrule}{1pt}%
\fbox{\includegraphics[page=2, scale=0.45]{J0800_443.pdf} \hfill \includegraphics[page=1, scale=0.4]{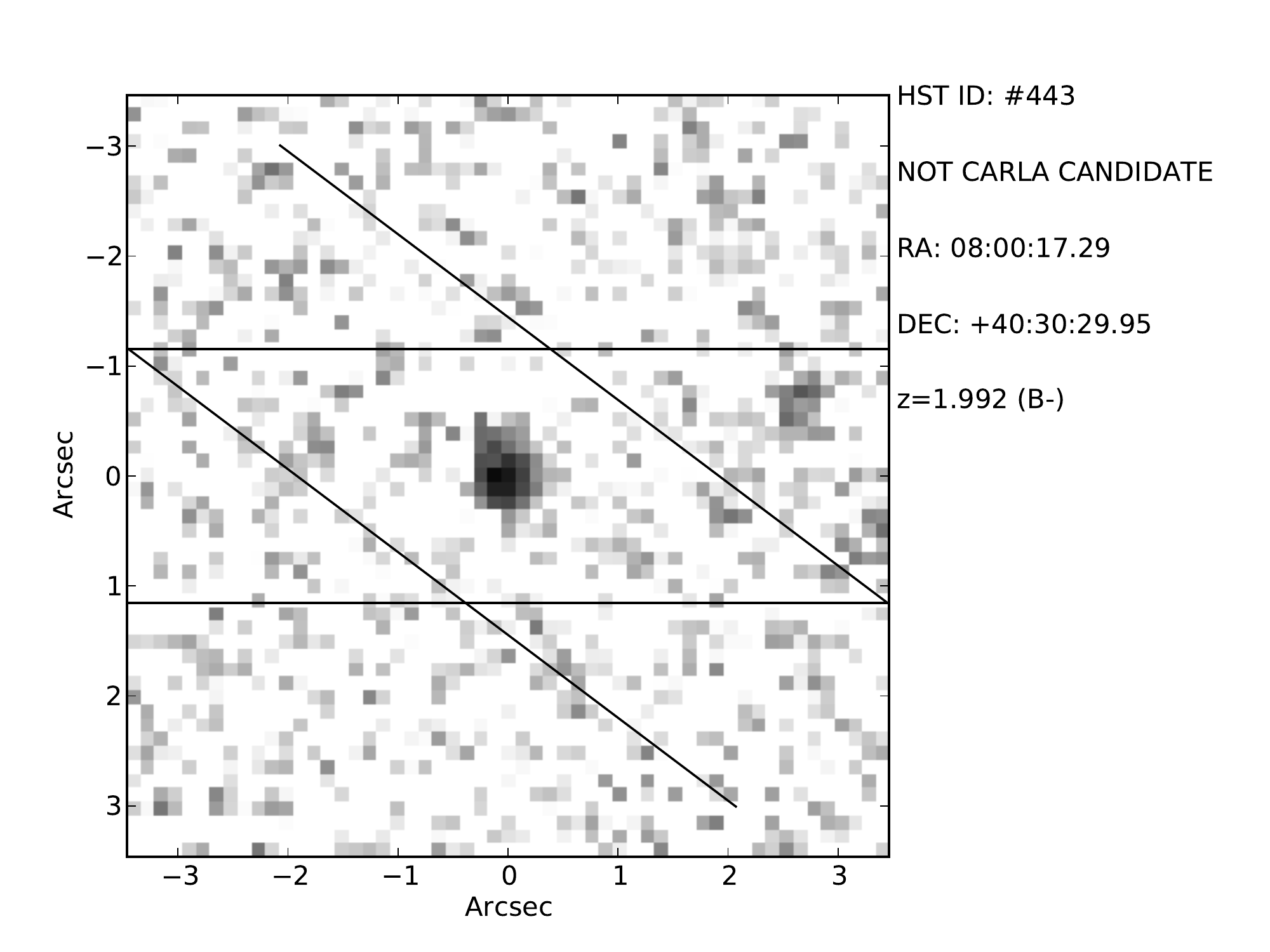}  \mbox{ (p)}}%
}\\
{%
\setlength{\fboxsep}{0pt}%
\setlength{\fboxrule}{1pt}%
\fbox{\includegraphics[page=2, scale=0.45]{J0800_542.pdf} \hfill \includegraphics[page=1, scale=0.4]{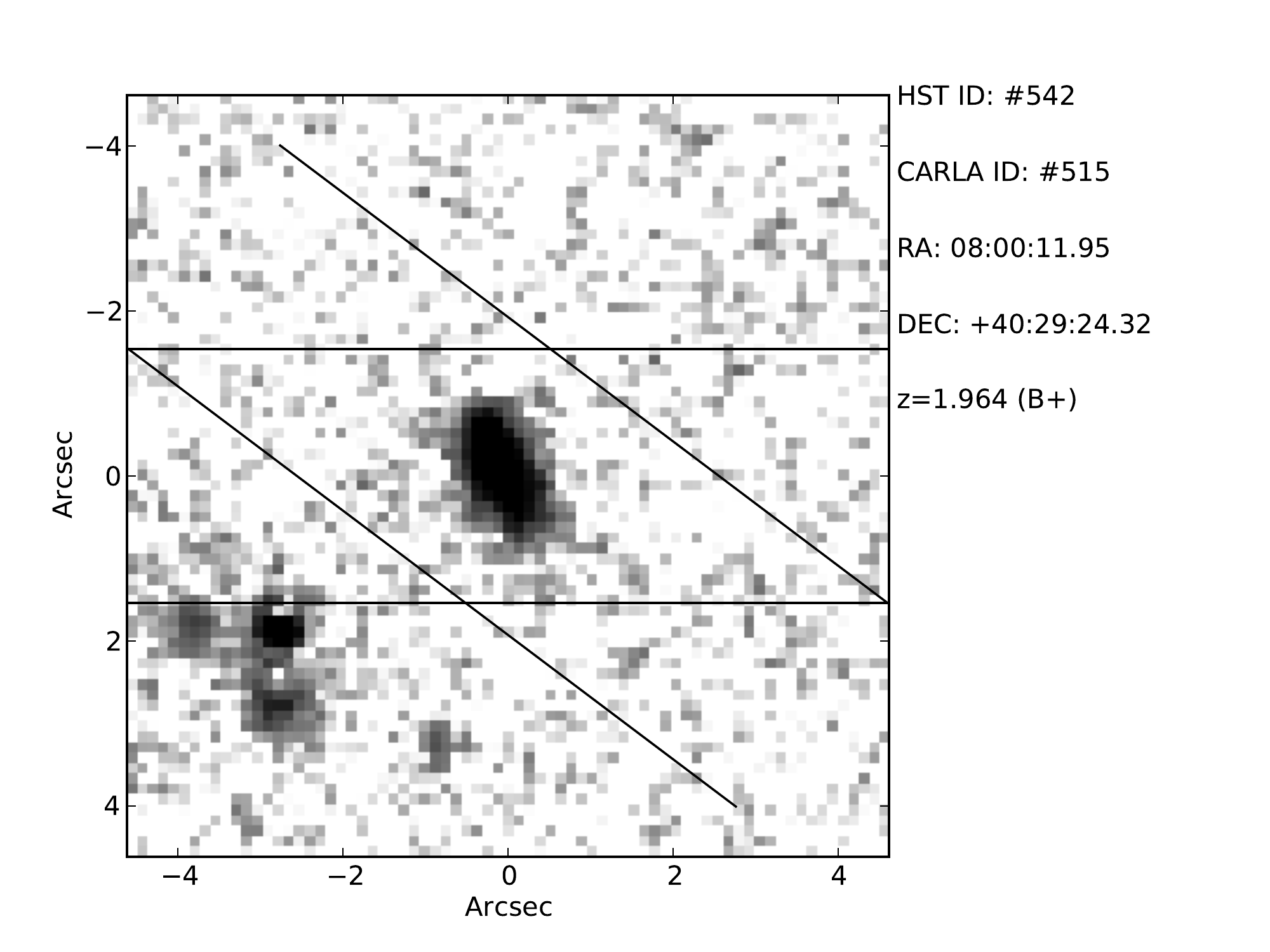}  \mbox{(q)}}%
}\\
{%
\setlength{\fboxsep}{0pt}%
\setlength{\fboxrule}{1pt}%
\fbox{\includegraphics[page=2, scale=0.45]{J0800_591.pdf} \hfill \includegraphics[page=1, scale=0.4]{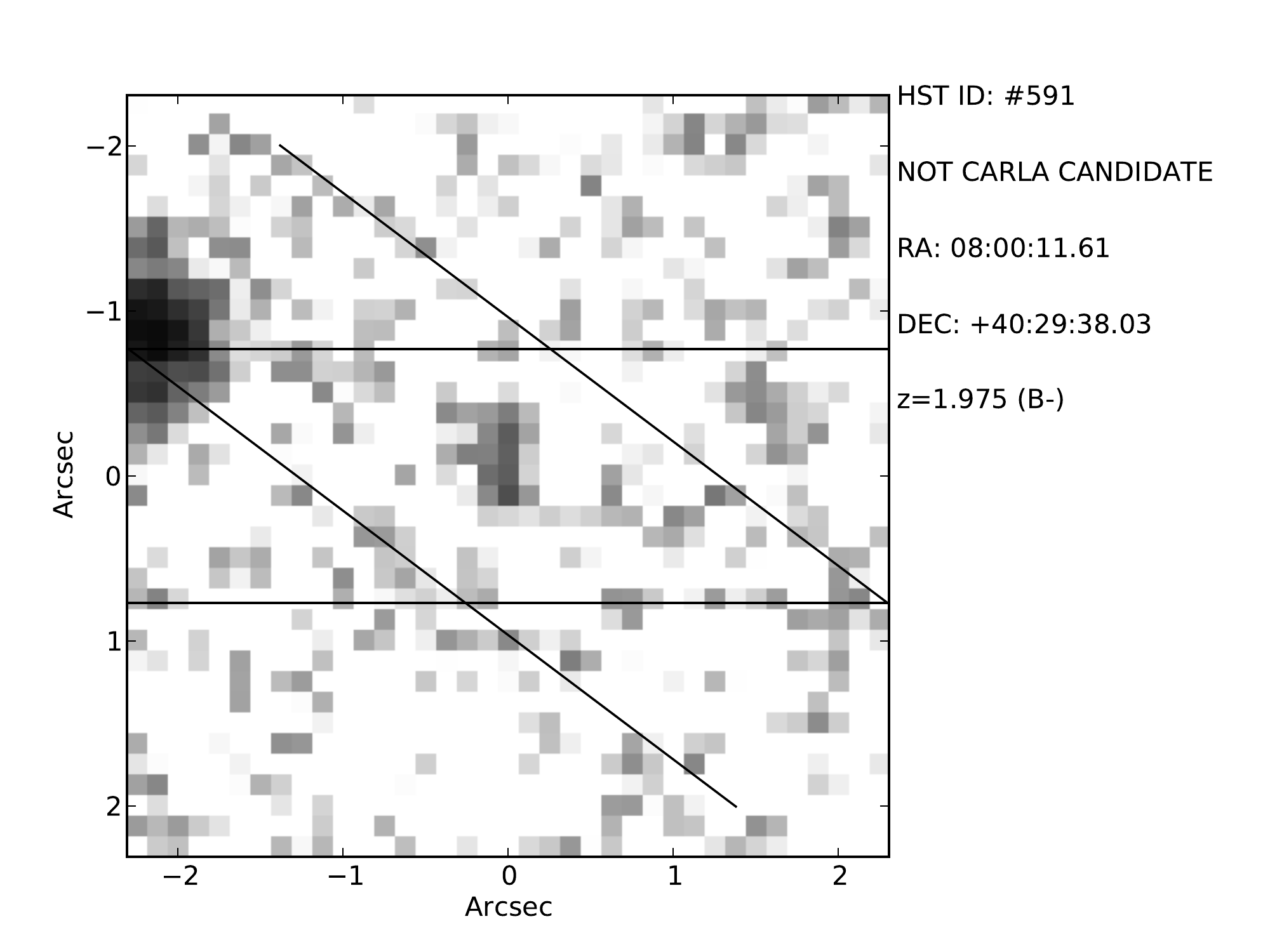}  \mbox{(r)}}%
}\\
\textbf{\mbox{}\\ Figure \ref{fig:spectra}} --- Continued.\\
\mbox{}\\ \mbox{}\\ \mbox{}\\ \mbox{}\\ \mbox{}\\ \mbox{}\\
\end{figure*}

\begin{figure*}[!ht]
{%
\setlength{\fboxsep}{0pt}%
\setlength{\fboxrule}{1pt}%
\fbox{\includegraphics[page=2, scale=0.45]{J0800_749.pdf} \hfill \includegraphics[page=1, scale=0.4]{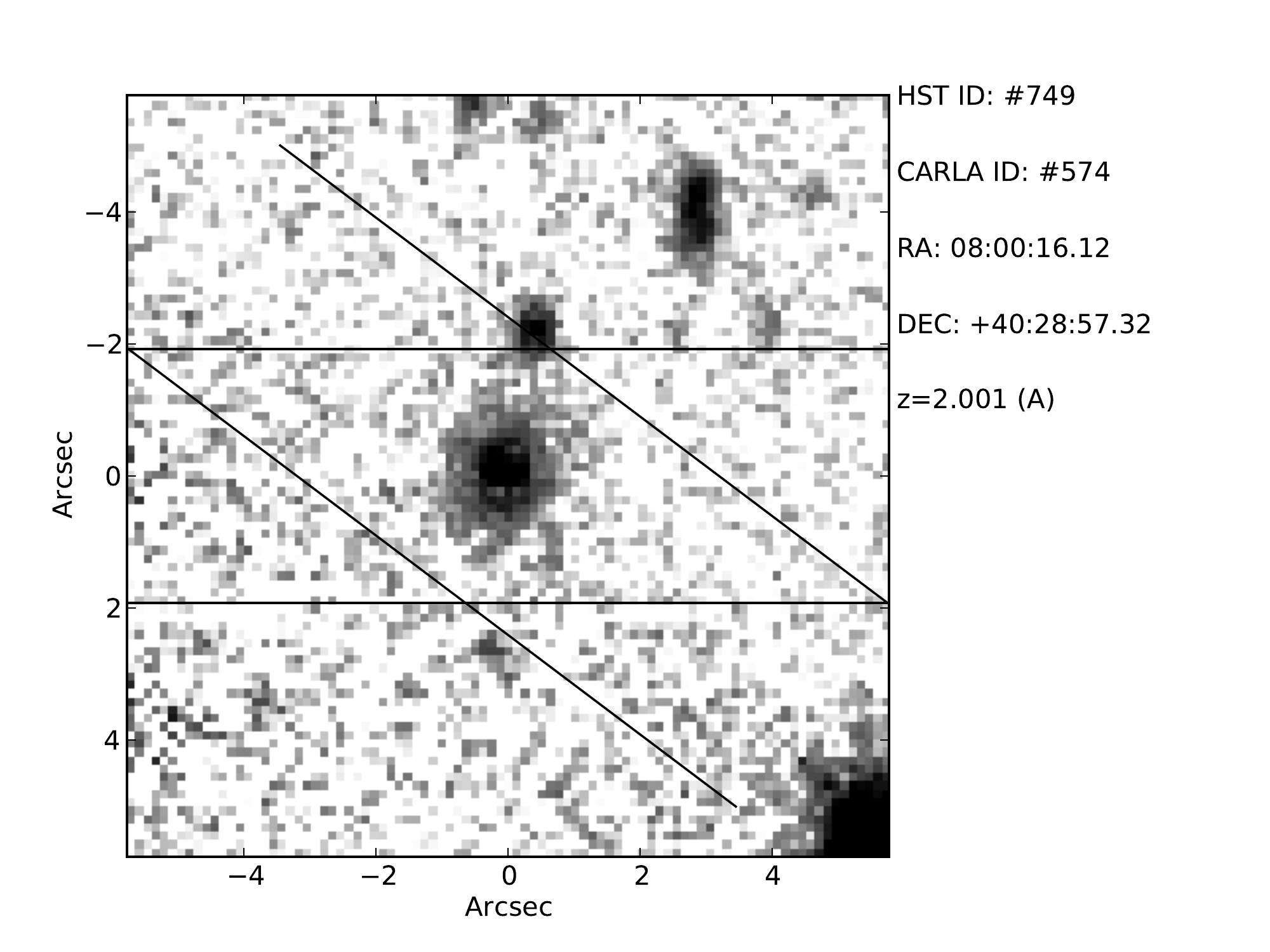}  \mbox{(s)}}%
}\\
{%
\setlength{\fboxsep}{0pt}%
\setlength{\fboxrule}{1pt}%
\fbox{\includegraphics[page=2, scale=0.45]{J0800_767.pdf} \hfill \includegraphics[page=1, scale=0.4]{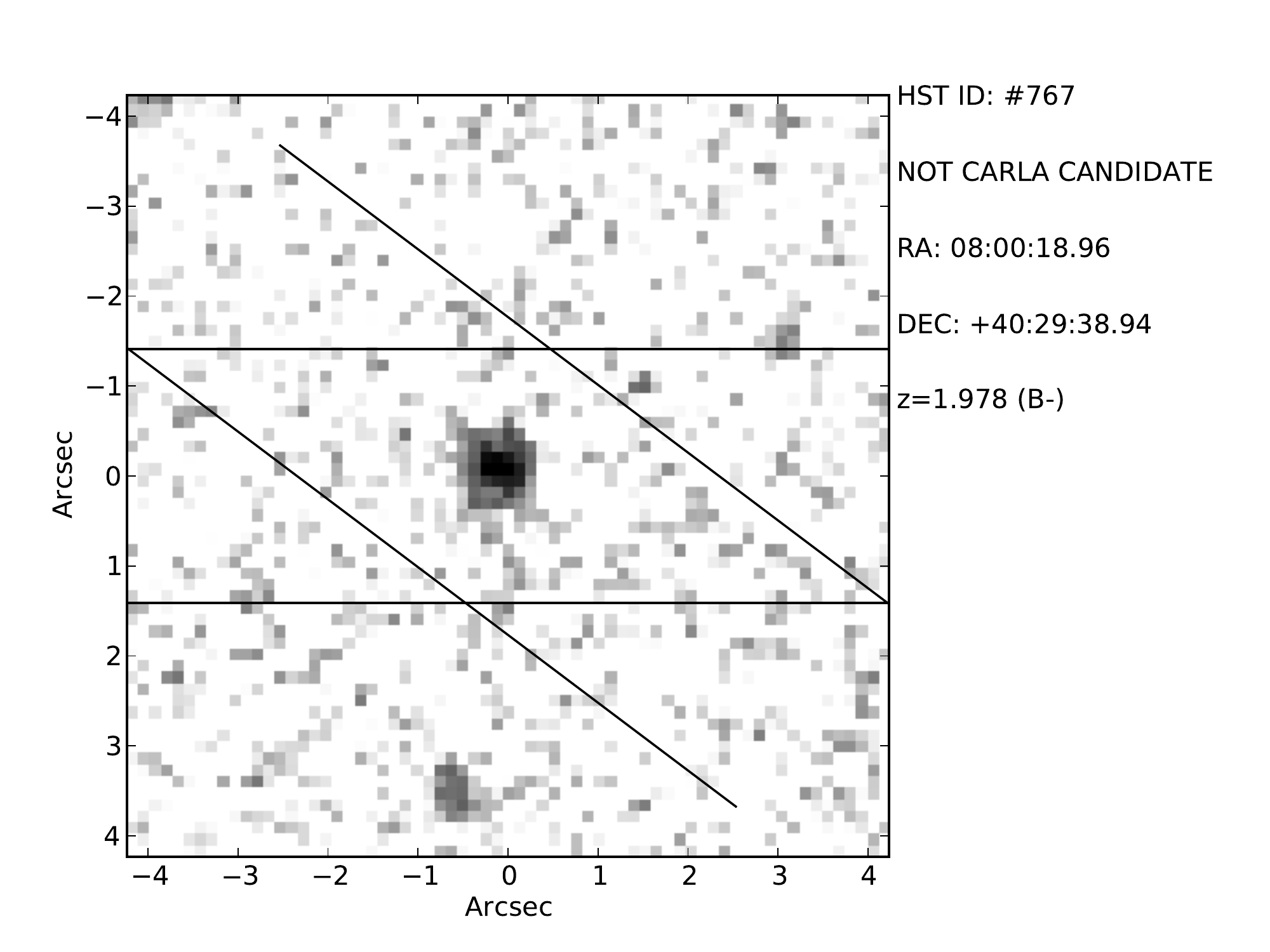}  \mbox{(t)}}%
}\\
\textbf{\mbox{}\\ Figure \ref{fig:spectra}} --- Continued.
\end{figure*}

\end{document}